\documentclass[12pt]{article}
\usepackage{amsmath}
\usepackage[dvips]{epsfig}
\usepackage{amsmath, amsthm, amssymb,bm}
\usepackage{xcolor}
\usepackage{multirow}
\usepackage{graphicx}
\usepackage{subfigure}
\usepackage{epstopdf}
\usepackage[round,sort&compress]{natbib}
\usepackage{caption}
\usepackage{listings}
\usepackage{makecell}
\usepackage[bottom]{footmisc}
\usepackage[toc,title]{appendix}
\usepackage{xr}
\newtheorem{lemma}{Lemma}

\numberwithin{equation}{section}
\newtheorem{thm}{Theorem}[section]

\usepackage{geometry}
\begin{document}

\def\spacingset#1{\renewcommand{\baselinestretch}%
{#1}\small\normalsize} \spacingset{1.2}

\title{A Pairwise Hotelling Method for Testing High-Dimensional Mean Vectors}
\author{Zongliang Hu$^1$, Tiejun Tong$^{2}$ and Marc G. Genton$^{3}$ \\
\small $^1$College of Mathematics and Statistics, Shenzhen University, Shenzhen, China \\
\small $^2$Department of Mathematics, Hong Kong Baptist University, Hong Kong \\
\small $^3$Statistics Program,
King Abdullah University of Science and Technology, Thuwal, \\
\small  Saudi Arabia \\
}

\date{}
\maketitle

\begin{abstract}
\noindent
For high-dimensional small sample size data, Hotelling's $T^2$ test is not applicable for testing mean vectors due to the singularity problem in the sample covariance matrix.
To overcome the problem, there are three main approaches in the literature.
Note, however, that each of the existing approaches may have serious limitations and only works well in certain situations.
Inspired by this, we propose a pairwise Hotelling method for testing high-dimensional mean vectors, which, in essence, provides a good balance between the existing approaches.
To effectively utilize the correlation information, we construct the new test statistics as the summation of Hotelling's test statistics for the covariate pairs with strong correlations and the squared $t$ statistics for the individual covariates that have little correlation with others. We further derive the asymptotic null distributions and power functions  for the proposed  Hotelling tests under some regularity conditions.
Numerical results show that our new tests are able to control the type I error rates, and can  achieve a higher statistical power compared to existing methods,
especially when the covariates are highly  correlated.
Two real data examples are also analyzed and they both demonstrate the efficacy of our pairwise Hotelling tests.
\end{abstract}

\newpage

\spacingset{1.2}

\section{Introduction}\label{sec1}

A fundamental problem in multivariate statistics  is to
test whether a mean vector is equal to a given vector for the one-sample test,
or to test whether two mean vectors are equal for the two-sample test.
To start with, let $\bm{\mu}$ and $\Sigma$ be the mean vector and covariance matrix of
a random vector ${\bm{X}}$, respectively.
For the one-sample case, we are interested in testing the hypothesis
\begin{align}\label{onesampleT}
H_0:  {\bm \mu}={\bm \mu}_0
~~~~~{\rm versus}~~~~~  H_1: {\bm \mu} \neq {\bm \mu}_0,
\end{align}
where ${\bm \mu}_0=(\mu_{01}, \ldots, \mu_{0p})^T$ is a given vector,
$p$ is the dimension, and the superscript $T$ denotes the transpose of a vector or a matrix.
Assume that  ${\bm{X}}_{k}=\left(X_{k1},\ldots,X_{kp}\right)^T $ for $k=1,\ldots,n$
are independent copies of ${\bm{X}}=(X_1,\ldots,X_p)^T$,
where $n$ is the sample size.
Then for testing hypothesis (\ref{onesampleT}), under the assumption of data normality,
the classical  Hotelling's $T^2$ test \citep{Hotelling1931}  is given as
\begin{align*}\label{HT-square}
T^2= n( \bm{\bar{X}}-{\bm \mu}_0 )^T S^{-1}( \bm{\bar{X}}-{\bm \mu}_0 ),
\end{align*}
where $\bm{\bar{X}}=\sum_{k=1}^{n}\bm{X}_k/n$ is the sample mean vector, and
$S=\sum_{k=1}^{n} (\bm{X}_k- \bm{\bar{X}})
(\bm{X}_k- \bm{\bar{X}})^T/(n-1)$ is the sample covariance matrix.

In the era of big data, high-dimensional data are increasingly
collected from various fields with a wide range of applications.
For high-dimensional  data, the dimension is
usually larger or much larger  than the sample size,
 and the resulting ``large $p$ small $n$" paradigm poses new
challenges for the testing problem (\ref{onesampleT}).
As an example, when testing whether two gene sets,
or pathways, have equal expression levels under two
experimental conditions, one may encounter the scenario
in which  the number of genes ($p$)  is much larger than the number of samples ($n$).
For high-dimensional small sample size data, as  pointed out by  \citet{bai1996},
Hotelling's $T^2$ test will not be applicable due to
the singularity problem in the sample covariance matrix.

To overcome the singularity problem in high-dimensional settings, a number of methods have been developed in  recent literature for remedying  Hotelling's $T^2$ test. In view of the different approaches in handling the non-invertible sample covariance matrix $S$, there are three main categories of testing methods as follows:

\begin{enumerate}
  \item[(1)] In the first category, researchers substituted  the sample covariance matrix $S$ with the $p\times p$ identity matrix $I_p$. This leads to the {\it unscaled Hotelling's tests} (UHT) with the test statistic as $$T_{\rm UHT}^2 = n(\bm{\bar{X}}-{\bm \mu}_0)^T (\bm{\bar{X}}- {\bm \mu}_0).$$ For references, see, for example, \citet{bai1996}, \citet{chen2010}, \citet{ahmad2014}
      and \citet{park2017}. In addition, \citet{Xu2016} considered an adaptive testing procedure  with the  test statistic as
      $T(\gamma)=\sum_{j=1}^{p} ( \bar{X}_j- \mu_{0j} )^{\gamma}$, and   \citet{He2019} also follows the idea of UHT and proposed  a unified $U$-statistic for testing  mean vectors, covariance matrices and regression coefficients.

  \item[(2)] In the second category, researchers replaced the sample covariance matrix by a diagonal or block diagonal covariance matrix, and for which we refer to them as the   {\it diagonal Hotelling's tests} (DHT). Specifically, by letting $D ={\rm diag}(S)$ be the diagonal covariance matrix,  \citet{wu2006} introduced the test statistic
      $$T_{\rm DHT}^2 = n(\bm{\bar{X}}- {\bm \mu}_0 )^T D^{-1} ( \bm{\bar{X}}-{\bm \mu}_0 ).$$
      \citet{srivastava2008}  studied the limiting behaviors  of this test statistic under data normality.
      \citet{tony2014} considered a test based on the maximum of the  squared marginal $t$ statistics. 
      \citet{Hu2018test} proposed a likelihood ratio test based on a diagonal covariance matrix structure.
      \citet{feng2017} grouped the covariates into many small blocks, and then under the assumption that there is little correlation between these blocks, 
      they constructed their test statistic as the summation of Hotelling's $T^2$ statistics within each small block. 
      More studies on DHT include, for example, \citet{srivastava2009}, \citet{park2013test}, \citet{srivastava2013}, \citet{feng2015},  \citet{Carroll2015}, and \citet{Dong2015}.

  \item[(3)]  In the third category, researchers applied regularization methods for estimating the covariance matrix to overcome the singularity problem in the sample covariance matrix. We refer to these methods  as the {\it regularized  Hotelling's tests} (RHT).
      To name a few,  \citet{chen2011} proposed a ridge-type regularization with the test statistic as
$$T_{\rm RHT, 1}^2 = n( \bm{\bar{X}}- {\bm \mu}_0 )^T (S+ \lambda I)^{-1}
(\bm{\bar{X}} - {\bm \mu}_0 ).$$
This test statistic was also considered by \citet{li2016} for the two-sample testing problem.
\citet{lopes2011} proposed another regularized test statistic based on the random projection technique,  $$T_{\rm RHT, 2}^2 = n(\bm{\bar{X}}- {\bm \mu}_0  )^T P_R^T
(P_R S P_R^T)^{-1} P_R (\bm{\bar{X}}- {\bm \mu}_0 ),$$
where $P_R$ is a random matrix of size $k\times p$.
Further developments on the projection-based techniques include, for example, \citet{thulin2014}, \citet{srivastava2015}, and \citet{zoh2017powerful}.

\end{enumerate}

The UHT and DHT tests in the first two categories do not account for the correlations
among the covariates. When there are highly correlated covariates in the data, neither of the two methods may provide a valid test with a controlled type I error rate  and/or an acceptable statistical power.
In contrast, RHT in the third category is a universal way that attempts to  account for all the correlations within the covariance matrix. In other words, the ridge-type and projection-based statistics
did not take into account  the sparsity of the covariance matrix. Consequently, RHT may  not be able to provide a satisfactory performance when the sample size $n$ is relatively small compared to the dimension $p$ \citep{Dong2015}.
Besides,  \citet{li2017} considered a composite Hotelling's test (CHT) to account for the correlations. The author extracted the 2-dimensional pairs $(X_{i},X_{j})^T$ with $i < j$ from the $p$-dimensional vector ${\bm X}$, and then took the average of the classical Hotelling's test statistics for all the bivariate sub-vectors. Similarly  to others, when the covariance matrix is sparse and the sample size is small, CHT may not provide a satisfactory performance either.
This phenomenon was also reported in \citet{bickel2004}, where,
if the estimated correlations  are very noisy due to the small sample size, it is probably better not to estimate them at all.

To overcome the drawbacks in the aforementioned  tests, we propose a new category of testing methods to further advance the existing literature on testing high-dimensional mean vectors.
Our main idea is to take the advantages of the second and third categories and provide a good balance between them. Specifically, to effectively utilize the correlation information,
we first construct the classical Hotelling's statistics for the  covariate pairs with strong correlations,
whereas, for the individual covariates that have little correlation with others, we apply the squared $t$ statistics
to account for their respective contributions to the multivariate testing problem.
Our new test statistics are the summation over all the Hotelling's statistics and the squared $t$ statistics.
Consequently, they  are able to capture sufficient dependence information among the components and, at the same time, account for the sparsity of covariance matrices. We further derive the asymptotic null distributions and power functions of the new statistics, and  investigate the regularity conditions that are needed for establishing the asymptotic results of the proposed test statistics. Simulation studies and real data analyses show that our proposed tests outperform the existing methods in a wide range of settings.

The rest of the paper is organized as follows. In Section 2, we propose the pairwise Hotelling's testing method for the one-sample test. The asymptotic distributions of the test statistic are also derived under, respectively, the null and local alternative hypotheses. In Section 3, we propose the pairwise Hotelling's testing method for the two-sample test, and derive the asymptotic results including the asymptotic null distribution and power function. In Section 4, we conduct simulation studies to evaluate the proposed tests and compare them with the existing methods. We then apply the proposed tests to two real data examples in Section 5, and conclude the paper in Section 6 with a brief summary and some future work. The technical details are provided in the Appendices.

\section{One-Sample Test}\label{one-sample}

In this section, we consider the one-sample testing  problem (\ref{onesampleT})
under the ``large $p$ small $n$" paradigm.
Recall that Hotelling's $T^2$ test is not applicable when the dimension is larger than the sample size.  To overcome the singularity problem, one possible approach is to downsize the dimension of the sample covariance matrix.

To achieve this, we decompose the $p$-dimensional vector $\bm{X}$ into a series of  bivariate sub-vectors  $(X_{i},X_{j})^T$ with $i < j$.
We then  apply the bivariate  Hotelling's  test statistic  to account for their pairwise correlation as
\begin{align*}
T_{ij}^2
&=(\bar{X}_{i}-\mu_{0i}, \bar{X}_{j}-\mu_{0j})
\left(
\begin{array}{cc}
    s_{ii} & s_{ij}\\ s_{ji}& s_{jj}
\end{array}
\right)^{-1}
(\bar{X}_{i}-\mu_{0i}, \bar{X}_{j}-\mu_{0j})^T  \\
&=(\bar{\bm {X}}-\bm{\mu}_0)^T P_{ij}^T
( P_{ij} S P_{ij}^T )^{-1}
P_{ij} (\bar{\bm {X}}-\bm{\mu}_0),
\end{align*}
where $\bar{X}_i=\sum_{k=1}^{n}X_{ki}/n$ is the sample mean of the
$i$th covariate, $s_{ij}$ is the sample covariance of the $i$th and $j$th covariates,
and
$P_{ij}=
 \left(
   \begin{array}{ccccccc}
     0 &\cdots&1&\cdots &0& \cdots &0\\
     0 &\cdots&0&\cdots &1& \cdots &0\\
   \end{array}
 \right)$  is a $2\times p$ matrix with the $(1,i)$ and $(2,j)$ components being 1 and all others  being $0$.
Finally,  the following $U$-type test statistic
can be applied to accumulate  all the pairwise correlations among the covariates:
\begin{equation}\label{W_stat}
W_1 = n\sum_{j=2}^{p}\sum_{i=1}^{j-1}T_{ij}^2
   =n ({\bm{\bar X}} - \bm{\mu}_0 )^T \left(  \sum_{j=2}^{p}\sum_{i=1}^{j-1}
P_{ij}^T
( P_{ij} S P_{ij}^T )^{-1}
P_{ij}\right) ({\bm{\bar X}} - \bm{\mu}_0).
\end{equation}

The pairwise idea in the test statistic $W_1$ can be traced back to the pairwise likelihood methods.
For likelihood-based inference involving distributions with high-dimensional dependencies,
it can be a powerful approach to apply the approximate likelihoods based
on the bivariate marginal distributions
(\citealp{cox2004note}, \citealp{varin2011}, \citealp{li2017}).
It is also worth noting that, as long as $n\geq 3$, the pairwise method  in (\ref{W_stat}) is always applicable and so it resolves the singularity problem in the original Hotelling's $T^2$ test.

\subsection{Pairwise Hotelling's test statistic}
For high-dimensional data, it is often the case that the covariance matrix is sparse, in which  only a small  proportion of the correlations will be non-zero. In such settings, the $U$-type test  statistic $W_1$ will involve many noisy terms.  Consequently, the test  may not provide a sufficiently large power, in particular when $n$ is relatively small compared to $p$.

To further improve the test statistic (\ref{W_stat}), we propose a thresholding method  by  shrinking  the small estimates of correlations to zero
to reduce the noise level in $W_1$.
To be more specific, we consider a screening procedure based on Kendall's tau correlation matrix. Let $R =(r_{ij})_{1\leq i,j \leq p}\in \mathbb{R}^{p\times p}$ be Kendall's tau correlation matrix,
and $\Gamma =(\tau_{ij})_{1\leq i,j \leq p}\in \mathbb{R}^{p\times p}$ with $\tau_{ij}=|r_{ij}|$,
where $|\cdot|$ is the absolute value function.
Let also
$$A_1=\{(i,j): \tau_{ij}> \tau_0 {~\rm and~} i<j  \}
~~{\rm and}~~A_2=\{i:   \tau_{ij}<\tau_0 {\rm~ for~all~} j\neq i  \}$$
be two sets of indices, where $\tau_0\in [0,1]$ is a pre-specified threshold.
Clearly, the covariate pairs with strong correlations fall into $A_1$,
and the individual covariates with little correlation with others
fall into $A_2$.  In practice,  $R$, $A_1$ and $A_2$ are  all unknown and need to be estimated from the sample data.

Assume that  $\hat{R}$ is Kendall's tau sample correlation matrix.
Then with a given  $\tau_0$,  the sample estimates of  $A_1$ and $A_2$  are, respectively,
$$\hat{A}_1=\{(i,j): \hat{\tau}_{ij}> \tau_0 {~\rm and~} i<j\}~~
{\rm and}~~\hat{A}_2=\{i: \hat{\tau}_{ij}<\tau_0 {\rm~ for~all~} j\neq i  \},$$
where $\hat{\tau}_{ij}=|\hat{r}_{ij}|$.
In addition, let $\bm{X}_{ij;k}=(X_{ki}, X_{kj})^T\in \mathbb{R}^2$
be the $k$th sample of $(X_{i}, X_{j})^T$,  $\bar{\bm{X}}_{\{i,j\}}$
be the sample mean  vector,
and $S_{\{i, j\}}$  be the sample covariance matrix of ${\bm{X}}_{ij;k}$.
Then for testing hypothesis (\ref{onesampleT}),
the thresholding test statistic can be represented as
\begin{equation*}
W_1(\tau_0)=n\sum_{(i,j) \in \hat{A}_1}
\Big(\bar{\bm{X}}_{\{i,j\}}-{\bm \mu}_{0,\{i,j\}}\Big)^T
 S_{\{i,j\}}^{-1}\Big(\bar{\bm{X}}_{\{i,j\}}-{\bm \mu}_{0,\{i,j\}}\Big)
+n\sum_{i \in \hat{A}_2} \frac{{(\bar{x}_i-\mu_{0i})^2}}{s_{ii}},
\end{equation*}
where ${\bm \mu}_{0,\{i,j\}}=(\mu_{0i},\mu_{0j})^T$.
The test statistic $W_1(\tau_0)$ has fully taken into account the pairwise  correlations among the covariates.
Specifically, we  apply  Hotelling's test statistics
to account for the contributions from the covariate pairs with strong correlations (i.e., for all  $(i,j) \in \hat{A}_1$),
and apply the squared $t$ statistics to account for the contributions from
the individual covariates with little correlation with others
(i.e., for all $i \in \hat{A}_2$).

Let $P_{i}=(0,\ldots,1,\ldots,0)$, where the $i$th component is 1 and all others are $0$. Let also $\widehat{P}_{\mathcal{O}}= \sum_{(i,j) \in \hat{A}_1} P_{ij}^T
(P_{ij} S P_{ij}^T)^{-1}P_{ij}
+\sum_{i \in \hat{A}_2} P_{i}^T (P_{i}S P_{i}^T)^{-1} P_{i}$.
With the new notations, we can rewrite  $W_1(\tau_0)$ as
 \begin{eqnarray*}
W_1(\tau_0)= n(\bar{\bm{X}}-{\bm \mu}_0)^T\widehat{P}_{\mathcal{O}} (\bar{\bm{X}}-{\bm \mu}_0).
\end{eqnarray*}
For simplicity, we also let  $P_{\mathcal{O}} =\sum_{(i,j) \in A_1} P_{ij}^T (P_{ij}\Sigma P_{ij}^T)^{-1}P_{ij}
 + \sum_{i \in A_2} P_{i}^T ( P_{i}\Sigma P_{i}^T)^{-1} P_{i}$ be the unknown population value of $\widehat {P}_{\mathcal{O}}$.
Note that $W_1(\tau_0)$ involves the terms
$(\bm{X}_s-{\bm \mu}_{0})^T \widehat{P}_{\mathcal{O}} (\bm{X}_s-{\bm \mu}_{0})$, $s=1,\ldots, n$, and they will introduce higher order moments in the centering and scaling parameters when establishing the limiting distributions.
Hence  to stabilize the test statistic, we apply  the leave-one-out method and propose the new test statistic as
\begin{eqnarray}\label{PHT1}
T_1(\tau_0)=\frac{1}{n(n-1)}\sum_{s=1}^{n}\sum_{t \neq s }^{n} (\bm{X}_s-{\bm \mu}_{0})^T
\widehat{P}_{\mathcal{O}}^{(s,t)}(\bm{X}_t-{\bm \mu}_{0}),
\end{eqnarray}
where
$\widehat{P}_{\mathcal{O}}^{(s,t)}=\sum_{(i,j) \in \hat{A}_1 }
P_{ij}^T (P_{ij} S^{(s,t)} P_{ij}^T)^{-1}P_{ij}
+\sum_{i \in \hat{A}_2 }P_{i}^T ( P_{i}S^{(s,t)} P_{i}^T)^{-1} P_{i},$
and $S^{(s,t)}$ is the sample covariance matrix without observations $\bm{X}_s$ and $\bm{X}_t$.
We refer to our new test statistic in (\ref{PHT1}) as the pairwise
Hotelling's test statistic, or, for short,  the PHT statistic.
As a special case, if we set $\tau_0=1$,
then $\hat{A}_1=\emptyset$ and $\hat{A}_2=\{1,\ldots,p\}$
so that the PHT statistic reduces  to the diagonal Hotelling's test in \citet{park2013test}.
On the other hand, if we set $\tau_0=0$,
then $\hat{A}_1=\{(i,j): i<j\}$ for $i,j=1,\ldots,p$ and $\hat{A}_2=\emptyset$;
that is, the PHT statistic accounts for all the correlations in the covariance matrix so that it is the same as  the test statistic $W_1$ in (\ref{W_stat}).

\subsection{Asymptotic results}
Following the assumptions in \citet{chen2010},
we assume that the random vector ${\bm X}=(X_1,\dots,X_p)^T$ follows the linear model:
\begin{eqnarray}\label{modelChen1}
{\bm X}=C_1 {\bm Z} + \bm{\mu},
\end{eqnarray}
where $C_1\in \mathbb{R}^{p \times q}$ with $q \geq p$ such that $\Sigma=C_1 C_1^T$, $\bm{\mu}=(\mu_1,\ldots,\mu_p)^T$,
and the random vector ${\bm Z}$ satisfies that $E({\bm Z})={\bm 0}$
and $\text{Var}({\bm Z})=I_q$.
In addition for  ${\bm Z}=(Z_{1},\ldots,Z_{q})^T$,
we assume that the following moment  conditions hold:
$E(Z_{j}^4)=3+\Delta_1< \infty$ where $\Delta_1$ is a positive constant,
and
\begin{eqnarray*}
E(Z_{l_1}^{\alpha_1} Z_{l_2}^{\alpha_2} \cdots Z_{l_k}^{\alpha_k} )
=E(Z_{l_1}^{\alpha_1})E(Z_{l_2}^{\alpha_2}) \cdots E(Z_{l_k}^{\alpha_k}),
\end{eqnarray*}
where $k$ is a positive integer such that $\alpha_1+\cdots+\alpha_k\leq 8$,
and $l_1 \neq l_2 \neq \cdots \neq l_k$.

We further assume that $\{(X_i,X_j): i, j=1,2,\ldots,p {\rm~with~} i\neq j\}$ is a two-dimensional random field,
and define the $\rho$-mixing coefficient for $X=\{X_j, j=1,2,\ldots,p\}$ as
$$\rho(s)=\sup \Big\{ |{\rm Corr}(g_1,g_2)|: g_1 \in \mathcal{L}_2(X(A_3)),
g_2\in \mathcal{L}_2(X(A_4)), {\rm dist}(A_3,A_4)\geq s \Big\}$$
over any possible sets $A_3, A_4 \subset \{1,2,\ldots,p \}$ with
$\text{card}(A_3)\leq 2$ and $\text{card}(A_4)\leq 2$,
where  $\text{card}(\cdot)$ is the operator that counts the number of elements in
a given set,
${\rm dist}(A_3,A_4)=\min_{i\in A_3, j \in A_4}|i-j|$ is the distance
between $A_3$ and $A_4$,  ${\rm Corr}(g_1,g_2)$ is the correlation between $g_1$ and $g_2$, and $\mathcal{L}_2(X(E))$ is the set of all measurable functions defined on the $\sigma$-algebra generated by $X$ over $E \subset \{1,2,\ldots,p\}$ with the existence of the second moment.

To establish the asymptotic null and alternative distributions of the proposed test statistic,
we also need the following conditions:
\begin{itemize}
\item[(C1)] There exists a finite positive number $\bar{K}_1$ such that $1/\bar{K}_1  \leq \lambda_{p}(\Sigma) \leq \cdots \leq  \lambda_{1}(\Sigma) \leq \bar{K}_1$, where $\lambda_{i}(\Sigma)$ is the $i$th largest eigenvalue of $\Sigma$.
\item[(C2)] Assume that $\{X_{j}: j \geq 1\}$
is a $\rho$-mixing sequence such that $\rho(s)\leq \varpi_0\exp{(-s)}$,
where $\varpi_0>0$ is a constant.

\item[(C3)] There exists an oracle constant $\tau^* \in (0,1)$ such that,
for a finite positive integer ${K}_0$,
${\sup}_{i\leq p}{\text{card}(A_{i}^*)} \leq K_0$,
where  $A_{i}^*=\{j: \tau_{ij}>\tau^* \}$.
In addition,
we assume that $\underset{i,j=1,\ldots,p}{\lim\inf}\{\tau_{ij} | \tau_{ij}>\tau^* \}>\tau^*$ and $\underset{i,j=1,\ldots,p}{\lim\sup}\{\tau_{ij} | \tau_{ij}<\tau^* \}<\tau^*$.

\item[(C4)] There exists a positive integer $m_0\geq 4$ such that
the higher order moments, $E({X}_{1}^{4 m_0+2}),\ldots, E({X}_{p}^{4 m_0+2})$, are bounded uniformly,
which indicates that there exists a constant $\varpi_1>0$, such that $E({X}_{kj}^{4 m_0+2})<\varpi_1$ holds for
$j=1,\ldots,p$. In addition, we assume that
$E\big\|S_{\{i,j\}}^{-1} \big\|^8$ for $(i,j)\in A_1$
and $E(s_{jj}^{-8})$ for $j \in A_2$ are bounded uniformly,
where $\big\| \cdot  \big\|$ is the Frobenius norm.

\item[(C5)]  Assume that $\bm{\mu}^T  P_{\mathcal{O}}  \bm{\mu} =o(\sqrt{p/n})$.
There exists a constant $\varpi_2>0$ such that $|\mu_j-\mu_{0j}|^2\leq \varpi_2 /\sqrt{n}$.
\end{itemize}

Condition (C1) assumes that the eigenvalues are bounded uniformly away from
0 and $\infty$, which  is the same condition as in  \citet{tony2014} and \citet{Xu2016}.
Condition (C2) is the so-called $\rho$-mixing condition, which follows from \citet{zhengyan1997limit} that implies a weak dependence structure of the data.
The weak dependence structure is commonly assumed  in many genome-wide association studies.
As an  example, single nucleotide polymorphisms (SNPs) have a local dependence structure in which the correlations between SNPs
often decay rapidly as the distances between gene locus increase.
Condition (C3) assumes  that our PHT statistic allows the number of covariate pairs with strong correlations to increase at the same order of $p$.
Conditions  (C4) and (C5) are two technical conditions that are needed for deriving  the asymptotic results of the proposed  test statistic.

\vskip 10pt
\begin{thm}\label{pro1}
Assume that $\tau_0$ satisfies
$\underset{i,j=1,\ldots,p}{\lim\inf}\{\tau_{ij} | \tau_{ij}>\tau_0 \}>\tau_0$
and $\underset{i,j=1,\ldots,p}{\lim\sup}\{\tau_{ij} | \tau_{ij}<\tau_0 \}<\tau_0$.
Let $\hat{A}_1$ and $\hat{A}_2$ be the two sets based on the threshold $\tau_0$
in the screening procedure.
Then for any given positive integer $m_0$, if $p=O(n^{m_0})$,
we have $$P(\hat{A}_2= A_2)\geq  P(\hat{A}_1 = A_1)  \to 1
{\rm ~~as~~}n \to \infty.$$
\end{thm}

\vskip 10pt
The proof of Theorem \ref{pro1} is given in Appendix \ref{Prof.pro1}.
This theorem  shows  that, when the sample size becomes large, the selected  sets
$\hat{A}_1$ and $\hat{A}_2$ based on the sample data are consistent
estimates of $A_1$ and $A_2$, respectively.

\vskip 10pt
\begin{thm}\label{th1}
Assume that $\tau_0 \geq \tau*$,  $\underset{i,j=1,\ldots,p}{\lim\inf}\{\tau_{ij} | \tau_{ij}>\tau_0 \}>\tau_0$,
and $\underset{i,j=1,\ldots,p}{\lim\sup}\{\tau_{ij} | \tau_{ij}<\tau_0 \}<\tau_0$.
Then under conditions (C1)--(C5) and if $p=o(n^{m_0-1})$
with $m_0$ defined in (C4), we have
\begin{equation*}
\frac{T_1(\tau_0)-{\bm \delta}_1^T P_{\mathcal{O}} {\bm \delta}_1}{\sqrt{2n^{-2}{\rm tr}(\Lambda_1^2)}}
\stackrel{D}{\longrightarrow}N(0,1) {\rm ~~as~~} n\to \infty,
\end{equation*}
where
${\bm \delta}_1={\bm \mu}-{\bm \mu}_0$,
$\Lambda_1  =\Sigma^{1/2} P_{\mathcal{O}} \Sigma^{1/2}$,
 $\rm{tr}(\cdot)$ is the trace function, and $\stackrel{D}{\longrightarrow}$ denotes convergence in distribution.
\end{thm}

\vskip 10pt
The proof of Theorem \ref{th1} is given in Appendix  \ref{Prof.th1}.
This theorem shows that, despite  the exact threshold $\tau*$ being unknown,
we can select a larger threshold $\tau_0 > \tau*$,  such that
if $\tau_0$ satisfies $\underset{i,j=1,\ldots,p}{\lim\inf}\{\tau_{ij} | \tau_{ij}>\tau_0 \}>\tau_0$ and $\underset{i,j=1,\ldots,p}{\lim\inf}\{\tau_{ij} | \tau_{ij}<\tau_0 \}<\tau_0$, then the  test statistic $T_1(\tau_0)$ still converges to the standard normal distribution after proper centering and scaling.

To apply  Theorem \ref{th1}  for practical use, we need a ratio consistent estimator for the unknown  ${\rm tr}(\Lambda_1^2)$.
For this purpose, we establish  the following lemma,
with the proof in Appendix \ref{Prof.lemma1}.

\vskip 10pt
\begin{lemma}\label{lemma1.1}
Assume that $p=o(n^{3})$, and   $\tau_0$ satisfies the assumptions in Theorem \ref{th1}.
Then under conditions (C1)--(C5),
\begin{eqnarray*}
\widehat{{\rm tr}(\Lambda_1^2)}
&=& \frac{1}{n(n-1)} \sum_{s \neq t}^{n} ( {\bm X_{s}} -{\bar{\bm X}^{(s,t)}})^T \widehat{P}_{\mathcal{O}}^{(s,t)} {\bm X}_{t}
( {\bm X_{t}} -{\bar{\bm X}^{(s,t)}})^T \widehat{P}_{\mathcal{O}}^{(s,t)} {\bm X}_{s}
\end{eqnarray*}
is a ratio consistent estimator of ${\rm tr}(\Lambda_1^2)$,
where ${\bar{\bm X}^{(s,t)}}$ is the sample mean vector without  observations $\bm{X}_s$ and $\bm{X}_t$. Consequently, under the null hypothesis in (\ref{onesampleT}),
\begin{equation*}
\frac{T_1(\tau_0)}{\sqrt{2n^{-2}\widehat{{\rm tr}(\Lambda_1^2)} }}
\stackrel{D}{\longrightarrow}N(0,1) {\rm ~~as~~} n\to \infty.
\end{equation*}
\end{lemma}

\vskip 10pt

Also by  Theorem \ref{th1}, the power function of the PHT statistic for the one-sample test is given as
\begin{equation}\label{U.power}
{\rm Power}(\bm{\delta}_1)=\Phi \Big(-z_{\alpha}+
\frac{{\bm \delta}_1^T P_{\mathcal{O}} {\bm \delta}_1 }{\sqrt{2n^{-2}
{{\rm tr}(\Lambda_1^2)}}}\Big),
\end{equation}
where  $\Phi(x)$ is the cumulative distribution
function of the standard normal distribution.
The performance of the new test depends on the quantities
${\bm \delta}_1^T P_{\mathcal{O}}{\bm \delta}_1$
and ${{\rm tr}(\Lambda_1^2)}$.
Theoretically, a reasonable choice of the threshold $\tau_0$
can be to maximize
${\rm Power}(\bm{\delta}_1)$ so that the PHT statistic achieves
the highest asymptotic power.
However, this maximization procedure is infeasible in practice,
since ${\bm \delta}_1^TP_{\mathcal{O}}{\bm \delta}_1/ \sqrt{ {\rm tr}({\Lambda}_1^2)}$ involves unknown quantities including  ${\bm \delta}_1$ and $\Sigma$. We further the practical choice  of $\tau_0$ in Section 4.3.

\section{Two-Sample Test}\label{two-sample}
This section considers  the two-sample test for  mean vectors with equal covariance matrices.
Let $\{\bm{X}_s=(X_{s1},\ldots, X_{sp})^T\}_{s=1}^{n_1}$ and
$\{\bm{Y}_t=(Y_{t1},\ldots, Y_{tp})^T\}_{t=1}^{n_2}$ be
two groups of independent and identically distributed (i.i.d.) random vectors from two independent  multivariate populations.
Let also  $E(\bm{X}_s)=\bm{\mu}_1=(\mu_{11},\ldots,\mu_{1p})^T$
be the mean vector of the first population, $E(\bm{Y}_t)=\bm{\mu}_2=(\mu_{21},\ldots,\mu_{2p})^T$
be the mean vector of the second population,
and $\Sigma$ be  the common covariance matrix
 for both populations.
For the two-sample test, we are interested in testing the hypothesis
\begin{equation}\label{H0}
  H_0: \bm{\mu}_1= \bm{\mu}_2{\rm~~versus~~} H_1: \bm{\mu}_1 \neq  \bm{\mu}_2.
\end{equation}

\subsection{Pairwise Hotelling's test statistic}
Following the similar notations as those for the one-sample test,
we let Kendall's tau correlation matrix be $R =(r_{ij})_{1\leq i,j \leq p}\in \mathbb{R}^{p\times p}$,
and  $\Gamma =(\tau_{ij})_{1\leq i,j \leq p}\in \mathbb{R}^{p\times p}$ with $\tau_{ij}=|r_{ij}|$.
Let also
$$A_1=\{(i,j): \tau_{ij}> \tau_0 {~\rm and~} i<j  \}
~~{\rm and}~~A_2=\{i:   \tau_{ij}<\tau_0 {\rm~ for~all~} j\neq i  \}$$
be two sets of indices, where $\tau_0\in [0,1]$ is a pre-specified threshold, and denote
$P_{\mathcal{O}} =\sum_{(i,j) \in A_1} P_{ij}^T (P_{ij}\Sigma P_{ij}^T)^{-1}P_{ij}
 + \sum_{i \in A_2} P_{i}^T ( P_{i}\Sigma P_{i}^T)^{-1} P_{i}.$

Assume that  $\hat{R}_1=(\hat{r}_{ij,1})_{1\leq i,j \leq p}\in \mathbb{R}^{p\times p}$
and $\hat{R}_2=(\hat{r}_{ij,2})_{1\leq i,j \leq p}\in \mathbb{R}^{p\times p}$
are Kendall's tau sample correlation matrices of the two groups,  respectively.
For simplicity, let  $N=n_1+n_2$, and assume that $n_1/N \to \varphi_0 \in (0,1)$ as $N \to \infty$.
Then with a given  $\tau_0$, the sample estimates of
$A_1$ and $A_2$  are, respectively,
$$\hat{A}_1=\{(i,j): \hat{\tau}_{ij}> \tau_0 {~\rm and~} i<j\}
~~{\rm and}~~
\hat{A}_2=\{i: \hat{\tau}_{ij}<\tau_0 {\rm~ for~all~} j\neq i  \},$$
where $\hat{\tau}_{ij}=({n_1} \hat{\tau}_{ij,1}+{n_2}\hat{\tau}_{ij,2})/N$,
$\hat{\tau}_{ij,1}=|\hat{r}_{ij,1}|$
and $\hat{\tau}_{ij,2}=|\hat{r}_{ij,2}|$.
In addition, we need the following notations related to the sample covariance matrices:
\begin{itemize}
  \item[(1)] Let $S_1$ (or $S_2$) be the sample covariance matrix of group 1 (or group 2),  $S_{1}^{(s)}$ (or $S_{2}^{(s)}$) be the sample covariance matrix of group 1 (or group 2) without observation $\bm{X}_s$ (or $\bm{Y}_s$), and $S_{1}^{(s,t)}$ (or $S_{2}^{(s,t)}$) be the sample covariance matrix of group 1 (or group 2)  without observations $\bm{X}_s$ and $\bm{X}_t$ (or $\bm{Y}_s$ and $\bm{Y}_t$).
  \item[(2)] Let $s_{1,jj}$ (or $s_{2,jj}$) be the sample variance of $X_{kj}$
      (or $Y_{kj}$), and $S_{1,\{ij\}}$ and $S_{2,\{ij\}}$ be the sample covariance matrix of $(X_{ki},X_{kj})^T$ and $(Y_{ki},Y_{kj})^T$, respectively.
       Let also  $s_{1,jj}^{(s,t)}$ (or $s_{2,jj}^{(s,t)}$) be the sample variance of $X_{kj}$
      (or $Y_{kj}$) without observations ${X}_{sj}$ and ${X}_{tj}$ (or ${Y}_{sj}$ and ${Y}_{tj}$).
  \item[(3)]  Let  $S_{1*}^{(s,t)}= [(n_1-2)S_1^{(s,t)}+  n_2 S_2]/(N-2)$
     be the pooled sample covariance matrix without observations $\bm{X}_s$ and $\bm{X}_t$ in group 1, and
    $S_{2*}^{(s,t)}= [n_1 S_1 + (n_2-2) S_2^{(s,t)}] /(N-2)$ be the pooled sample covariance matrix without observations $\bm{Y}_s$ and $\bm{Y}_t$ in group 2.
  \item[(4)] Let $S_{12}=[(n_1-1)S_1 +(n_2-1)S_2]/(N-2)$ be the pooled sample covariance matrix of the two groups, and $S_{12,*}^{(s,t)}=[(n_1-1)S_1^{(s)} + (n_2-1)S_2^{(t)}]/(N-2)$  be the pooled sample covariance matrix without $\bm{X}_s$ in group 1 and $\bm{Y}_t$ in group 2.
\end{itemize}

Following the similar arguments as in (\ref{W_stat}), we can propose the
$U$-type test statistic for the two-sample test as
\begin{equation}\label{W_stat2}
W_2 =\frac{n_1+n_2}{n_1n_2} ({\bm{\bar X}} - {\bm{\bar Y}} )^T \left(  \sum_{j=2}^{p}\sum_{i=1}^{j-1}
P_{ij}^T
( P_{ij} S_{12} P_{ij}^T )^{-1}
P_{ij}\right) ({\bm{\bar X}} - {\bm{\bar Y}}),
\end{equation}
where ${\bm{\bar X}}$ and ${\bm{\bar Y}}$ are the sample mean vectors of the two groups.
Further by  the screening procedure and the leave-one-out method, our PHT statistic
for the two-sample test can be proposed  as
\begin{eqnarray}\label{stat_2sam}
T_2(\tau_0)
&=&
\frac{1}{n_1(n_1-1)}\sum_{s=1}^{n_1}\sum_{t\neq s}^{n_1}{\bm{X}}_{s}^T \widehat{P}_{1,\mathcal{O}}^{(s,t)}{\bm{X}}_{t}
+\frac{1}{n_2(n_2-1)}\sum_{s=1}^{n_2}\sum_{t\neq s}^{n_2}{\bm{Y}}_{s}^T \widehat{P}_{2,\mathcal{O}}^{(s,t)}{\bm{Y}}_{t}   \nonumber  \\
&&-\frac{2}{n_1n_2}\sum_{s=1}^{n_1}\sum_{t=1}^{n_2}
{\bm{X}}_{s}^T
\widehat{P}_{12,\mathcal{O}}^{(s,t)}{\bm{Y}}_{t},
\end{eqnarray}
where $\widehat{P}_{1,\mathcal{O}}^{(s,t)}$,
$\widehat{P}_{2,\mathcal{O}}^{(s,t)}$ and
$\widehat{P}_{12,\mathcal{O}}^{(s,t)}$
 are three sample-based estimates of ${P}_{\mathcal{O}}$ with
 $$\widehat{P}_{1,\mathcal{O}}^{(s,t)}=\sum_{(i,j) \in \hat{A}_1}
P_{ij}^T (P_{ij} S_{1*}^{(s,t)} P_{ij}^T)^{-1}P_{ij}
+\sum_{i \in \hat{A}_2}P_{i}^T ( P_{i} S_{1*}^{(s,t)} P_{i}^T)^{-1} P_{i},$$
$$\widehat{P}_{2,\mathcal{O}}^{(s,t)}=\sum_{(i,j) \in \hat{A}_1}
P_{ij}^T (P_{ij} S_{2*}^{(s,t)} P_{ij}^T)^{-1}P_{ij}
+\sum_{i \in \hat{A}_2}P_{i}^T ( P_{i} S_{2*}^{(s,t)} P_{i}^T)^{-1} P_{i},$$
$$\widehat{P}_{12,\mathcal{O}}^{(s,t)}=\sum_{(i,j) \in \hat{A}_1}
P_{ij}^T (P_{ij} S_{12,*}^{(s,t)} P_{ij}^T)^{-1}P_{ij}
+\sum_{i \in \hat{A}_2}P_{i}^T ( P_{i} S_{12,*}^{(s,t)} P_{i}^T)^{-1} P_{i}.$$
When $\tau_0=1$, we have $\hat{A}_1=\emptyset$ and $\hat{A}_2=\{1,\ldots,p\}$
so that the PHT statistic reduces  to the diagonal Hotelling's test in \citet{park2013test}.
In contrast, when $\tau_0=0$, we have $\hat{A}_1=\{(i,j): i<j\}$ for $i,j=1,\ldots,p$ and $\hat{A}_2=\emptyset$, and so   the PHT  statistic is indeed
the $U$-type test statistic (\ref{W_stat2}) for the  two-sample test.


\subsection{Asymptotic results}
For ease of notation,
we assume that the random vectors ${\bm X}=(X_1, \ldots, X_p)^T$
and ${\bm Y}=(Y_1, \ldots, Y_p)^T$ follow the following  two models:
\begin{eqnarray}\label{model1}
{\bm X}=C_2 {\bm Z}^{(1)} + \bm{\mu}_1
\text{~~and~~}
{\bm Y}=C_2 {\bm Z}^{(2)} + \bm{\mu}_2,
\end{eqnarray}
where
$C_2\in \mathbb{R}^{p \times q}$ with $q \geq p$ such that $\Sigma=C_2 C_2^T$, and
the random vector ${\bm Z}^{(i)}$ satisfies that $E({\bm Z}^{(i)})={\bm 0}$
and $\text{Var}({\bm Z}^{(i)})=I_q$ for $i=1,2$.
In addition, we assume that the following moment conditions hold:
$E\big(Z_{j}^{(i)}\big)^4=3+\Delta_2< \infty$ where
$\Delta_2$ is a positive constant, and
\begin{eqnarray}\label{con1}
E\left( ( Z_{l_1}^{(i)} )^{\alpha_1} ( Z_{l_2}^{(i)} )^{\alpha_2}  \cdots ( Z_{l_k}^{(i)} )^{\alpha_k} \right)
=E\Big( ( Z_{l_1}^{(i)} )^{\alpha_1} \Big)
E\Big( ( Z_{l_2}^{(i)} )^{\alpha_2} \Big)
 \cdots E\Big( ( Z_{l_k}^{(i)} )^{\alpha_k} \Big),
\end{eqnarray}
where $k$ a positive integer such that $\alpha_1+\cdots+\alpha_k\leq 8$,
and $l_1 \neq l_2 \neq \cdots \neq l_k$.

We further assume that
$\{(X_i, X_j): i,j=1,2,\ldots,p  {\rm ~with~} i\neq j \}$
and $\{(Y_i, Y_j): i,j=1,2,\ldots,p  {\rm ~with~} i\neq j \}$
are two random fields.
To derive the asymptotic null and alternative distributions of the proposed two-sample PHT  statistic, we need the following conditions:
\begin{itemize}
\item[(C$1'$)] There exists a finite positive number $\bar{K}_2$ such that $1/\bar{K}_2  \leq \lambda_{p}(\Sigma) \leq \cdots \leq  \lambda_{1}(\Sigma) \leq \bar{K}_2$.

\item[(C$2'$)] Assume that $\{X_{j}: j \geq 1\}$
and $\{Y_{j}: j \geq 1\}$ are two $\rho$-mixing sequences,
with the corresponding $\rho$-mixing coefficients
$\rho_X(s)$ and $\rho_Y(s)$, respectively.
There exists a constant $\varpi_3>0$ such that $\rho_X(s) \leq \varpi_3\exp{(-s)}$
and $\rho_Y(s)\leq \varpi_3\exp{(-s)}$.

\item[(C$3'$)]
There exists an oracle constant $\tau^*>0$ such that,
for a finite positive integer ${K}_0$,
${\sup}_{i\leq p}{\text{card}(A_{i}^*)} \leq K_0$,
where $A_{i}^*=\{j: \tau_{ij}>\tau^* \}$.
In addition, we assume that $\underset{i,j=1,\ldots,p}{\lim\inf}\{\tau_{ij} | \tau_{ij}>\tau^* \}>\tau^*$ and $\underset{i,j=1,\ldots,p}{\lim\sup}\{\tau_{ij} | \tau_{ij}<\tau^* \}<\tau^*$.

\item[(C4$'$)] There exists a positive integer $m_0\geq 4$ such that
the higher order moments, $E({X}_{j}^{4 m_0+2})$ and $E({Y}_{j}^{4 m_0+2})$, are bounded uniformly
for $j=1,\ldots,p$. In addition, we assume that
$E\big\|S_{1,{\{ij\}}}^{-1} \big\|^8$ and $E\big\|S_{2,{\{ij\}}}^{-1} \big\|^8$  are
 bounded uniformly for $(i,j)\in A_1$, and $E(s_{1,jj}^{-8})$
and $E(s_{2,jj}^{-8})$ are  bounded uniformly for $j \in A_2$.

\item[(C5$'$)] Assume that $(\bm{\mu}_1- \bm{\mu}_2)^T  P_{\mathcal{O}} (\bm{\mu}_1- \bm{\mu}_2) =o(\sqrt{p/N})$  and $\bm{\mu}_1^T  P_{\mathcal{O}} \bm{\mu}_1 =o(\sqrt{p/N})$.
There exists  a constant $\varpi_4>0$ such that $\mu_{1j}^2+\mu_{2j}^2\leq \varpi_4 /\sqrt{N}$.
\end{itemize}

It is noteworthy that conditions (C1$'$)-(C5$'$) are analogous to conditions (C1)-(C5), respectively.
Condition (C1$'$) assumes that the eigenvalues are bounded uniformly away from 0 and $\infty$.
Condition (C2$'$) implies a weak dependence structure among  the data.
Condition (C3$'$) assumes  that our PHT statistic allows the number of covariate pairs with strong correlations to increase at the same order of $p$.
Conditions (C4$'$) and (C5$'$) are two technical conditions that are needed for deriving the asymptotic results of the proposed  test statistic.

\vskip 10pt
\begin{thm}\label{pro2sam}
~
Assume that  $\tau_0$ satisfies
$\underset{i,j=1,\ldots,p}{\lim\inf}\{\tau_{ij} | \tau_{ij}>\tau_0 \}>\tau_0$
and $\underset{i,j=1,\ldots,p}{\lim\sup}\{\tau_{ij} | \tau_{ij}<\tau_0 \}<\tau_0$.
Let $\hat{A}_1$ and $\hat{A}_2$ be the two sets based on the threshold $\tau_0$
in the screening procedure.
Then  for any given positive integer $m_0$, if  $p=O(N^{m_0})$,
we have $$P(\hat{A}_2= A_2)\geq  P(\hat{A}_1 = A_1)  \to 1
{\rm ~~as~~} N \to \infty.$$
\end{thm}
\vskip 10pt

The proof of Theorem \ref{pro2sam} is given in Appendix  \ref{Prof.pro2sam}. This theorem shows that, the selected sets $\hat{A}_1$ and $\hat{A}_2$ based on the sample data will converge to $A_1$ and $A_2$, respectively, when the sample sizes tend to infinity.

\vskip 10pt
\begin{thm}\label{th2_1}
Assume that $\tau_0 \geq \tau*$, $\underset{i,j=1,\ldots,p}{\lim\inf}\{\tau_{ij} | \tau_{ij}>\tau_0 \}>\tau_0$,
and $\underset{i,j=1,\ldots,p}{\lim\sup}\{\tau_{ij} | \tau_{ij}<\tau_0 \}<\tau_0$.
Then under conditions  (C1$'$)--(C5$'$)
and if $p=o(N^{m_0-1})$ with $m_0$ defined in (C4$'$),  we have
\begin{equation*}
\frac{T_2(\tau_0)-{\bm \delta}_2^T P_{\mathcal{O}} {\bm \delta}_2 }{\sqrt{\phi(n_1,n_2){\rm tr}({\Lambda}_1^2)}}
\stackrel{D}{\longrightarrow}N(0,1) {\rm ~~as~~} N\to \infty,
\end{equation*}
where ${\bm \delta}_2={\bm \mu}_2-{\bm \mu}_1$ and $\phi(n_1,n_2)={2}/\{n_1(n_1-1)\}+{2}/\{n_2(n_2-1)\}+{4}/{(n_1n_2)}.$
\end{thm}

The proof of Theorem \ref{th2_1} is given in Appendix  \ref{Prof.th2_1}.
This theorem shows that, for a larger threshold $\tau_0 > \tau*$, if  $\tau_0$ satisfies $\underset{i,j=1,\ldots,p}{\lim\inf}\{\tau_{ij} | \tau_{ij}>\tau_0 \}>\tau_0$ and $\underset{i,j=1,\ldots,p}{\lim\inf}\{\tau_{ij} | \tau_{ij}<\tau_0 \}<\tau_0$, then the test statistic
$T_2(\tau_0)$ still converges to the standard  normal distribution after proper centering and scaling. Hence, despite the exact threshold $\tau*$ that satisfies condition (C3$'$) being unknown in
practice, we can always  select a larger threshold when performing the test.

To apply Theorem \ref{th2_1} for practical use, we have the following lemma for deriving a ratio consistent estimator for  ${\rm tr}(\Lambda_1^2)$,  where the proof  is given in Appendix \ref{Prof.lemma2_1}.

\vskip 10pt
\begin{lemma}\label{lemma2_1}
Assume that $p=o(N^{3})$, and  $\tau_0$ satisfies the assumptions in Theorem~\ref{th2_1}.
Then under conditions (C1$'$)--(C5$'$),
\begin{eqnarray*}
\widehat{{\rm tr}(\Lambda_1^2)}
&=& \frac{1}{2n_1 (n_1-1)} \sum_{s=1}^{n_1} \sum_{t\neq s}^{n_1}
( {\bm X_{s}} -{\bar{\bm X}^{(s,t)}})^T \widehat{P}_{1,\mathcal{O}}^{(s,t)} {\bm X}_{t}
( {\bm X_{t}} -{\bar{\bm X}^{(s,t)}})^T \widehat{P}_{1,\mathcal{O}}^{(s,t)} {\bm X}_{s}\\
&&
+\frac{1}{2n_2 (n_2-1)} \sum_{s=1}^{n_2} \sum_{t\neq s}^{n_2}
( {\bm Y_{s}} -{\bar{\bm Y}^{(s,t)}})^T \widehat{P}_{2,\mathcal{O}}^{(s,t)} {\bm Y}_{t}
( {\bm Y_{t}} -{\bar{\bm Y}^{(s,t)}})^T \widehat{P}_{2,\mathcal{O}}^{(s,t)} {\bm Y}_{s}
\end{eqnarray*}
is a ratio consistent estimator of ${\rm tr}(\Lambda_1^2)$,
where $\bar{\bm X}^{(s,t)}$ (or $\bar{\bm Y}^{(s,t)}$) is the sample mean vector of group 1 (or group 2)  without observations $\bm{X}_s$ and $\bm{X}_t$ (or $\bm{Y}_s$ and $\bm{Y}_t$).
Consequently, under the null hypothesis of (\ref{H0}),
\begin{equation*}
\frac{T_2(\tau_0) }{\sqrt{\phi(n_1,n_2)\widehat{{\rm tr}(\Lambda_1^2)} }}
\stackrel{D}{\longrightarrow}N(0,1) {\rm ~~as~~} N\to \infty.
\end{equation*}
\end{lemma}

\vskip 10pt
By Theorem \ref{th2_1},
the power function of the PHT statistic for the two-sample test  is given as
\begin{equation}\label{T2.power}
{\rm Power}(\bm{\delta}_2)=
\Phi \Big(-z_{\alpha}+\frac{ {\bm \delta}_2^T P_{\mathcal{O}} {\bm \delta}_2 }
{\sqrt{\phi(n_1,n_2) { {\rm tr}({\Lambda}_1^2)}} }\Big).
\end{equation}
The performance of the new test depends on the quantities
${\bm \delta}_2^TP_{\mathcal{O}}{\bm \delta}_2$
and  ${ {\rm tr}({\Lambda}_1^2)}$.
Theoretically,  to achieve the highest asymptotic power for the PHT statistic, a reasonable choice for the threshold  $\tau_0$ can be to maximize ${\rm Power}(\bm{\delta}_2)$.
However, this maximization procedure may not be feasible in practice, since ${\bm \delta}_2^TP_{\mathcal{O}}{\bm \delta}_2/ \sqrt{ {\rm tr}({\Lambda}_1^2)}$ involves unknown quantities including ${\bm \delta}_2$ and $\Sigma$.
In Section 4.3, we provide a data driven procedure for the selection of $\tau_0$ when there is no prior information available
for the signals and the structure of the covariance matrix.

\section{Monte Carlo Simulation Studies}
The purpose of this section is to  assess the finite sample performance of our proposed testing method.
For ease of presentation, we conduct simulation studies   for the two-sample test only.
We also consider seven other tests for comparison:
the unscaled Hotelling's tests including CQ from  \citet{chen2010} and  aSUP from \citet{Xu2016}; the diagonal Hotelling's tests including  PA  from \citet{park2013test},  GCT from \citet{Carroll2015}, and  DLRT from \citet{Hu2018test};  the composite Hotelling's test CHT from \citet{li2017}; and the regularized Hotelling's test RMPBT from \citet{zoh2017powerful}.

For each simulation, we  generate observations $\bm{X}_s$ for $s=1,\dots, n_1$ and
$\bm{Y}_t $ for $t=1,\ldots, n_2$, respectively, from   model (\ref{model1}).
Without loss of generality,
we let  $\bm{\mu}_1={\bm 0}$ and  $\Sigma \in \mathbb{R}^{p\times p}$ be the common covariance matrix.
Then,  ${\bm X}_s=\Sigma^{1/2} {\bm Z}_s^{(1)}$
and ${\bm Y}_t=\Sigma^{1/2} {\bm Z}_t^{(2)} + \bm{\mu}_2$,
where all  the components of $\bm{Z}_s^{(1)}$ and $\bm{Z}_t^{(2)}$
are i.i.d. random variables with zero mean and unit variance.
Under the null hypothesis, we set $\bm{\mu}_2={\bm 0}$.
Under the alternative hypothesis, we set $\bm{\mu}_2=(\mu_{21},\dots,\mu_{2p_0},0,\dots,0)^T$, where $p_0=\lfloor{\beta p}\rfloor$ with
$\beta \in [0,1]$ being a tuning parameter that controls the degree of sparsity in the signals, and $\lfloor{x}\rfloor$ is the largest integer that is equal to or less than $x$.

\subsection{Normal data}\label{normal.data}
In the first simulation, $\bm{Z}_s^{(1)}$ and $\bm{Z}_t^{(2)}$ are generated
from the $p$-dimensional multivariate normal distribution $N_p(\bm{0},I_p)$.
Let $D_p= {\rm diag}{(d_{11}^2,\ldots, d_{pp}^2)}$ be a diagonal matrix with $d_{ii}$  randomly sampled from
the uniform distribution on $[0.5,1.5]$.
For the common covariance matrix $\Sigma$, we consider four different structures  as follows.

\begin{enumerate}
  \item[(1)] $\Sigma_1= D_p^{1/2}R_1 D_p^{1/2}$, where $R_1=(0.9^{|i-j|})_{p \times p}$.

  \item[(2)]  $\Sigma_2= D_p^{1/2}R_2 D_p^{1/2}$, where $R_2=((-0.9)^{|i-j|})_{p \times p}$.

  \item[(3)] $\Sigma_3= D_p^{1/2} R_{3} D_p^{1/2}$, where $R_{3}$ is a block diagonal matrix
  with the same block as $B=(0.9^{I(i \neq j)})_{5 \times 5}$, and $I(\cdot)$ is the identity function.
  \item[(4)]  $\Sigma_4=D_p^{1/2} R_4 D_p^{1/2}$, where $R_4=I_p$ is the identity matrix.
\end{enumerate}

Table \ref{TypeI.1} summarizes the empirical sizes for the eight tests over 2000 simulations with the given covariance matrices.  The threshold for PHT is set as $\tau_0=0.8$.  As shown in Table \ref{TypeI.1}, PHT is able to provide  a more stable test statistic with a better controlled type I error rate under most settings.
When the dimension is large and the correlations among the covariates  are strong, DLRT, GCT and RMPBT suffer from significantly inflated type I error rates compared to those for  PA, aSUP and CQ. When the covariates are weakly correlated, e.g. the diagonal structure, most tests have a reasonable type I error rate except for CHT. To be more specific,  CHT
often has an unacceptably large type I error rate compared to the nominal level at $\alpha=0.05$, and hence  it does not provide  a valid test.

\begin{figure}[!htbp]
    \centering
    {\includegraphics[width=16cm,height=19cm]{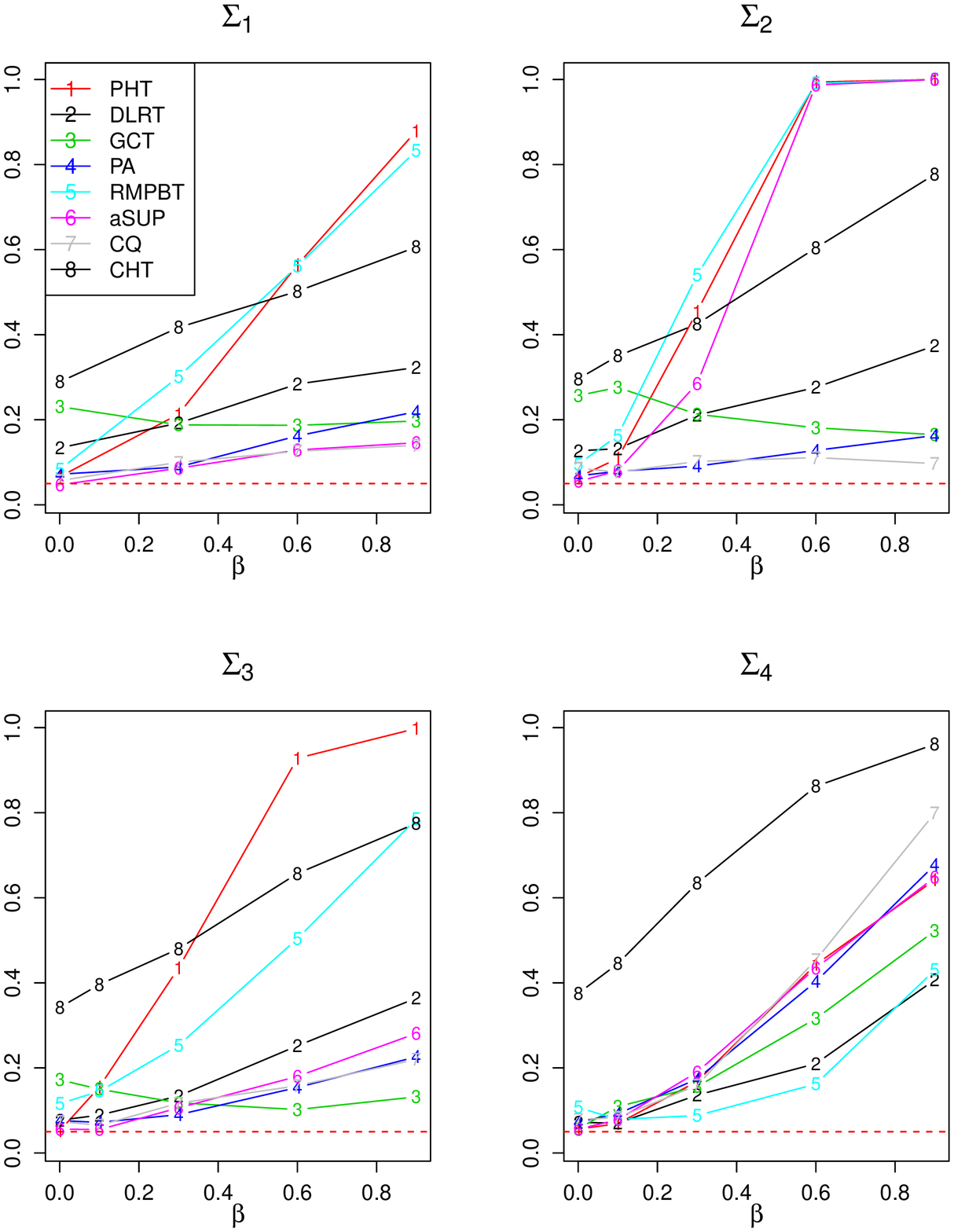}}
    \caption{\label{powerp=100} Power comparison between  PHT, DLRT, GCT, PA, RMPBT, aSUP, CQ and CHT with $n_1=30, n_2=25$, and $p=100$. The horizontal dashed lines represent the nominal level of $\alpha=0.05$, and the results are based on normal data with 1000 simulations.}
\end{figure}

\begin{figure}[!htbp]
    \centering
    {\includegraphics[width=16cm,height=19cm]{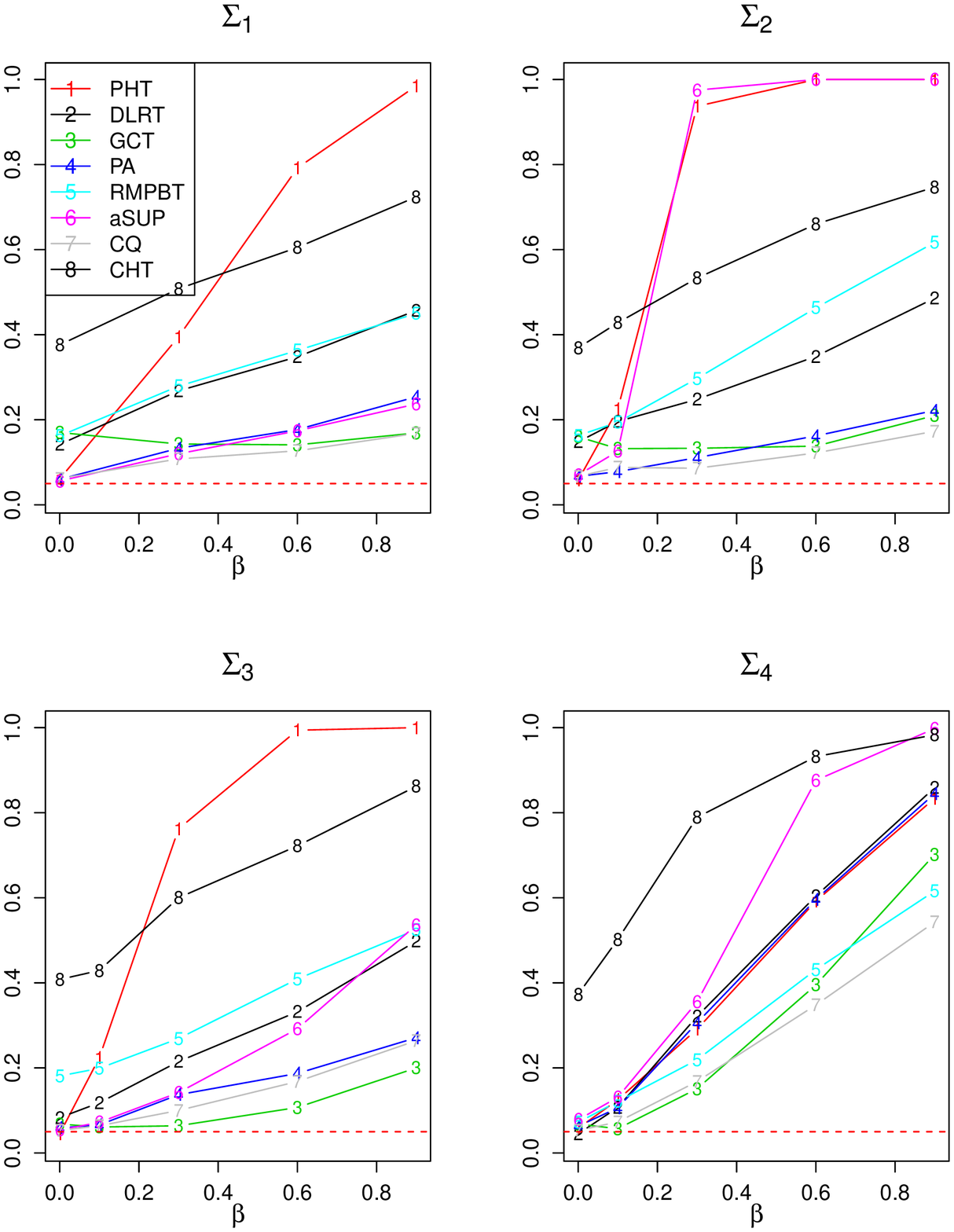}}
    \caption{\label{powerp=500} Power comparison between PHT, DLRT,  GCT, PA, RMPBT, aSUP, CQ and CHT  with $n_1=30, n_2=25$, and $p=500$.
    The horizontal dashed lines represent the nominal level of $\alpha=0.05$, and the results are based on  normal data with 1000 simulations.}
\end{figure}

To assess the power performance of the eight tests, we set the $j$th nonzero component in $\bm{\mu}_2$ as $\mu_{2j}=\kappa \delta_j$ where $\kappa$ controls the signal strength,
and  $\delta_j \sim  N(1.5,1)$ for $j=1,\ldots, p_0$. The other  parameters are
$n_1=30, n_2=25$, $(\kappa=0.1, p=100)$ or $(\kappa=0.075, p=500)$, respectively.
We then randomly generate  1000  data sets under each scenario, and plot the simulation results in Figures \ref{powerp=100} and \ref{powerp=500}.

\begin{center}
{
\begin{table}[htbp!]
\caption
{\label{TypeI.1} Type I error rates  for  PHT, DLRT,  GCT, PA, RMPBT, aSUP, CQ and CHT with normal data, where the sample sizes are $n_1=30$ and $n_2=25$, respectively,
and the nominal level is $\alpha=0.05$. }
{ {
\begin{center}
\begin{tabular}{c|c|c|c|c|c|c|c|c|c}
\hline\hline
\multirow{2}{*}{ }& \multirow{2}{*}{{$p$}}
& \multirow{2}{*}{PHT}
& \multirow{2}{*}{DLRT}& \multirow{2}{*}{GCT}
& \multirow{2}{*}{PA}& \multirow{2}{*}{RMPBT}
& \multirow{2}{*}{aSUP}& \multirow{2}{*}{CQ} &  \multirow{2}{*}{CHT} \\
&  & & & & & & & &  \\ \hline
\multirow{2}{*}{ $\Sigma_1$ }
&$100$ &0.066&	0.134&	0.231&	0.072&	0.083&	0.047&	0.057&	0.289 \\\cline{2-10}
&$500$  &0.061 &0.142 &0.170  &0.060  &0.162 &0.056  &0.062 &0.376\\ \hline
\multirow{2}{*}{ $\Sigma_2$ }
&$100$ &0.065&	0.127&	0.257&	0.068&	0.095&	0.055&	0.086&	0.296  \\\cline{2-10}
&$500$  &0.058 &0.148&0.160   &0.067   &0.163  & 0.070&0.067&   0.369\\ \hline
\multirow{2}{*}{ $\Sigma_3$}
&$100$  &0.052&	0.077&	0.172&	0.076&	0.116&	0.056&	0.074&	0.342
\\\cline{2-10}
&$500$  &0.045  &0.093  &0.068  &0.059&0.181&0.054&0.053&
0.408 \\ \hline
\multirow{2}{*}{ $\Sigma_4$ }
&$100$  &0.056 &	0.073	& 0.107	& 0.056	&0.057	&0.079	&0.072	&0.375  \\ \cline{2-10}
&$500$  &0.057  &0.045  &0.068  &0.048 &0.072 &0.078&0.053
&0.373   \\ \hline\hline
\end{tabular}\end{center}
} }
\end{table}
}
\end{center}

According to the figures, when the true covariance matrix has
a complex structure (inclduing $\Sigma_1$, $\Sigma_2$ and $\Sigma_3$), our proposed PHT has a significant improvement for the power performance.
Specifically, as long as the signals are not too sparse, PHT always has a higher power compared to the other tests.
When the covariates are independent of each other,  aSUP achieves the highest power when the dimension is large. Meanwhile, PHT also exhibits a high power for detection which is  nearly as good as PA.  RMPBT has a good power performance when the dimension is not large.  However, if  the dimension becomes large, RMPBT suffers from a low power, especially when the covariance matrix follows a diagonal structure. DLRT, GCT and CQ also suffer from a low power for detection especially when some covariates are highly correlated.

\subsection{Heavy-tailed data}\label{heavy.data}
In the second simulation, $\bm{Z}_s^{(1)}$ and $\bm{Z}_t^{(2)}$ are generated from a heavy-tailed
distribution to examine the robustness of the proposed tests.
Following \citet{Carroll2015} and \citet{Hu2018test}, we consider a ``double" Pareto distribution with parameters
$a>0$ and $b>0$. The detailed algorithm is as follows:
\begin{description}
\item Step 1: Generate two independent random variables $U$ and $V$, where $U$ is from the Pareto distribution with  the cumulative distribution function $F(x)=1-(1+x/b)^{-a}$ for $x\geq0$, and  $V$ is a binary random variable with $P(V=1)=P(V=-1)=0.5$. Then $Z=UV$ follows the double Pareto distribution with parameters $a$ and $b$.
\item[]Step 2: Generate random vectors
$\widetilde{ \bm{Z} }_s^{(1)}=(\tilde{z}_{s1}^{(1)},\tilde{z}_{s2}^{(1)},\ldots,\tilde{z}_{sp}^{(1)})^T$
for $s=1,\ldots,n_1$  and
$\widetilde{ \bm{Z} }_t^{(2)}=(\tilde{z}_{t1}^{(2)},\tilde{z}_{t2}^{(2)},\ldots,\tilde{z}_{tp}^{(2)})^T$
for $t=1,\ldots,n_2$,
where all the components of $\widetilde{ \bm{Z} }_s^{(1)}$ and $\widetilde{ \bm{Z} }_t^{(2)}$ are sampled independently from the double Pareto distribution with parameters  $a=16.5$ and $b=8$.
  \item[]Step 3: Let ${\textstyle { \bm{Z}_s^{(1)} =\widetilde{ \bm{Z} }_s^{(1)} /c_0} }$ and ${\textstyle { \bm{Z}_t^{(2)} =\widetilde{ \bm{Z} }_t^{(2)} /c_0} }$
      where $c_0^2={512}/{899}$ is the variance of the double Pareto  distribution with parameters $a=16.5$ and $b=8$.
\end{description}
Once ${\bm Z}_s^{(1)}$ and ${\bm Z}_t^{(2)}$ are given, we adopt all other settings  the same as  those in Section 4.1 to generate the observations of ${\bm X}_s$ and ${\bm Y}_t$ for each simulation.

\begin{figure}[!htbp]
    \centering
    {\includegraphics[width=16cm,height=19cm]{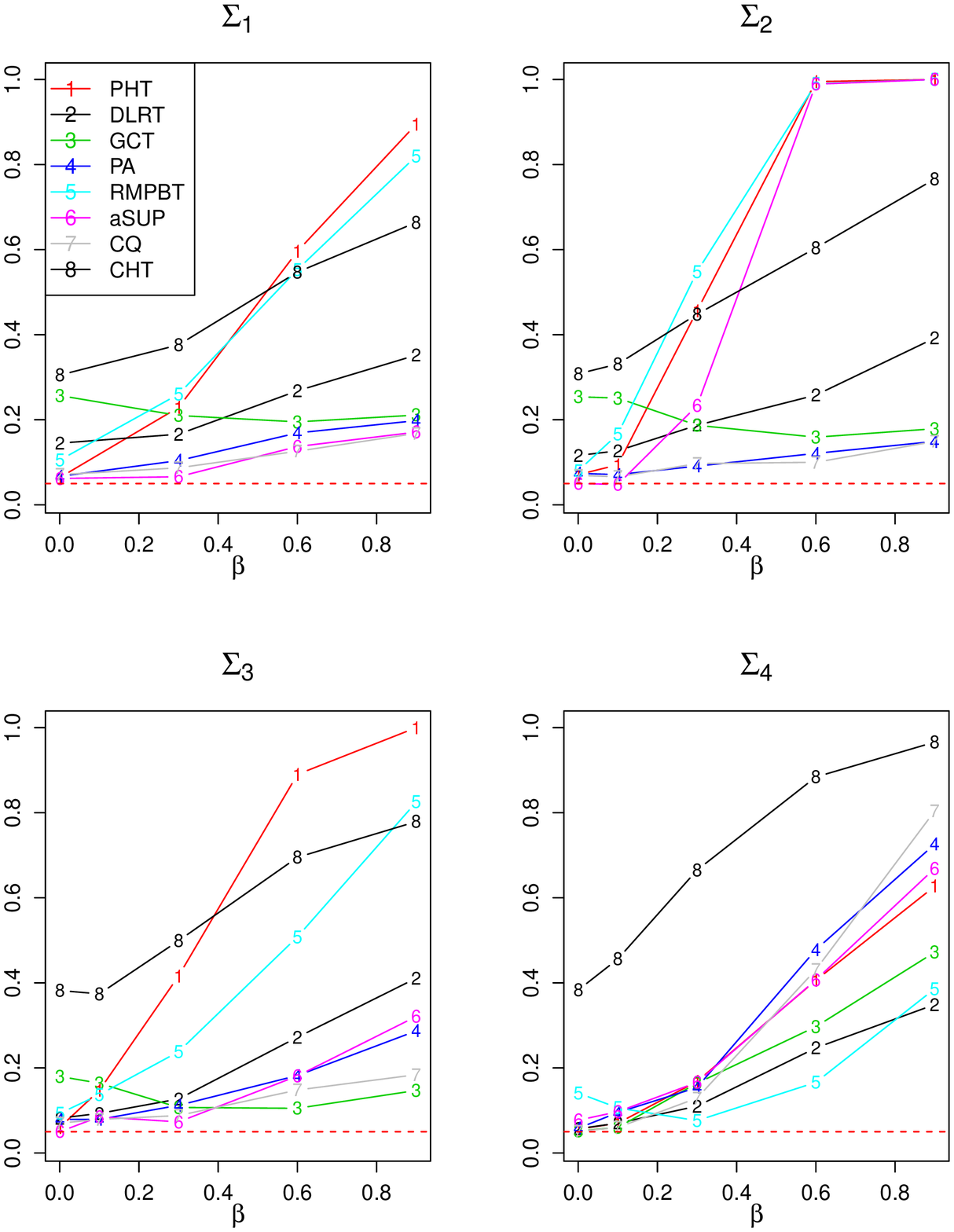}}
    \caption{\label{heavy_powerp=100} Power comparison between PHT, DLRT, GCT, PA, RMPBT, aSUP, CQ and CHT  with $n_1=30, n_2=25$, and $p=100$. The horizontal dashed lines represent the nominal  level of $\alpha=0.05$, and the results are based on heavy-tailed data with 1000 simulations.}
\end{figure}

\begin{figure}[!htbp]
    \centering
    {\includegraphics[width=16cm,height=19cm]{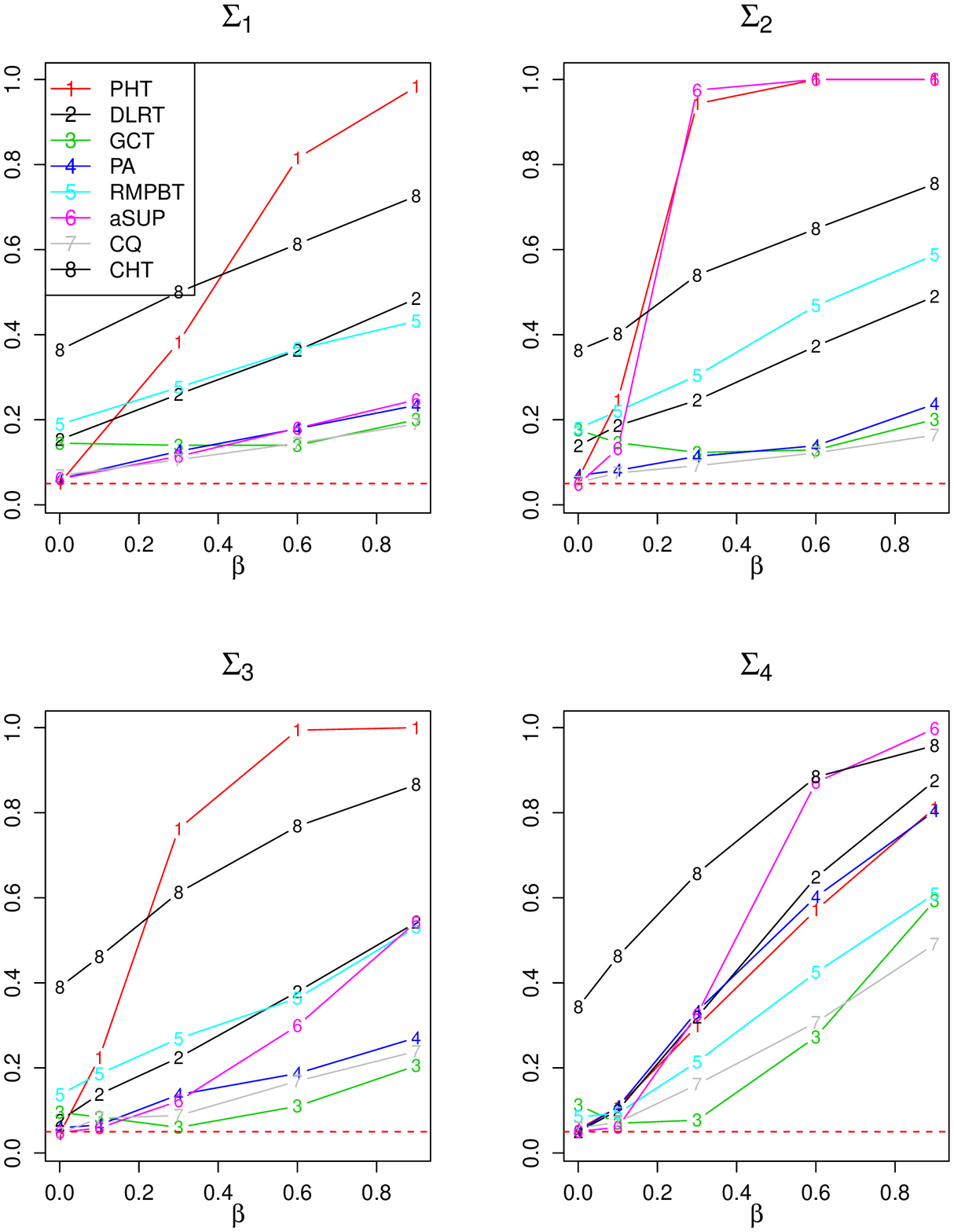}}
    \caption{\label{heavy_powerp=500} Power comparison between
    PHT, DLRT, GCT, PA, RMPBT, aSUP, CQ and CHT with $n_1=30, n_2=25$, and $p=500$.
    The horizontal dashed lines represent the nominal level of $\alpha=0.05$, and the results are based on heavy-tailed data with 1000 simulations.}
\end{figure}

Table \ref{TypeI.2} and Figures \ref{heavy_powerp=100} and \ref{heavy_powerp=500}  present the empirical sizes and power  for the eight tests with heavy-tailed data at the nominal level of $\alpha=0.05$.  The simulations for computing the empirical sizes and power are over 2000 and 1000 simulations, respectively.
In particular, when the dimension is large and the correlations among the covariates are strong,
our PHT is able to control the type I error rate, and can achieve  a higher power for detection.
RMPBT exhibits a good power performance when the dimension is not large and the covariance matrix has a complex structure, but it suffers from a slightly inflated type I error rate. When the dimension is large and the covariance matrix has a complex structure (including $\Sigma_1$, $\Sigma_2$ and $\Sigma_3$), RMPBT exhibits a substantially inflated type I error rate, and suffers from a low power. In addition, aSUP also exhibits a well controlled type I error rate in most settings; however, its power performance may be sensitive to the
structure of covariance matrix, e.g., aSUP suffers from a low power under $\Sigma_1$ and $\Sigma_3$ but has a good power performance under $\Sigma_2$.
DLRT has a well controlled type I error rate under the diagonal covariance matrix, but suffers from a low power. Finally, GCT and CHT always suffer from a significantly inflated type I error rate.

\begin{center}
{
\begin{table}[htbp!]
\caption
{\label{TypeI.2} Type I error rates for  PHT, DLRT,  GCT, PA, RMPBT, aSUP, CQ and CHT with  heavy-tailed data, where the sample sizes are $n_1=30$ and $n_2=25$, respectively, and the nominal level is $\alpha=0.05$. }
{ {
\begin{center}
\begin{tabular}{c|c|c|c|c|c|c|c|c|c}
\hline\hline
\multirow{2}{*}{}& \multirow{2}{*}{{$p$}} & \multirow{2}{*}{PHT}
& \multirow{2}{*}{DLRT}& \multirow{2}{*}{GCT}
& \multirow{2}{*}{PA}& \multirow{2}{*}{RMPBT}
& \multirow{2}{*}{aSUP}& \multirow{2}{*}{CQ}&\multirow{2}{*}{CHT}\\
&  & & & & & & & & \\\hline
\multirow{2}{*}{$\Sigma_1$ }
&$100$ & 0.064 &	0.145 &	0.257  &0.067 &0.105  &0.062 &0.071	 &0.305
\\\cline{2-10}
&$500$   & 0.051 &  0.153 &0.145   &0.061 &0.189  &0.057 &0.069  &0.364 \\ \hline
\multirow{2}{*}{$\Sigma_2$ }
&$100$  &0.071 &0.116 &0.254	 &0.074	 &0.080 &0.050	&0.069	&0.308   \\\cline{2-10}
&$500$  &0.061  & 0.139 & 0.176 &0.069   &0.180  & 0.048 &0.053  &0.363  \\ \hline
\multirow{2}{*}{$\Sigma_3$ }
&$100$  &0.060	&0.082	&0.180	&0.079	&0.093	&0.050	&0.070	&0.382  \\\cline{2-10}
&$500$  &0.051  &0.076  &0.096  &0.054 &0.137 &0.048 &0.046 &0.390  \\ \hline
\multirow{2}{*}{$\Sigma_4$ }
&$100$
&0.056	&0.058	&0.141	&0.077	&0.052	&0.051	&0.057	&0.383 \\\cline{2-10}
&$500$  &0.055  &0.050  &0.113  &0.051 &0.085  &0.051  &0.059     &0.356\\ \hline\hline
\end{tabular}\end{center}
} }
\end{table}
}
\end{center}

\subsection{A data-driven choice of the threshold $\tau_0$ }
In this section, we provide a data-driven method for selecting the threshold $\tau_0$. When there is no prior information on the covariance matrix structure, a reasonable choice for the threshold $\tau$ can be to maximize the empirical estimator for the signal-to-noise ratio  that determines the  power for the PHT statistic.
According to (\ref{U.power}) and
(\ref{T2.power}), we have
${\rm SNR}_1(\tau_0)= {({\bm \mu}-{\bm \mu}_0 )^T P_{\mathcal{O}} ({\bm \mu}-{\bm \mu}_0 )}/{\sqrt{{\rm tr}(\Lambda_1^2)} }$
 and
${\rm SNR}_2(\tau_0) ={({\bm \mu}_2-{\bm \mu}_1 )^T P_{\mathcal{O}} ({\bm \mu}_2-{\bm \mu}_1 )}/
{ \sqrt{{\rm tr}(\Lambda_1^2)} } $
for the one- and two-sample tests, respectively.
We then estimate the  two ratios by
\begin{equation*}
\widehat{ {\rm SNR} }_1(\tau_0)=
\frac{T_1(\tau_0)}{  \sqrt{\widehat{{\rm tr}(\Lambda_1^2)}} }
{\rm~~ and~~}
\widehat{ {\rm SNR} }_2(\tau_0)=
\frac{T_2(\tau_0)}{  \sqrt{\widehat{{\rm tr}(\Lambda_1^2)}} }.
\end{equation*}
According to Lemmas  \ref{lemma1.1} and \ref{lemma2_1}, we have
$\widehat{ {\rm SNR} }_1(\tau_0) \stackrel{P}{\longrightarrow}  {\rm SNR}_1(\tau_0)$  as $n \to \infty$,
and $\widehat{ {\rm SNR} }_2(\tau_0) \stackrel{P}{\longrightarrow}  {\rm SNR}_2(\tau_0)$ as $N \to \infty$,
where $\stackrel{P}{\longrightarrow}$ denotes convergence in probability.
For simplicity,  we present  the selection procedure of the threshold $\tau_0$ for the two-sample test only, whereas the same procedure can be readily adapted   to the one-sample test as well.

\begin{description}
  \item []Step 1: Randomly generate two subsets
  ${\rm Set}_X^{*}=\{ {\bm X}_k, k=1,\ldots, n_1^{*}\}$
  and ${\rm Set}_Y^{*}=\{ {\bm Y}_l, l=1,\ldots, n_2^{*}\}$,
  where $n_1^{*}<n_1$ and $n_2^{*}<n_2$; ${\bm X}_{l}^{*}$ and ${\bm Y}_{k}^{*}$  are randomly selected without replacement from $\{ {\bm X}_1,\ldots, {\bm X}_{n_1}\}$ and $\{ {\bm Y}_1,\ldots, {\bm Y}_{n_2}\}$, respectively.
  \item[]Step 2: Given the grid points $\mathcal{T}_{\tau_0}=\{ \tau_{01}, \ldots, \tau_{0H} \}$, for each point
  $\tau_{0h} \in \mathcal{T}_{\tau_0}$, compute $\widehat{ {\rm SNR} }_2(\tau_{0h})$ using ${\rm Set}_X^{*}$ and ${\rm Set}_Y^{*}$ and then select $\hat{\tau}_{0}={\rm argmax}_{\tau_{0h} \in {\mathcal{T}}_{\tau_0}}\widehat{ {\rm SNR} }_2(\tau_{0h})$.
  \item[]Step 3: Repeat Steps 1-2 for $B$ times, and denote the selected
  $\hat{\tau}_{0}$ as $\hat{\tau}_{0}^{(b)}$ for the $b$th time. The optimal ${\tau}_{0}$
  is defined as the median of $\{ \hat{\tau}_{0}^{(1)},\ldots,\hat{\tau}_{0}^{(B)} \}$.
\end{description}
In addition, to  balance the computation time and the  detection ability of our PHT statistic, we recommend to take  $n_1^{*}=\lfloor{2n_1/3}\rfloor$, $n_2^{*}=\lfloor{2n_2/3}\rfloor$,
$B=10$, and $\mathcal{T}_{\tau_0}=\{0.7, 0.8, 0.9, 1 \}$.

 \begin{figure}[!htbp]
    \centering
    {\includegraphics[width=16cm,height=19cm]{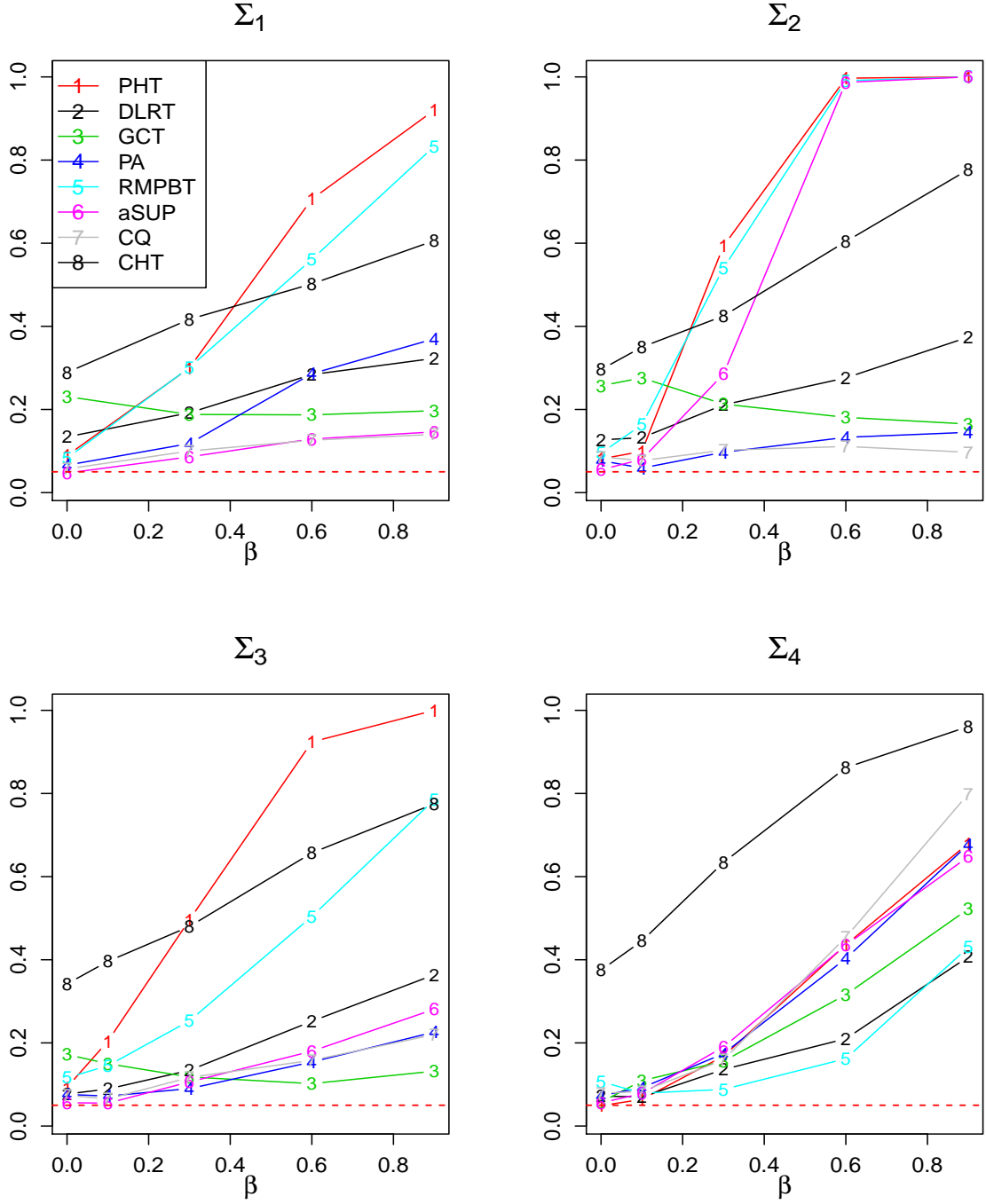}}
    \caption{\label{adp_powerp=100} Power comparison between  PHT, DLRT, GCT, PA, RMPBT, aSUP, CQ and CHT with $n_1=30, n_2=25$, and $p=100$. The horizontal dashed lines represent the nominal level of $\alpha=0.05$, and the results are based on normal data with 1000 simulations.}
\end{figure}

 \begin{figure}[!htbp]
    \centering
    {\includegraphics[width=16cm,height=19cm]{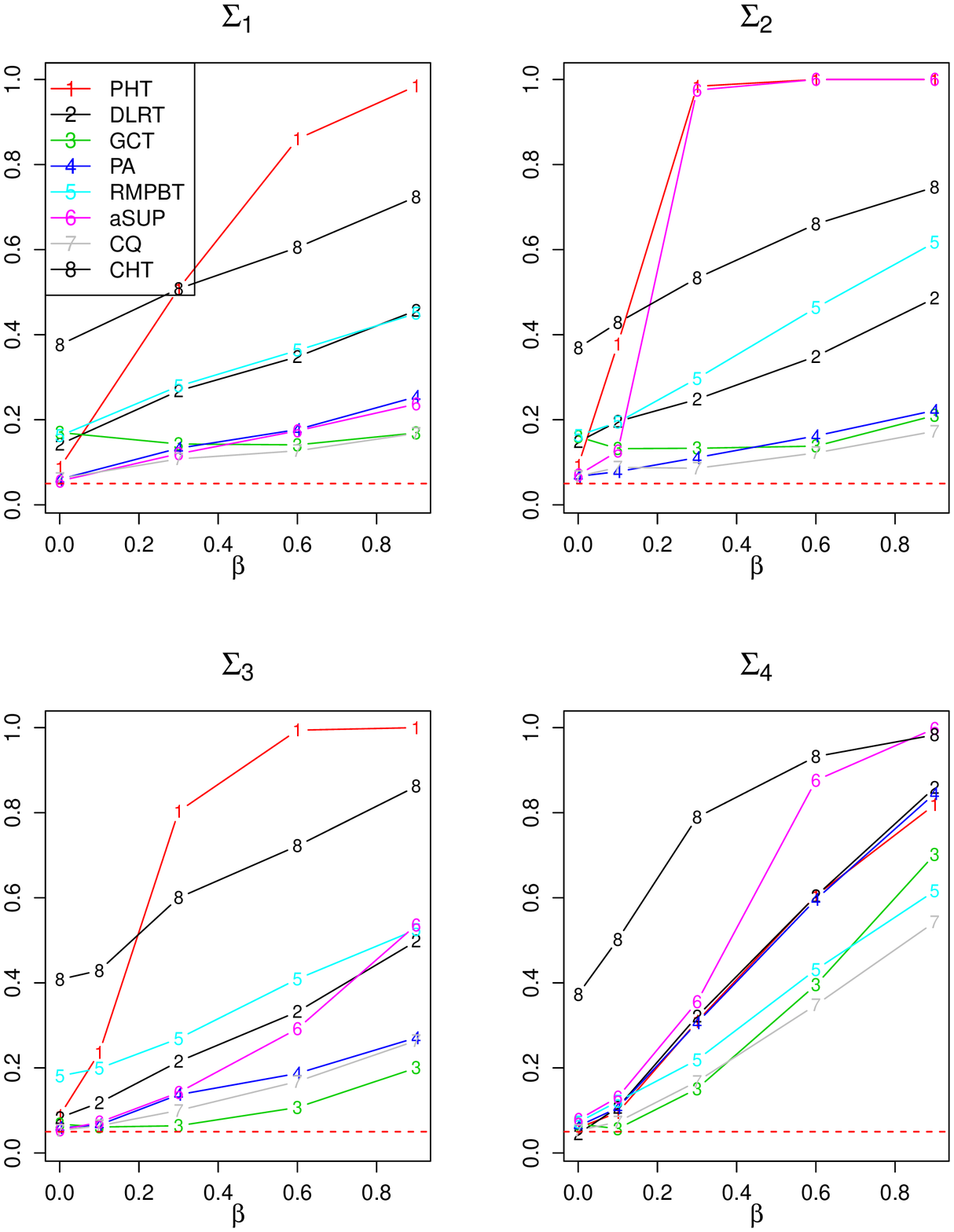}}
    \caption{\label{adp_powerp=500} Power comparison between PHT, DLRT,  GCT, PA, RMPBT, aSUP, CQ and CHT  with $n_1=30, n_2=25$, and $p=500$.
    The horizontal dashed lines represent the nominal level of $\alpha=0.05$, and the results are based on  normal data with 1000 simulations.}
\end{figure}

To assess the usefulness of the  selection procedure for $\tau_0$,   we compare PHT with the other seven tests.
For the common covariance matrix, we also consider the four structures, $\Sigma_1$, $\Sigma_2$, $\Sigma_3$ and $\Sigma_4$,
and remain the other parameters the same as in the previous simulations.
Figures \ref{adp_powerp=100} and \ref{adp_powerp=500} display the power performance for the eight tests with normal data
at the nominal level $\alpha=0.05$.
Specifically,  if the covariance matrix has a complex structure (including $\Sigma_1$, $\Sigma_2$ and $\Sigma_3$),  PHT always possesses a higher power compared to  all other methods  as long as the signals are not too sparse. When the covariance matrix follows a diagonal  structure, aSUP achieves the highest power as the dimension becomes  large, whereas PHT also exhibits a high power for detection which is nearly the same as PA.
We also note that  PMPBT suffers from a low power, especially when the dimension is large.
In addition, when the correlations among the covariates are strong, DLRT, GCT and CQ usually also  from a low power for detection.

\section{Applications}

\subsection{Small round blue cell tumors data }
We apply our proposed PHT to analyze two microarray data sets.
The first data set is the small round blue cell tumors (SRBCTs)
from \citet{Khan2001}, which contains the expression of 2308 genes for four types of
childhood tumors. The data set can be downloaded from the link:  http://www.biolab.si/supp/bi-cancer/projections/info/
SRBCT.html.
As in \citet{zoh2017powerful}, we are also interested in testing the
 differential expression of genes between the Burkitt
lymphoma (BL) tumor and the neuroblastoma (NB) tumor.
The sample sizes of the BL and NB tumors are 11 and 18, respectively.
We compare our PHT with the other seven tests as  DLRT, GCT, PA, RMPBT, aSUP, CQ and CHT.
The $p$-values of the eight tests are all smaller than 0.0001, which indicates that all eight tests significantly reject the null hypothesis of the two-sample test at  the nominal level of $\alpha=0.05$.

\subsection{Leukemia data}

The second data set contains the leukemia data from two different groups  of patients: one is the acute lymphoblastic leukemia (ALL),
and the other is  the acute myeloid leukemia (AML).
The data set contains 7,129 genes and 72 samples
with 47 ALL patients and 25 AML patients, which is publicly available in the R package ``golubEsets".
To compare the performance of the tests, we first perform the two-sample $t$ tests to screen
the top $250$ significant genes.  We then apply the eight tests
to the selected gene set, respectively. The $p$-values of the eight tests are all smaller than 0.0001. This indicates that the mean expression levels of the gene set
between the ALL and AML groups are significantly different.

To further compare the performance of the eight tests,
we select the top $50$ significant genes and the last $200$ nonsignificant genes
to form a new gene set.
In such a way, the signal strength  of the new gene set
will be  weaker than that with the top $250$ significant genes.
We then apply  the permutation method to create two artificial groups for the new gene set
to mimic the null and alternative hypotheses, respectively.
Specifically, we randomly sample two distinct subclasses without replacement from the pooled data
with the sample sizes $30$ and $17$, respectively.
Since both classes are partitioned from the pooled data, the null hypothesis can be regarded as to be true.
Finally, we repeat the procedure 1000 times,
and perform the eight tests  at the nominal level of 0.05.
The rejection rate is computed to represent the false positive rate.
Similarly, to mimic the alternative hypothesis, we randomly sample one class from
the ALL  group with the sample size 30 and another class from the AML group
with the sample size 17. Then, with 1000 simulations for each test method,
we compute the rejection rates at the nominal level of 0.05.
For each test method, the rejection rate is computed to represent   the true positive rate.

Table \ref{PHTtab1} shows that DLRT, RMPBT, aSUP, GCT and CHT all suffer from inflated false positive rates, in particular for GCT and CHT. In contrast, PHT, PA and CQ provide a reasonable type I error rate and can serve as valid tests. Finally, PHT is nearly as high as PA, and provides a higher true positive rate than CQ.

\begin{center}
{
\begin{table}[htbp!]
\caption{\label{PHTtab1}
False and true positive rates of the eight tests with PHT, DLRT, GCT, PA, RMPBT, aSUP, CQ and CHT. The nominal level is 0.05.
}
\begin{center}
\begin{tabular}{c|c|c|c|c|c|c|c|c}
\hline\hline
    &PHT    &DLRT   &GCT      &PA     &RMPBT   &aSUP   &CQ    &CHT
    \\ \hline
false positive rate &0.081  &0.229  &0.587   &0.082   &0.101   & 0.221    &0.078 &0.573        \\ \hline
true positive rate & 0.811 &0.968  &0.027   &0.841   &0.998   &0.972     &0.773 &1.000       \\ \hline\hline
\end{tabular}
\end{center}
\end{table}
}
\end{center}


\section{Conclusion}

We provide a pairwise Hotelling method for testing whether a mean vector is equal to a given vector for the one-sample test, or testing whether two mean vectors are equal for the two-sample test in the high-dimensional low-sample size setting.
Our proposed pairwise Hotelling' test (PHT) statistics are different from the existing tests, including  the unscaled Hotelling's tests
(UHT),  the diagonal Hotelling's tests (DHT) and
the  regularized Hotelling's tests (RHT). Specifically,  UHT and DHT both ignore the correlation information among the covariates.  When some covariates exhibit strong correlations in the data, neither of the two methods can provide a  satisfactory performance. On the contrary, RHT does account for the correlations, yet it involves a regularized covariance matrix. When the sample size is relatively small compared to the dimension, the regularized covariance matrix can be very noisy, especially when  the  covariance matrix is sparse. Consequently, the test statistics involving the sample covariance matrix may lead to inflated type I error rates and/or suffer from low statistical power.
Our proposed pairwise Hotelling method  overcomes the drawbacks of  DHT and RHT. Specifically,
we first perform a screening procedure and pick up the  covariate pairs that exhibit strong correlations, then construct the classical Hotelling's test statistics for those covariate pairs. For the remaining covariates that are  weakly correlated with others, we construct the squares of componentwise $t$ statistics for each of the individual covariates.
Our proposed PHT statistics are then the summation over all the Hotelling's test statistics and the squared $t$ statistics. Simulation results show that our new tests can significantly improve the statistical power when some covariates are highly correlated; and even for the settings when most covariates are weakly correlated, our proposed tests are still able to maintain a high power compared to the existing tests in the literature.

%
 \spacingset{1.08}

\bibliographystyle{dcu}

\bibliography{reference-all}
\clearpage

\begin{appendices}

\renewcommand{\thesection}{ \Alph{section}:}
\setcounter{section}{0}
\section{Some preliminary results}
\renewcommand{\thesection}{\Alph{section}}

\begin{lemma}\label{tr_inequ}
Under the assumptions of model (\ref{modelChen1}),
let $\bm{Z}_1$ and $\bm{Z}_2$ be independent copies of $\bm{Z}$.
Then for any symmetric matrices $\Gamma_{1}, \Gamma_2 \in \mathbb{R}^{q \times q}$,
we have
\begin{eqnarray}\label{lemma2.0}
E\{(\bm{Z}_1^T \Gamma_1 \bm{Z}_2)^4\}
&=&3{\rm tr}^2(\Gamma_1^2)+6{\rm tr}(\Gamma_1^4)
+6 \Delta {\rm tr}(\Gamma_1 ^2 \odot \Gamma_1  ^2)
+\Delta^2 \sum_{i,i=1}^{q} ({\Gamma_1})_{ij}^4 \nonumber \\
&\leq&
3{\rm tr}^2(\Gamma_1 ^2)+6{\rm tr}(\Gamma_1^4)+
6 \Delta {\rm tr} (\Gamma_1^4)+\Delta^2 {\rm tr}^2(\Gamma_1^2),
\end{eqnarray}
where  $\odot$  is the Hadamard product,
and $({\Gamma_1})_{ij}$ is the $(i,j)$th element of ${\Gamma_1}$.
Furthermore,
\begin{equation}\label{lemma2.1}
E\{(\bm{Z}_1^T \Gamma_1 \bm{Z}_1) (\bm{Z}_1^T \Gamma_2 \bm{Z}_1)\}
={\rm tr}(\Gamma_1){\rm tr}(\Gamma_2)+2{\rm tr}(\Gamma_1 \Gamma_2)
+\Delta {\rm tr}(\Gamma_1 \odot \Gamma_2).
\end{equation}
\end{lemma}

\vskip12pt
\proof
Note that ${\rm tr}(\Gamma_1^2 \odot \Gamma_1^2)\leq {\rm tr}(\Gamma_1^4)$
and $\sum_{i,j=1}^{q}({\Gamma_1})_{ij}^4 \leq (\sum_{i,j=1}^{q}({\Gamma_1})_{ij}^2 )^2
={\rm tr}^2({\Gamma_1}^2)$.
Hence, the inequality in (\ref{lemma2.0}) holds.
The rest of the proof follows the same as that for
Proposition A.1 in \citet{chen2010matrices}.
\hfill$\Box$

\vskip12pt

\begin{lemma}\label{Lemmaeigen}
Under conditions (C1), (C5) and (C5$'$), we have
\begin{description}
\item[$(i)$]
the eigenvalues of $P_\mathcal{O}$ are  bounded away from 0 and $\infty$, which indicates that
$ 1/\lambda_1(\Sigma) \leq \lambda_p(P_\mathcal{O}) \leq \lambda_1(P_\mathcal{O})\leq K_0/\lambda_p(\Sigma)$;
\item[$(ii)$]
${\rm tr}(\Lambda_1^4)/{\rm tr}^2(\Lambda_1^2)=o(1)$
 and  ${\rm tr}(\Lambda_1^2)=O(p)$, where
$\Lambda_1  =\Sigma^{1/2} P_{\mathcal{O}} \Sigma^{1/2}$;
\item[$(iii)$]
${\bm \mu}^T P_{\mathcal{O}} \Sigma P_{\mathcal{O}}{\bm \mu}/{\rm tr}(\Lambda_1^2)=o(n^{-1})$
and
$({\bm \mu}_1-{\bm \mu}_2)^T P_{\mathcal{O}} \Sigma P_{\mathcal{O}}({\bm \mu}_1-{\bm \mu}_2)/{\rm tr}(\Lambda_1^2)
=o(N^{-1})$.
\end{description}
\end{lemma}

\vskip12pt
\proof  We first show $(i)$.
For any ${\bm \zeta }=(\zeta_1,\ldots,\zeta_p)^T \in \mathbb{R}^{p}$ with $\| {\bm \zeta } \|=1$,
we have
\begin{eqnarray*}
{\bm \zeta}^T P_\mathcal{O}{\bm \zeta}
&=&\sum_{(i,j) \in A_1}
{\bm \zeta}^TP_{ij}^T (P_{ij} \Sigma P_{ij}^T)^{-1}P_{ij}{\bm \zeta}
 +\sum_{i \in A_2} {\bm \zeta}^T P_{i}^T ( P_{i} \Sigma P_{i}^T)^{-1} P_{i} {\bm \zeta}\\
&=&\sum_{(i,j) \in A_1}
{\bm \zeta}_{\{i,j\}}^T \Sigma_{\{i,j\}}^{-1} {\bm \zeta}_{\{i,j\}}
 +\sum_{i \in A_2} {{\zeta}_i^2}/{\sigma_{ii}},
\end{eqnarray*}
where ${\bm \zeta}_{\{i,j\}}=(\zeta_i, \zeta_j)^T$.

Note that $\lambda_p(\Sigma) \leq \underset{i=1, \dots, p}{\min}\sigma_{ii}
\leq \underset{i=1, \dots, p}{\max}\sigma_{ii} \leq \lambda_1(\Sigma)$
 and $\lambda_p(\Sigma) \leq \lambda_2(\Sigma_{\{i,j\}}) \leq \lambda_1(\Sigma_{\{i,j\}}) \leq \lambda_1(\Sigma)$,
where $\Sigma_{\{i,j\}}=P_{ij} \Sigma P_{ij}^T$.
Hence,  $\sigma_{ii}^{-1}$ are  bounded uniformly, and
for any $(i,j)\in A_1$,
\begin{align}\label{uniform_Sigma}
1/\lambda_1(\Sigma) \leq \lambda_2(\Sigma_{\{i,j\}}^{-1}) \leq \lambda_1(\Sigma_{\{i,j\}}^{-1}) \leq 1/\lambda_p(\Sigma).
\end{align}
Consequently,
$ 1/\lambda_1(\Sigma) \leq {\bm \zeta}^T P_\mathcal{O}{\bm \zeta}\leq
K_0/\lambda_p(\Sigma)$. This shows that the eigenvalues of $P_\mathcal{O}$ are also  bounded uniformly by $1/\lambda_1(\Sigma)$
and $K_0/\lambda_p(\Sigma)$.

To show $(ii)$, we  note that
$\lambda_1(\Lambda_1)\leq \lambda_1(P_{\mathcal{O}}) \lambda_1(\Sigma) $
and
$\lambda_p(\Lambda_p) \geq \lambda_p(P_{\mathcal{O}}) \lambda_p(\Sigma)$.
This indicates that
${\rm tr}(\Lambda_1^4)=O(p)$, and
\begin{eqnarray}\label{ord_trace}
 {\rm tr}(\Lambda_1^2)=O(p).
\end{eqnarray}
Hence,
${\rm tr}(\Lambda_1^4)/{\rm tr}^2(\Lambda_1^2) =o(1)$.

Note also that
$\lambda_1(P_{\mathcal{O}} \Sigma P_{\mathcal{O}})$
$\leq \lambda_1^2(P_{\mathcal{O}}) \lambda_1(\Sigma)$
and $\lambda_p(P_{\mathcal{O}} \Sigma P_{\mathcal{O}})
\geq \lambda_p^2(P_{\mathcal{O}}) \lambda_p(\Sigma)$,
which are bounded away from $0$ and $\infty$.
Therefore,
$({\bm \mu}-{\bm \mu}_0)^T P_{\mathcal{O}} \Sigma P_{\mathcal{O}}({\bm \mu}-{\bm \mu}_0)/[
({\bm \mu}-{\bm \mu}_0)^T({\bm \mu}-{\bm \mu}_0)]
=O(1).$
Finally, by  (\ref{ord_trace}) and  condition (C5$'$), we obtain $(iii)$.
\hfill$\Box$

\vskip12pt
\begin{lemma}\label{boundlemma5}
For the one-sample test, let
${\bm{\mu}}_{ij}=P_{ij}{\bm{\mu}}=(\mu_{i}, \mu_j)^T,$
$\tilde{\bm{{X}}}_{ij;s}=\Sigma_{\{i,j\}}^{-\frac{1}{2}}({\bm{X}}_{ij;s}-
{\bm{\mu}}_{ij})=(\tilde{X}_{ki},\tilde{X}_{kj})^T,$
and $\tilde{\bm{\mu}}_{ij}=\Sigma_{\{i,j\}}^{-\frac{1}{2}}{\bm{\mu}_{ij}}.$
Let  $\widetilde{S}_{\{i,j\}}^{(s,t)}$ be the sample covariance matrix without observations $\tilde{\bm{{X}}}_{ij;s}$ and $\tilde{\bm{{X}}}_{ij;t}$.
Under conditions (C3)--(C5), we have
\begin{description}
\item[$(i)$]
$E( \tilde{\bm{{X}}}_{ij;s} )=\bm{0}$ and
$\text{Var}( \tilde{\bm{{X}}}_{ij;s} )=I_2$. In addition, $\widetilde{S}_{\{i,j\}}^{(s,t)}=\Sigma_{\{i,j\}}^{-\frac{1}{2}}S_{\{i,j\}}^{(s,t)} \Sigma_{\{i,j\}}^{-\frac{1}{2}}$,
 and
$E \big(\widetilde{S}_{\{i,j\}}^{(s,t)} \big)=I_2$;
\item[$(ii)$]
for any positive integer $m_0$ satisfying condition (C4),
the higher order moments
$E(\tilde{X}_{i}^{4 m_0+2})$ and $E(\tilde{X}_{j}^{4 m_0+2})$ are bounded uniformly
over $(i,j)\in A_1$,
where $\tilde{\bm{{X}}}_{ij}=
(\tilde{X}_i,\tilde{X}_j)^T
=\Sigma_{\{i,j\}}^{-\frac{1}{2}}({\bm{X}}_{ij}-{\bm{\mu}}_{ij})$.
In addition,
$E\big\| (\widetilde{S}_{\{i,j\}}^{(s,t)})^{-1} \big\|^8$
are bounded uniformly over $(i,j)\in A_1$,
and consequently,
$E \big\| \big(  ( \widetilde{S}_{\{i,j\}}^{(s,t)} )^{-1}-I_2  \big) \big\|^4
=O(n^{-2})$ hold uniformly over $(i,j)\in A_1$.
\end{description}
\end{lemma}

\vskip12pt
\proof
By direct calculation for the mean, covariance matrix and sample covariance matrix of $\tilde{\bm{{X}}}_{ij;s}$, it is easy to verify that $(i)$ holds.

To show $(ii)$, we note that
\begin{align*}
E(\widetilde{X}_{i}^{4 m_0+2})
&=E\Big(\big|(1,0) \times  \tilde{\bm{{X}}}_{ij}\big| \Big)^{4 m_0+2}
\leq
E\Big(\big\|(1,0) \Sigma_{\{i,j\}}^{-\frac{1}{2}}({\bm{X}}_{ij}-{\bm{\mu}}_{ij})  \big\|\Big)  \\
&\leq
\big\|\Sigma_{\{i,j\}}^{-\frac{1}{2}}\big\|^{4 m_0+2}
\times E\Big( \big| {X}_{i}-{\mu}_{i}  \big| \Big)^{4 m_0+2}.
\end{align*}
Then by (\ref{uniform_Sigma}) and  condition (C4),
we can conclude that $E(\widetilde{X}_{i}^{4 m_0+2})$ are bounded uniformly.
Similarly, $E(\widetilde{X}_{j}^{4 m_0+2})$ are also bounded uniformly.

Besides,
$E\big\| (\widetilde{S}_{\{i,j\}}^{(s,t)})^{-1} \big\|^8
=E\big\| ( S_{\{i,j\}}^{(s,t)})^{-1} \Sigma_{\{i,j\}} \big\|^8
\leq E\big\| ( S_{\{i,j\}}^{(s,t)})^{-1}\big\|^8 \times\big\| \Sigma_{\{i,j\}} \big\|^8.$
By condition (C4),
$E\big\| ( \widetilde{S}_{\{i,j\}}^{(s,t)})^{-1} \big\|^4$
are bounded uniformly  over $(i,j)\in A_1$.
In addition,
\begin{align*}
E \big(  \big\|  ( \widetilde{S}_{\{i,j\}}^{(s,t)} )^{-1}-I_2  \big\|^4\big)
= &~
E\Big\{  \big\| ( \widetilde{S}_{\{i,j\}}^{(s,t)} )^{-1}  \big\|^4
\times \big\| I_2 -\widetilde{S}_{\{i,j\}}^{(s,t)} \big\|^4 \Big\}     \\
\leq&~
\Big\{E  \big\| ( \widetilde{S}_{\{i,j\}}^{(s,t)} )^{-1}  \big\|^8 \Big\}^{1/2}
\times
\Big\{E \big\| I_2 -\widetilde{S}_{\{i,j\}}^{(s,t)} \Big\|^8 \big\}^{1/2}.
\end{align*}
Note also that for $(i,j) \in A_1$,
$E \| I_2-\widetilde{S}_{\{i,j\}}^{(s,t)} \|^8$
 are finite combination of higher order moments
and the highest order terms are
$E(\tilde{X}_{ki}^{16})$ and $E(\tilde{X}_{kj}^{16})$ for $k\neq s,t$. Then by condition (C4), we can see that
\begin{align} \label{Bounded.ord1}
E \big\| I_2-\widetilde{S}_{\{i,j\}}^{(s,t)} \big\|^8 =O(n^{-4})
\end{align}
hold uniformly  over $(i,j) \in A_1$.
Together with  the fact that
$E \big\| ( \widetilde{S}_{\{i,j\}}^{(s,t)} )^{-1} \big\|^8$ are bounded uniformly,
we complete the proof of $(ii)$.
\hfill$\Box$

\vskip12pt
\begin{lemma}\label{boundlemma5.2sam}
For the two-sample test, let
${\bm{\mu}}_{1,ij}=P_{ij}{\bm{\mu}_1}=(\mu_{1i}, \mu_{1j})^T,$
${\bm{\mu}}_{2,ij}=P_{ij}{\bm{\mu}_2}=(\mu_{2i}, \mu_{2j})^T,$
$\tilde{\bm{{X}}}_{ij;s}=\Sigma_{\{i,j\}}^{-\frac{1}{2}}({\bm{X}}_{ij;s}-
{\bm{\mu}}_{1,ij})=(\tilde{X}_{ki},\tilde{X}_{kj})^T,$
$\tilde{\bm{{Y}}}_{ij;s}=\Sigma_{\{i,j\}}^{-\frac{1}{2}}({\bm{Y}}_{ij;s}-
{\bm{\mu}}_{2,ij})=(\tilde{Y}_{ki},\tilde{Y}_{kj})^T,$
$\tilde{\bm{\mu}}_{1,ij}=\Sigma_{\{i,j\}}^{-\frac{1}{2}}{\bm{\mu}_{1, ij}},$
and
$\tilde{\bm{\mu}}_{2,ij}=\Sigma_{\{i,j\}}^{-\frac{1}{2}}{\bm{\mu}_{2, ij}}.$
Let  $\widetilde{S}_{1, \{i,j\}}^{(s,t)}$ be the pooled sample covariance matrix for the observations $\{\tilde{\bm{{X}}}_{ij;k}\}_{k \neq s,t}^{n_1}$ and $\{\tilde{\bm{{Y}}}_{ij;k}\}_{k=1}^{n_2}$,
and
$\widetilde{S}_{2, \{i,j\}}^{(s,t)}$ be the pooled sample covariance matrix for the observations $\{\tilde{\bm{{X}}}_{ij;k}\}_{k =1}^{n_1}$ and $\{ \tilde{\bm{{Y}}}_{ij;k} \}_{k \neq s,t}^{n_2}$.
In addition,  let  ${S}_{12; \{i,j\}}^{(s,t)}$=$P_{ij}S_{12,*}^{(s,t)}P_{ij}^T$
and
$\widetilde{S}_{12;\{i,j\}}^{(s,t)}$=
$\Sigma_{ij}^{-1/2}S_{12;\{i,j\}}^{(s,t)}\Sigma_{ij}^{-1/2}$.
Under conditions (C3$'$)--(C5$'$), we have
\begin{description}
\item[$(i)$]
$E( \tilde{\bm{{X}}}_{ij;s} )= E( \tilde{\bm{{Y}}}_{ij;k} )=\bm{0}$ and
$\text{Var}( \tilde{\bm{{X}}}_{ij;s} )=
\text{Var}( \tilde{\bm{{Y}}}_{ij;k} )= I_2$.
In addition,
$\widetilde{S}_{1, \{i,j\}}^{(s,t)}
=\Sigma_{\{i,j\}}^{-\frac{1}{2}}S_{1, \{i,j\}}^{(s,t)} \Sigma_{\{i,j\}}^{-\frac{1}{2}}$,
$\widetilde{S}_{2, \{i,j\}}^{(s,t)}=\Sigma_{\{i,j\}}^{-\frac{1}{2}}S_{2, \{i,j\}}^{(s,t)} \Sigma_{\{i,j\}}^{-\frac{1}{2}}$,
and
$E (\widetilde{S}_{1, \{i,j\}}^{(s,t)} )=E (\widetilde{S}_{2, \{i,j\}}^{(s,t)} )=I_2$;


\item[$(ii)$]
for any positive integer $m_0$ satisfying condition (C4$'$),
the higher order moments
$E(\tilde{X}_{i}^{4 m_0+2})$,
$E(\tilde{X}_{j}^{4 m_0+2})$,
$E(\tilde{Y}_{i}^{4 m_0+2})$,
$E(\tilde{Y}_{j}^{4 m_0+2})$
are bounded uniformly over $(i,j)\in A_1$,
where $\tilde{\bm{{X}}}_{ij}= (\tilde{X}_i,\tilde{X}_j)^T
=\Sigma_{\{i,j\}}^{-\frac{1}{2}}({\bm{X}}_{ij}-{\bm{\mu}}_{1,ij})$
and $\tilde{\bm{{Y}}}_{ij}= (\tilde{Y}_i,\tilde{Y}_j)^T
=\Sigma_{\{i,j\}}^{-\frac{1}{2}}({\bm{Y}}_{ij}-{\bm{\mu}}_{2,ij})$;

\item[$(iii)$]
$E\big\| ( \widetilde{S}_{1, \{i,j\}}^{(s,t)})^{-1} \big\|^8$,
$E\big\| ( \widetilde{S}_{2, \{i,j\}}^{(s,t)})^{-1} \big\|^8$
and $E\big\| ( \widetilde{S}_{12, \{i,j\}}^{(s,t)})^{-1} \big\|^8$
are bounded uniformly over $(i,j)\in A_1$.
Furthermore,
$E \big\| \big(  ( \widetilde{S}_{1, \{i,j\}}^{(s,t)} )^{-1}-I_2  \big) \big\|^4
=O(N^{-2})$, $E \big\| \big(  ( \widetilde{S}_{2, \{i,j\}}^{(s,t)} )^{-1}-I_2  \big) \big\|^4
=O(N^{-2})$ and
$E \big\| \big(  ( \widetilde{S}_{12, \{i,j\}}^{(s,t)} )^{-1}-I_2  \big) \big\|^4
=O(N^{-2})$
hold uniformly over $(i,j)\in A_1$.
\end{description}

\end{lemma}

\vskip12pt
\proof

The proof is similar as that for Lemma \ref{boundlemma5} and so we omit it.
\hfill$\Box$
\end{appendices}

\begin{appendices}
\renewcommand{\thesection}{\Alph{section}:}
\section{ Proofs of Theorem 2.1, Theorem 2.2 and Lemma 1}
\renewcommand{\thesection}{\Alph{section}}

\subsection{Proof of Theorem \ref{pro1}}\label{Prof.pro1}

By the definition of event $A_1$, the following  events
are equivalent:
\begin{align*}
\{\hat{A}_1 = A_1\}
&=   \Big(  \cap_{i=1}^{p} \cap_{j=i+1}^{p} \{ \hat{\tau}_{ij} < \tau_0 \big| \tau_{ij} < \tau_0 \} \Big)   \cap
   \Big( \cap_{i=1}^{p}  \cap_{j=i+1}^{p} \{ \hat{\tau}_{ij}> \tau_0  \big|
\tau_{ij}  >\tau_0 \}\Big).
\end{align*}
Therefore,
\begin{align}\label{set1}
P\Big(\{\hat{A}_1 \neq A_1\}\Big)
  &\leq  P\Big(\cup_{i=1}^{p}  \cup_{j=i+1}^{p}
  \{ \hat{\tau}_{ij} > \tau_0 | \tau_{ij}< \tau_0 \} \Big)
  + P\Big( \cup_{i=1}^{p}
  \cup_{j=i+1}^{p} \{ \hat{\tau}_{ij}< \tau_0 | \tau_{ij}>\tau_0\}\Big)\nonumber\\
  &\leq \sum_{i=1}^{p} \sum_{j=i+1}^{p}
  P\Big(\{ \hat{\tau}_{ij}> \tau_0 | \tau_{ij}<\tau_0 \}\Big) +
  \sum_{i=1}^{p}  \sum_{j=i+1}^{p} P\Big( \{ \hat{\tau}_{ij} < \tau_0 | \tau_{ij} > \tau_0 \}\Big).
\end{align}

Under the assumptions of  Theorem \ref{pro1}, we have
$\underset{i,j=1,\ldots,p}{\lim\inf}\{ \tau_{ij}  \big|  \tau_{ij} >\tau_0 \}=c_1>\tau_0$
and $\underset{i,j=1,\ldots,p}{\lim\inf}\{ \tau_{ij}  \big|  \tau_{ij} <\tau_0 \}=c_2<\tau_0$.
Let $\epsilon_0=\min\{c_1-\tau_0, \tau_0-c_2\}$, then
\begin{eqnarray*}
\{ \hat{\tau}_{ij} >\tau_0 |  \tau_{ij} <\tau_0\} \subseteq
\{|\hat{\tau}_{ij} -\tau_{ij}| >\epsilon_0 |  \tau_{ij} <\tau_0\}
\subseteq
\{|\hat{r}_{ij} -r_{ij}| >\epsilon_0 |  \tau_{ij} <\tau_0\},
\end{eqnarray*}
\begin{eqnarray*}
\{ \hat{\tau}_{ij} <\tau_0 |  \tau_{ij} >\tau_0\} \subseteq
\{|\hat{\tau}_{ij} -\tau_{ij}| >\epsilon_0 |  \tau_{ij} >\tau_0\}
\subseteq
\{|\hat{r}_{ij} -r_{ij}| >\epsilon_0 |  \tau_{ij} >\tau_0\}.
\end{eqnarray*}
Further by Hoeffding's inequality (see, e.g., Lemma 1  in  \cite{li2012robust}) for  Kendall's tau statistics,
\begin{align*}
&P\Big(\{\hat{\tau}_{ij}>\tau_0 | \tau_{ij}<\tau_0\}\Big)
 \leq P\Big(\{|\hat{r}_{ij} -r_{ij}| >\epsilon_0 | \tau_{ij}<\tau_0\}\Big)\leq 2 \exp\Big(-\frac{n\epsilon_0^2}{4}\Big),\\
 &P\Big(\{\hat{\tau}_{ij}<\tau_0 | \tau_{ij}>\tau_0\}\Big)
 \leq P\Big(\{|\hat{r}_{ij} -r_{ij}| >\epsilon_0 | \tau_{ij}>\tau_0\}\Big)\leq 2
 \exp\Big(-\frac{n\epsilon_0^2}{4}\Big).
\end{align*}
Plugging them into (\ref{set1}), we have
\begin{equation}\label{ineqA_1}
P\Big(\{\hat{A}_1 \neq A_1\}\Big)\leq 2{p(p-1)}\exp\Big(-\frac{n\epsilon_0^2}{4}\Big)
\to 0.
\end{equation}

Next, by the definition of event $A_2$, we have
\begin{align}\label{ineqA_2.1}
\{\hat{A}_2 \neq A_2\}
=\cup_{i=1}^{p} \Big(
\{  i \not\in \hat{A}_2 \big| i \in A_2  \}
\cup \{  i \in \hat{A}_2 \big| i \not\in A_2 \} \Big).
\end{align}
Accordingly,
\begin{align}\label{ineqA_2}
&~~~~\big\{  i \not\in  \hat{A}_2 \big| i \in A_2  \big\}
    \cup \big\{  i \in \hat{A}_2 \big| i \not\in A_2  \big\} \nonumber\\
&= \Big( \cup_{j \neq i}^{p} \{ \hat{\tau}_{ij} > \tau_0 \big| i \in  A_2  \} \Big)
   \cup
   \Big(  \cap_{j \neq i}^{p} \{ \hat{\tau}_{ij} < \tau_0  \big| i \not\in  A_2 \}  \Big)  \nonumber\\
&\subseteq   \Big(  \cup_{j \neq i}^{p}
   \{  |\hat{\tau}_{ij} - {\tau}_{ij}| > \epsilon_0  \big| {\tau}_{ij}<\tau_0 \} \Big)
  \cup
   \Big(  \cup_{j \neq i}^{p} \{ |\hat{\tau}_{ij} - {\tau}_{ij}| > \epsilon_0  \big| {\tau}_{ij}> \tau_0 \}  \Big)
\end{align}
This indicates that, as $n \to \infty$,
\begin{align}\label{final_set2}
P\Big(\{\hat{A}_2\neq A_2\}\Big)\leq 2{p(p-1)} \exp\Big(-\frac{n\epsilon_0^2}{4}\Big) \to 0.
\end{align}
Combining (\ref{ineqA_1}) and (\ref{final_set2}), we complete the proof of Theorem  \ref{pro1}.
\hfill$\Box$

\subsection{Proof of Theorem \ref{th1}} \label{Prof.th1}
Without loss of generality, we assume $\bm{\mu}_0={\bm 0}$.
Let
\begin{eqnarray*}
U_{n1}  &=&  \frac{1}{n(n-1)}\sum_{s=1}^{n}\sum_{t \neq s}^{n}{\bm{X}}_{s}^T P_{\mathcal{O}} {\bm{X}}_{t},\\
U_{n2}  &=&  \frac{1}{n(n-1)}\sum_{s=1}^{n}\sum_{t \neq s }^{n}{\bm{X}}_{s}^T \big(\widehat{P}_{\mathcal{O}}^{(s,t)}-P_{\mathcal{O}}\big) {\bm{X}}_{t}.
\end{eqnarray*}
Then to show Theorem \ref{th1}, it suffices to show that

\begin{eqnarray}\label{asy_Un1}
\frac{U_{n1}-{\bm \mu}^T P_{\mathcal{O}} {\bm \mu}}{\sqrt{2n^{-2}{\rm tr}(\Lambda_1^2)}} \stackrel{D}{\longrightarrow}N(0,1) {\rm ~~as~~} n\to\infty,
\end{eqnarray}
and
\begin{eqnarray}\label{asy_Un2}
\frac{U_{n2}}{\sqrt{2n^{-2}{\rm tr}(\Lambda_1^2)}} \stackrel{P}{\longrightarrow}0 {\rm ~~as~~} n\to\infty.
\end{eqnarray}
%


\vskip 12pt
\noindent{\it  {\bf Part I: Proof of (\ref{asy_Un1})}}
\vskip 12pt
Let
$U_{n11}=\sum_{s=1}^{n}\sum_{t\neq s}^{n}
({\bm{X}}_{s}-{\bm\mu})^T P_{\mathcal{O}} ({\bm{X}}_{t}-{\bm \mu})/[n(n-1)]$
and
$U_{n12}=2 \sum_{t=1}^{n} {\bm\mu}^T P_{\mathcal{O}}({\bm{X}}_{t}-{\bm \mu})/n,$
then $U_{n1}-{\bm \mu}^T P_{\mathcal{O}} {\bm \mu}  = U_{n11}+U_{n12}.$
Note that $E(U_{n12})=0$, and also by  $(iii)$ in Lemma \ref{Lemmaeigen}, we have
$\text{Var}(U_{n12})=O({\bm \mu}^T P_{\mathcal{O}} \Sigma P_{\mathcal{O}}{\bm \mu} /n)
=o({\rm tr}(\Lambda_1^2)/n^2)$. Therefore,
${U_{n12}}/{\sqrt{2n^{-2}{\rm tr}(\Lambda_1^2)}}\stackrel{P}{\longrightarrow}0.
$

Next, we show that
\begin{eqnarray}
\frac{U_{n11}}{\sqrt{2n^{-2}{\rm tr}(\Lambda_1^2)}} \stackrel{D}{\longrightarrow}N(0,1) {\rm ~~as~~} n\to\infty.
\end{eqnarray}
Without loss of generality, let ${\bm \mu}={\bm 0}$. Define the sequence
$V_{nt}=\sum_{s=1}^{t-1}
{\bm{X}}_{s}^T P_{\mathcal{O}}{\bm{X}}_{t}/ [n(n-1)]$ and
$\Psi_{m}=\sum_{t=2}^{m}V_{nt}$, $m=2,\ldots,n$.
Let $\mathcal{F}_{m}({\bm X}_1,\ldots,{\bm X}_m)$ be the $\sigma$ algebra
generated by ${\bm X}_1,\ldots,{\bm X}_m$ for $m=2,\ldots,n$. Consequently, $U_{n1}=2\sum_{t=2}^{n}V_{nt}$.

For applying the central limit theorem in Corollary 3.1 of \cite{hall2014},
we need to verify three statements as follows:
\begin{description}
  \item[$(i)$] For each $n$, $\{\Psi_{m}, \mathcal{F}_{m} \}_{m=1}^n$ is the sequence of zero mean and a square integrable martingale;
  \item[$(ii)$] $ \eta_n/ \sigma_{n}^{2} \stackrel{P}{\longrightarrow} 1/4 $, where $\eta_n=\sum_{t=2}^{n}E(V_{nt}^2|\mathcal{F}_{n,t-1})$ and $\sigma_{n}^2=\text{Var}(U_{n11})$;
  \item[$(iii)$]
$\sum_{t=2}^{n}  E\big\{V_{nt}^2 I(|V_{nt}|>\epsilon \sigma_{n})|\mathcal{F}_{n,t-1}\big\}/\sigma_{n}^{2}\stackrel{P}{\longrightarrow}0$.
\end{description}

For $(i)$, it is easy to verify that
$V_{nt}$ is zero mean and square integrable.
Consequently, $\Psi_{m}$ is also zero mean and square integrable.
Thus, we only need to show that $\Psi_{m}$ is a martingale.
Note that for $l>m$,
\begin{eqnarray*}
E(\Psi_{l}|\mathcal{F}_{m}) &=& \sum_{t=2}^{l}E(V_{nt}|\mathcal{F}_{m})\\
&=&  \Psi_{m} +  \sum_{t=m+1}^{l}E(V_{nt}|\mathcal{F}_{m}) \\
&=&   \Psi_{m}+  \sum_{t=m+1}^{l}E\Big(\sum_{s=1}^{t-1}{\bm X}_s^T P_\mathcal{O}|\mathcal{F}_{m}\Big) E\Big({\bm X}_t \Big)=\Psi_{m}.
\end{eqnarray*}
This completes the proof of $(i)$.

For $(ii)$, we have
\begin{eqnarray*}
E(V_{nt}^2|\mathcal{F}_{n,t-1})
 &=&  \frac{1}{n^2(n-1)^2}
E\Big( \Big(\sum_{s=1}^{t-1}{\bm{X}}_{s}^T P_{\mathcal{O}} {\bm{X}}_{t} \Big)^2
|\mathcal{F}_{n,t-1} \Big)\\
 &=& \frac{1}{n^2(n-1)^2} E\Big(\sum_{s=1}^{t-1}
 {\bm{X}}_{s}^T P_{\mathcal{O}}{\bm{X}}_{t}{\bm{X}}_{t}^T P_{\mathcal{O}}{\bm{X}}_{s}
 |\mathcal{F}_{n,t-1} \Big) \\
 & &+ \frac{1}{n^2(n-1)^2} E\Big(\sum_{s \neq l}^{t-1}
 {\bm{X}}_{s}^T P_{\mathcal{O}}{\bm{X}}_{t}{\bm{X}}_{t}^T P_{\mathcal{O}}{\bm{X}}_{l}
  |\mathcal{F}_{n,t-1} \Big) \\
 &=& \frac{1}{n^2(n-1)^2} \Big(\sum_{s=1}^{t-1}
 {\bm{X}}_{s}^T P_{\mathcal{O}}\Sigma P_{\mathcal{O}}{\bm{X}}_{s}
 + \sum_{s \neq l}^{t-1}
 {\bm{X}}_{s}^T P_{\mathcal{O}}\Sigma P_{\mathcal{O}}{\bm{X}}_{l} \Big),
\end{eqnarray*}
and
\begin{eqnarray}\label{sigma2}
\sigma_{n}^2
= \frac{1}{n^2(n-1)^2}\text{Var}
\Big(\sum_{t\neq s}^{n}{\bm{X}}_{s}^T P_{\mathcal{O}}{\bm{X}}_{t}\Big)
=\frac{2}{n(n-1)}{\rm tr }(\Lambda_1^2).
\end{eqnarray}
Let
\begin{align*}
\eta_n
&= \sum_{t=2}^{n}E(V_{nt}^2|\mathcal{F}_{t-1})
= \frac{1}{n^2(n-1)^2}\sum_{t=2}^{n} \Big( \sum_{s=1}^{t-1}
{\bm{X}}_{s}^T P_{\mathcal{O}}\Sigma P_{\mathcal{O}}{\bm{X}}_{s}
+ \sum_{s \neq l}^{t-1}{\bm{X}}_{s}^T P_{\mathcal{O}}\Sigma P_{\mathcal{O}}{\bm{X}}_{l} \Big)\\
&= \frac{1}{n^2(n-1)^2} \Big(\sum_{s=1}^{n-1}
(n-s){\bm{X}}_{s}^T P_{\mathcal{O}}\Sigma P_{\mathcal{O}}{\bm{X}}_{s}
+ \sum_{t=2}^{n}\sum_{s \neq l}^{t-1}{\bm{X}}_{s}^T P_{\mathcal{O}}\Sigma P_{\mathcal{O}}{\bm{X}}_{l} \Big)\\
&= \frac{1}{n^2(n-1)^2} \Big( \sum_{s=1}^{n-1}
(n-s){\bm{X}}_{s}^T P_{\mathcal{O}}\Sigma P_{\mathcal{O}}{\bm{X}}_{s} \Big)
+ \frac{1}{n^2(n-1)^2} \Big(\sum_{t=2}^{n}\sum_{s \neq l}^{t-1}{\bm{X}}_{s}^T P_{\mathcal{O}}\Sigma P_{\mathcal{O}}{\bm{X}}_{l} \Big)\\
&= \eta_{n1}+\eta_{n2},
\end{align*}
Since
$E({\bm{X}}_{s}^T P_{\mathcal{O}}\Sigma P_{\mathcal{O}}{\bm{X}}_{s})={\rm tr}
(P_{\mathcal{O}} \Sigma P_{\mathcal{O}}\Sigma) = {\rm tr}(\Lambda_1^2),$
we have
\begin{eqnarray}\label{ii_1}
E\left( \frac{ \eta_{n1} }{ \sigma_{n} } \right)
&=&  \frac{1}{4 {\rm tr }(\Lambda_1^2)} E\Big({\bm{X}}_{s}^T P_{\mathcal{O}}\Sigma P_{\mathcal{O}}{\bm{X}}_{s}\Big)=\frac{1}{4}.
\end{eqnarray}

By the linear model (\ref{modelChen1})
and (\ref{lemma2.1}), we have
\begin{eqnarray}\label{U_n11_1}
E\big({\bm{X}}_{s}^T P_{\mathcal{O}}\Sigma P_{\mathcal{O}}{\bm{X}}_{s})^2\big)
&=&
E\big( ( {\bm{Z}}_{s}^T  \Gamma_3 {\bm{Z}}_{s})^2\big)   \nonumber\\
&=&
{\rm tr}(\Gamma_3){\rm tr}(\Gamma_3)+2 {\rm tr}(\Gamma_3^2)+
\Delta{\rm tr}(\Gamma_3 \odot \Gamma_3)                    \nonumber\\
&\leq&
{\rm tr}^2(\Gamma_3)+2 {\rm tr}(\Gamma_3^2)+\Delta{\rm tr}(\Gamma_3^2) \nonumber\\
&=&
{\rm tr}^2(\Lambda_1^2)+ 2 {\rm tr}(\Lambda_1^4)+\Delta{\rm tr}(\Lambda_1^4),
\end{eqnarray}
where $\Gamma_3=C^TP_{\mathcal{O}} \Sigma P_{\mathcal{O}}C$.
In what follows, we show that
$\text{Var}(\eta_{n1}/\sigma_{n}^2) \to 0$ as $n\to \infty$.
Note that
\begin{eqnarray*}
 \text{Var}\left( \frac{ \eta_{n1} }{\sigma_{n}^2} \right)
 &=& \frac{1}{\sigma_n^4} \frac{1}{n^4(n-1)^4} \text{Var} \Big(\sum_{s=1}^{n-1}
(n-s){\bm{X}}_{s}^T P_{\mathcal{O}}\Sigma P_{\mathcal{O}}{\bm{X}}_{s}\Big) \\
 &=& \frac{1}{\sigma_n^4} \frac{1}{n^4(n-1)^4} \sum_{s=1}^{n-1}
(n-s)^2 \text{Var} \Big\{\Big({\bm{X}}_{s}^T P_{\mathcal{O}}\Sigma P_{\mathcal{O}}{\bm{X}}_{s}\Big)^2 \Big\}\\
&=&  \frac{1}{4 {\rm tr}^2(\Lambda_1^2)}\frac{1}{n^2(n-1)^2}
\sum_{s=1}^{n-1}
(n-s)^2\Big\{ E\Big[\Big({\bm{X}}_{s}^T P_{\mathcal{O}}\Sigma P_{\mathcal{O}}{\bm{X}}_{s}\Big)^2 \Big]- {\rm tr}^2 (\Lambda_1^2)\Big\}\\
&=& O\Big(\frac{{\rm tr}(\Lambda_1^4)}{n{\rm tr}^2(\Lambda_1^2)}\Big).
\end{eqnarray*}
By $(ii)$ in Lemma \ref{Lemmaeigen}, we have
\begin{equation}\label{ii_2}
\text{Var}\left( \frac{ \eta_{n1} }{ \sigma_{n}^2 }\right)=  O\Big(\frac{{\rm tr}(\Lambda_1^4)}{n{\rm tr}^2(\Lambda_1^2)}\Big)\to 0.
\end{equation}
In addition, we can show that $E(\eta_{n2})=0$. Then similar to
the proof for (\ref{ii_2}),
\begin{eqnarray}\label{ii_3}
E\Big(\frac{\eta_{n2}}{\sigma_{n}^2}\Big)^2
&=&
\frac{1}{\sigma_{n}^4}E\Big(\frac{1}{n^2(n-1)^2} \Big(\sum_{t=2}^{n}\sum_{s \neq l}^{t-1}{\bm{X}}_{s}^T P_{\mathcal{O}}\Sigma P_{\mathcal{O}}{\bm{X}}_{l} \Big)\Big)^2 \nonumber \\
&=& \frac{1}{\sigma_{n}^4} \frac{4}{n^4(n-1)^4} \sum_{t=2}^{n}\sum_{s \leq l}^{t-1}
 E\Big( \Big({\bm{X}}_{s}^T P_{\mathcal{O}}\Sigma P_{\mathcal{O}}{\bm{X}}_{l} \Big)^2\Big)\nonumber \\
&=&O\Big(\frac{{\rm tr}(\Lambda_1^4)}{n{\rm tr}^2(\Lambda_1^2)}\Big)\to 0.
\end{eqnarray}
Combining (\ref{ii_1}), (\ref{ii_2}) and (\ref{ii_3}),
we complete the proof of $(ii)$.

For $(iii)$, since
$\sum_{t=2}^{n}  E\big\{V_{nt}^2 I(|V_{nt}|>\epsilon \sigma_{n})|\mathcal{F}_{n,t-1}\big\}/\sigma_{n}^{2}
\leq
\sum_{t=2}^{n}  E(V_{nt}^4 |\mathcal{F}_{t-1}) /(\epsilon^2 \sigma_{n}^{4}),
$
it suffices to show that
$\sum_{t=2}^{n}  E(V_{nt}^4 |\mathcal{F}_{t-1}) /(\epsilon^2 \sigma_{n}^{4})=o_p(1).$
By simple algebra, we can show that
\begin{eqnarray*}
E\Big(\sum_{t=2}^{n}E\big(V_{nt}^4 |\mathcal{F}_{t-1}\big)\Big)
  &=&  \frac{1}{n^4(n-1)^4} \sum_{t=2}^{n} E \Big( \Big( \sum_{s=1}^{t-1} {\bm{X}}_{s}^T P_{\mathcal{O}} {\bm{X}}_{t} \Big)^4\Big)
  =   3Q+P,
\end{eqnarray*}
where
$P=O(n^{-8})\sum_{t=2}^{n}\sum_{s=1}^{t-1}
E({\bm{X}}_{s}^T P_{\mathcal{O}} {\bm{X}}_{t} )^4$
and
$Q=O(n^{-8}) \sum_{t=3}^{n}\sum_{i\neq j}^{t-1}
E( ({\bm{X}}_{t}^T P_{\mathcal{O}} {\bm{X}}_{i} )$
$({\bm{X}}_{i}^T P_{\mathcal{O}}{\bm{X}}_{t} )
({\bm{X}}_{t}^T P_{\mathcal{O}} {\bm{X}}_{j} )
({\bm{X}}_{j}^T P_{\mathcal{O}} {\bm{X}}_{t} ))$.
Note that
\begin{align*}
Q &= O(n^{-8}) \sum_{t=3}^{n}\sum_{i\neq j}^{t-1}
E\Big(
\big({\bm{X}}_{t}^T P_{\mathcal{O}} {\bm{X}}_{i} \big)
\big({\bm{X}}_{i}^T P_{\mathcal{O}} {\bm{X}}_{t} \big)
\big({\bm{X}}_{t}^T P_{\mathcal{O}} {\bm{X}}_{j} \big)
\big({\bm{X}}_{j}^T P_{\mathcal{O}} {\bm{X}}_{t} \big)\Big)\\
&= O(n^{-5}) E\Big(
\big({\bm{X}}_{t}^T P_{\mathcal{O}} \Sigma P_{\mathcal{O}} {\bm{X}}_{t} \big)^2\Big).
\end{align*}
By (\ref{U_n11_1}), as $n \to \infty$ we can show that
$$\frac{Q}{\sigma_n^4}
=\frac{1}{4}\frac{n^2(n-1)^2Q}{{\rm tr }^2(\Lambda_1^2)}
=O\Big(\frac{1}{n}\big\{1+{{\rm tr }(\Lambda_1^4)}/{{\rm tr }^2(\Lambda_1^2)}\big\}\Big)
\to 0.$$
In addition, under model (\ref{model1}) we have
\begin{eqnarray*}
P= O(n^{-8})\sum_{t=2}^{n}\sum_{s=1}^{t-1}
E\Big({\bm{X}}_{s}^T P_{\mathcal{O}} {\bm{X}}_{t} \Big)^4=  O(n^{-8})\sum_{t=2}^{n}\sum_{s=1}^{t-1}
E\Big({\bm{Z}}_{s}^T C^T P_{\mathcal{O}} C {\bm{Z}}_{t} \Big)^4.
\end{eqnarray*}
Then by (\ref{lemma2.0}), we have
${P}/ {\sigma_n^4}=O(n^{-1})+ O(n^{-1}){\rm tr}(\Lambda_1^4)/{\rm tr}^2(\Lambda_1^2)\to 0$ as $n\to\infty$. This completes the proof of $(iii)$, and hence  (\ref{asy_Un1})
holds.

\vskip 12pt
\noindent{\bf Part II: Proof of (\ref{asy_Un2})}
\vskip 12pt

In the following, we show that ${U_{n2}}/{\sqrt{2n^{-2}{\rm tr}(\Lambda_1^2)}}\stackrel{P}{\longrightarrow}0$
as $n \to \infty$.
Let
{\small
\begin{align*}
U_{n21} &=\frac{1}{n(n-1)}\sum_{s=1}^{n}\sum_{t\neq s}^{n}{\bm{X}}_{s}^T
\Big(\sum_{(i,j) \in \hat{A}_1 } P_{ij}^T (P_{ij} S^{(s,t)} P_{ij}^T)^{-1}P_{ij}
-\sum_{(i,j) \in {A}_1} P_{ij}^T (P_{ij} S^{(s,t)} P_{ij}^T)^{-1}P_{ij}  \Big) {\bm{X}}_{t},             \\
U_{n22} &=\frac{1}{n(n-1)}\sum_{s=1}^{n}\sum_{t\neq s}^{n}{\bm{X}}_{s}^T
\Big( \sum_{i \in \hat{A}_2}P_{i}^T ( P_{i}S^{(s,t)} P_{i}^T)^{-1} P_{i}
-\sum_{i \in {A}_2 }P_{i}^T (P_{i}S^{(s,t)} P_{i}^T)^{-1} P_{i}
 \Big){\bm{X}}_{t},            \\
U_{n23} &=\frac{1}{n(n-1)}\sum_{s=1}^{n}\sum_{t\neq s}^{n}{\bm{X}}_{s}^T
\Big( \sum_{ (i,j) \in A_1 } P_{ij}^T (P_{ij} S^{(s,t)} P_{ij}^T)^{-1}P_{ij}
-\sum_{ (i,j) \in A_1 } P_{ij}^T (P_{ij} \Sigma P_{ij}^T)^{-1}P_{ij} \Big){\bm{X}}_{t},           \\
U_{n24} &=\frac{1}{n(n-1)}\sum_{s=1}^{n}\sum_{t\neq s}^{n}{\bm{X}}_{s}^T
\Big(\sum_{i \in {A}_2 }P_{i}^T (P_{i}S^{(s,t)} P_{i}^T)^{-1} P_{i}
-\sum_{i \in A_2} P_{i}^T (P_{i}\Sigma P_{i}^T)^{-1}P_{i}\Big){\bm{X}}_{t}.
\end{align*}}
By direct calculation, we have
\begin{align*}
U_{n2}
&=  \frac{1}{n(n-1)}\sum_{s=1}^{n}\sum_{t\neq s}^{n}{\bm{X}}_{s}^T \big(\widehat{P}_{\mathcal{O}}^{(s,t)} -P_{\mathcal{O}}\big) {\bm{X}}_{t}= U_{n21}+U_{n22}+U_{n23}+U_{n24}.
\end{align*}
Note  that for any $\epsilon_1>0$,
$\big\{ |U_{n21}| \geq  \epsilon_1 {\sqrt{2n^{-2}{\rm tr}(\Lambda_1^2)}} \big\}
\subseteq   \big\{ \hat{A}_1 \neq  A_1 \big\}.$
Then
\begin{eqnarray*}
P \big( |U_{n21}| \geq \epsilon_1 {\sqrt{2n^{-2}{\rm tr}(\Lambda_1^2)}} \big)
\leq  P \big( \hat{A}_1 \neq  A_1 \big)
\leq \frac{p(p-1)}{2}\exp(-\frac{n\epsilon_0^2}{4}),
\end{eqnarray*}
where the second inequality is based on (\ref{ineqA_1}). Hence,
$P ( |U_{n21}| \geq \epsilon_1 {\sqrt{2n^{-2}{\rm tr}(\Lambda_1^2)}} ) \to 0$
as $n \to \infty$.
Similarly, we can prove that
$P ( |U_{n22}| \geq  \epsilon_1 {\sqrt{2n^{-2}{\rm tr}(\Lambda_1^2)}} ) \to 0$
as $n \to \infty$.
This implies that
$U_{n21}/\sqrt{2n^{-2}{\rm tr}(\Lambda_1^2)}=o_p(1)$
and $U_{n22}/\sqrt{2n^{-2}{\rm tr}(\Lambda_1^2)}=o_p(1)$ as $n \to \infty$.
It remains to prove that $U_{n23}/\sqrt{2n^{-2}{\rm tr}(\Lambda_1^2)}=o_p(1)$
and  $U_{n24}/\sqrt{2n^{-2}{\rm tr}(\Lambda_1^2)}=o_p(1)$  as $n \to \infty$.
By (\ref{ord_trace}),
it is equivalent  to verifying that  $U_{n23}=o_p(p^{1/2}n^{-1})$ and
$U_{n23}=o_p(p^{1/2}n^{-1})$ as $n \to \infty$.

\vskip12pt
\noindent { \bf Part II-1: Proof of $U_{n23}=o_p(p^{1/2}n^{-1})$}
\vskip12pt

For simplicity, we omit the subscript $n$ hereafter.
Note that
{\small
\begin{eqnarray*}
U_{23}
&= &  \frac{1}{n(n-1)}\sum_{s=1}^{n}\sum_{t\neq s}^{n}
{\bm{X}}_{s}^T \Big(\sum_{ (i,j) \in A_1 } P_{ij}^T
\Big((P_{ij}S^{(s,t)} P_{ij}^T)^{-1}- (P_{ij}\Sigma P_{ij}^T)^{-1}
\Big) P_{ij}  \Big){\bm{X}}_{t}     \\
&=&  \frac{1}{n(n-1)}
\sum_{ (i,j) \in A_1 }\sum_{s=1}^{n}\sum_{t\neq s}^{n}{\bm{X}}_{ij;s}^T
 \Sigma_{\{i,j\}}^{-\frac{1}{2}}
\Big( \Sigma_{\{i,j\}}^{\frac{1}{2}}
({S}_{\{i,j\}}^{(s,t)})^{-1}\Sigma_{\{i,j\}}^{\frac{1}{2}}  - I_2 \Big)  \Sigma_{\{i,j\}}^{-\frac{1}{2}} {\bm{X}}_{ij;t}   \\
&=&  \frac{1}{n(n-1)}
\sum_{ (i,j) \in A_1 }\sum_{s=1}^{n}\sum_{t\neq s}^{n}
(\tilde{\bm{X}}_{ij;s}+\tilde{\bm{\mu}}_{ij})^T
\Big( (\widetilde{S}_{\{i,j\}}^{(s,t)})^{-1} - I_2 \Big)  (\tilde{\bm{X}}_{ij;t}+\tilde{\bm{\mu}}_{ij}).
\end{eqnarray*}}
\noindent Let $\Xi_{\{i,j\}}^{(s,t)}=(I_2-\widetilde{S}_{\{i,j\}}^{(s,t)})
+(I_2-\widetilde{S}_{\{i,j\}}^{(s,t)})^2
+\cdots+\big(I_2-\widetilde{S}_{\{i,j\}}^{(s,t)}\big)^{m_0}$.
By Taylor expansion for matrix functions (see, e.g., Theorem 4.8 in
\cite{higham2008}) and the fact that $\widetilde{S}_{\{i,j\}}^{(s,t)}$  is a
$\sqrt{n}$-consistent estimator of $I_2$,  we have
\begin{align*}
& \Big\| \big((\widetilde{S}_{\{i,j\}}^{(s,t)})^{-1} - I_2 \big)
-\Xi_{\{i,j\}}^{(s,t)} \Big\|=O_p(n^{-(m_0+1)/2}).
\end{align*}
 Together  with condition (C4) that
$E({X}_{s1}^{4m_0+2}), \ldots, E({X}_{sp}^{4m_0+2})$ are bounded uniformly, we have
{
\begin{align*}
U_{23}
& =   \frac{1}{n(n-1)}
\sum_{ (i,j) \in A_1 }\sum_{s=1}^{n}\sum_{t\neq s}^{n}
\tilde{\bm{X}}_{ij;s}^T
\Xi_{\{i,j\}}^{(s,t)}
\tilde{\bm{X}}_{ij;t}
+ \frac{2}{n(n-1)} \sum_{ (i,j) \in A_1 }
\sum_{s=1}^{n}\sum_{t\neq s}^{n}
\tilde{\bm{X}}_{ij;s}^T
\Xi_{\{i,j\}}^{(s,t)}
\tilde{\bm{\mu}}_{ij}   \\
&~~~~+\frac{1}{n(n-1)} \sum_{ (i,j) \in A_1 }
\sum_{s=1}^{n}\sum_{t\neq s}^{n}
\tilde{\bm{\mu}}_{ij}^T
\Big((\widetilde{S}_{\{i,j\}}^{(s,t)})^{-1} -I_2 \Big)
\tilde{\bm{\mu}}_{ij}
+\text{card}(A_1) O_p(n^{-{(m_0+1)}/{2}})\\
&= U_{231}+U_{232}+U_{233}+\text{card}(A_1) O_p(n^{-{(m_0+1)}/{2}}).
\end{align*}}
\noindent By (\ref{ord_trace}), we obtain ${\rm tr}(\Lambda_1^2)=O(p).$
Under condition (C3), we have $\text{card}(A_1) \leq K_0 p$,
and hence if $m_0\geq 4$, $\text{card}(A_1) O_p(n^{-(m_0+1)/2})=O_p(pn^{-(m_0+1)/2})
=o_p(p^{1/2}n^{-1})$ as $n \to \infty$.
Therefore, we only need to show that
$U_{231}=o_p (p^{1/2}n^{-1})$,
$U_{232}=o_p (p^{1/2}n^{-1})$
and $U_{233}=o_p (p^{1/2}n^{-1})$ as  $n \to \infty$.

\vskip12pt
\noindent { \bf Part II-1.1: Proof of $U_{231}=o_p (p^{1/2}n^{-1}) $ }
\vskip12pt
Since $E(U_{231})=0$, we only need to show that $E(U_{231}^2)=o(pn^{-2})$ as $n \to \infty$.
Noting that
\begin{align*}
& E(U_{231}^2)=\sum_{\substack{ (i_1,j_1) \in A_1 \\
                    (i_2,j_2) \in A_1 }}
E\Big( \frac{1}{n^2(n-1)^2} \sum_{s=1}^{n}\sum_{t\neq s}^{n}
\sum_{l=1}^{n}\sum_{m\neq l}^{n}
\big( \tilde{\bm{X}}_{i_1 j_1;s}^T \Xi_{\{i,j\}}^{(s,t)} \tilde{\bm{X}}_{i_1 j_1;t} \big)
\big( \tilde{\bm{X}}_{i_2 j_2;l}^T \Xi_{\{i,j\}}^{(l,m)} \tilde{\bm{X}}_{i_2 j_2;m} \big)
\Big)   \\
&=  \sum_{\substack{ (i_1,j_1) \in A_1 \\
                    (i_2,j_2) \in A_1 }}
\text{Cov}\Big(
\frac{1}{n(n-1)} \sum_{s=1}^{n}\sum_{t\neq s}^{n}
\tilde{\bm{X}}_{i_1 j_1;s}^T \Xi_{\{i_1,j_1\}}^{(s,t)}  \tilde{\bm{X}}_{i_1 j_1;t},\,
\frac{1}{n(n-1)} \sum_{l=1}^{n}\sum_{m \neq l}^{n}
\tilde{\bm{X}}_{i_2 j_2;l}^T \Xi_{\{i_2,j_2\}}^{(l,m)} \tilde{\bm{X}}_{i_2 j_2;m}\Big).
\end{align*}
Following the $\rho$-mixing inequality
(see, e.g., Theorem 1.1.2 in \citet{zhengyan1997limit}) and condition (C2), we have
\begin{align} \label{alpha.ineq}
&\Big|\text{Cov}\Big(\frac{1}{n(n-1)} \sum_{s=1}^{n}\sum_{t\neq s}^{n}
\tilde{\bm{X}}_{i_1 j_1;s}^T \Xi_{\{i_1,j_1\}}^{(s,t)} \tilde{\bm{X}}_{i_1 j_1;t},\,
\frac{1}{n(n-1)} \sum_{l=1}^{n}\sum_{m \neq l}^{n}
\tilde{\bm{X}}_{i_2 j_2;l}^T
\Xi_{\{i_2,j_2\}}^{(l,m)}
\tilde{\bm{X}}_{i_2 j_2;m}
\Big)\Big|   \nonumber  \\
& \leq \varpi_0 \rho\Big({\text{dist}(\{i_1,j_1\},\{i_2,j_2\})}\Big)
 \max_{(i,j)\in A_1}
 {\rm Var}\Big( \frac{1}{n(n-1)} \sum_{s=1}^{n}\sum_{t\neq s}^{n}
\tilde{\bm{X}}_{i j;s}^T \Xi_{\{i,j\}}^{(s,t)} \tilde{\bm{X}}_{i j;t}\Big) \nonumber  \\
& \leq \varpi_0 \exp\Big(-{\text{dist}(\{i_1,j_1\},\{i_2,j_2\})}\Big)
 \max_{(i,j)\in A_1}
 {\rm Var}\Big( \frac{1}{n(n-1)} \sum_{s=1}^{n}\sum_{t\neq s}^{n}
\tilde{\bm{X}}_{i j;s}^T \Xi_{\{i,j\}}^{(s,t)} \tilde{\bm{X}}_{i j;t}\Big),
\end{align}
where ${\text{dist}(\{i_1,j_1\},\{i_2,j_2\})}=\min\{|i_1-i_2|,|i_1-j_2|,
|j_1-i_2|, |j_1-j_2|\}$.
Then by condition (C3), we have
\begin{align*}
E(U_{231}^2)
&\leq \Big(2+\frac{\varpi_0}{1-\exp(-1)} \Big)K_0^2p
\max_{(i,j)\in A_1} {\rm Var}\Big( \frac{1}{n(n-1)} \sum_{s=1}^{n}\sum_{t\neq s}^{n}
\tilde{\bm{X}}_{i j;s}^T \Xi_{\{i,j\}}^{(s,t)} \tilde{\bm{X}}_{i j;t}\Big).
\end{align*}
If we can show that
\begin{equation}\label{asyU231}
{\rm Var}\Big( \frac{1}{n(n-1)} \sum_{s=1}^{n}\sum_{t\neq s}^{n}
\tilde{\bm{X}}_{i j;s}^T \Xi_{\{i,j\}}^{(s,t)} \tilde{\bm{X}}_{i j;t}\Big)=O(n^{-3})
\end{equation}
hold uniformly for $(i,j)\in A_1$, then $E(U_{231}^2)=O(pn^{-3})=o(pn^{-2}).$

We now show that (\ref{asyU231})  hold uniformly
for $(i,j)\in A_1$.
Note that
\begin{eqnarray*}
&&{\rm Var}\Big( \frac{1}{n(n-1)} \sum_{s=1}^{n}\sum_{t\neq s}^{n}
\tilde{\bm{X}}_{i j;s}^T \Xi_{\{i,j\}}^{(s,t)} \tilde{\bm{X}}_{i j;t}\Big)  \\
&=&  \frac{1}{n^2(n-1)^2}
\sum_{\substack{ s_1 =1, s_2=1  \\ t_1\neq s_1, t_2\neq s_2 }}^n
{E}\Big\{
\tilde{\bm{X}}_{i j;s_1}^T
\Big[(I_2-\widetilde{S}_{\{i,j\}}^{(s_1,t_1)})
+\cdots+\big(I_2-\widetilde{S}_{\{i,j\}}^{(s_1,t_1)}\big)^{m_0}\Big]
 \tilde{\bm{X}}_{i j;t_1}  \\
&&~~~~~~~~~~~~~~~~~~~~~~~~~~~~~~
 \times \tilde{\bm{X}}_{i j;s_2}^T
\Big[(I_2-\widetilde{S}_{\{i,j\}}^{(s_2,t_2)})
+\cdots+\big(I_2-\widetilde{S}_{\{i,j\}}^{(s_2,t_2)}\big)^{m_0}\Big]
 \tilde{\bm{X}}_{i j;t_2}
\Big\}.
\end{eqnarray*}
Then by letting
\begin{eqnarray}\label{J231}
 J_{\{i,j\}}(\nu_1,\nu_2)
& = &
\sum_{\substack{ s_1 =1, s_2=1  \\ t_1\neq s_1, t_2\neq s_2  }}^n
\frac{
{E} \big(
\tilde{\bm{X}}_{i j;s_1}^T
(I_2-\widetilde{S}_{\{i,j\}}^{(s_1,t_1)})^{\nu_1}
\tilde{\bm{X}}_{i j;t_1}
\tilde{\bm{X}}_{i j;s_2}^T
(I_2-\widetilde{S}_{\{i,j\}}^{(s_2,t_2)})^{\nu_2}
\tilde{\bm{X}}_{i j;t_2}\big)}
{{n^2(n-1)^2}},  \nonumber\\
\end{eqnarray}
we have
\begin{align}\label{lem6E_1}
{\rm Var}\Big( \frac{1}{n(n-1)} \sum_{s=1}^{n}\sum_{t\neq s}^{n}
\tilde{\bm{X}}_{i j;s}^T \Xi_{\{i,j\}}^{(s,t)} \tilde{\bm{X}}_{i j;t}\Big)
=\sum_{\nu_1=1}^{m_0}
\sum_{\nu_2=1}^{m_0}
J_{\{i,j\}}(\nu_1,\nu_2).
\end{align}

To verify (\ref{asyU231}), it suffices to show that
for any given $m_0\geq 4$,
\begin{align}\label{J_ij}
J_{\{i,j\}}(\nu_1,\nu_2)=O(n^{-3})
\end{align}
hold uniformly over $(i,j)\in A_1$.
To prove it, we let
\begin{align}\label{G1fun}
\widetilde{G}(\mathcal{C}_1, \mathcal{C}_1 )
=\frac{1}{(n-2)(n-3)}\Big[
\sum_{l_1 \in \mathcal{C}_1 } \sum_{l_2 \in \mathcal{C}_1}
\Big(I_2 -\frac{(\tilde{\bm{X}}_{i j;l_1}- \tilde{\bm{X}}_{i j;l_2} )
(\tilde{\bm{X}}_{i j;l_1}- \tilde{\bm{X}}_{i j;l_2} )^T}{2}
\Big)\Big].
\end{align}
In addition,  for  $\mathcal{C}_1 \cap \mathcal{C}_2 = \emptyset$, let
\begin{align}\label{G2fun}
\widetilde{G}(\mathcal{C}_1, \mathcal{C}_2 )
=\frac{2}{(n-2)(n-3)}\Big[
\sum_{l_1 \in \mathcal{C}_1 } \sum_{l_2 \in \mathcal{C}_2}
\Big(I_2 -\frac{(\tilde{\bm{X}}_{i j;l_1}- \tilde{\bm{X}}_{i j;l_2} )
(\tilde{\bm{X}}_{i j;l_1}- \tilde{\bm{X}}_{i j;l_2} )^T}{2}
\Big)\Big].
\end{align}
Let
$J_{231}(\nu_1,\nu_2|\{i,j\})=
{E} \{  \tilde{\bm{X}}_{i j;s_1}^T
(I_2-\widetilde{S}_{\{i,j\}}^{(s_1,t_1)})^{\nu_1}
\tilde{\bm{X}}_{i j;t_1}  \,
\tilde{\bm{X}}_{i j;s_2}^T
(I_2-\widetilde{S}_{\{i,j\}}^{(s_2,t_2)})^{\nu_2}
\tilde{\bm{X}}_{i j;t_2}  \}.$
In what follows, we decompose $J_{231}(\nu_1,\nu_2|\{i,j\})$ into
three exclusive sets.

\vskip12pt
\noindent{\bf(I)}
  Let $\mathcal{S}_1=\{(s_1, t_1, s_2, t_2) |
  s_1\neq t_1, s_2\neq t_2 \}$ such that $\{s_1, t_1\} \cap {\{s_2, t_2\}}=\emptyset$.
\vskip12pt

For easy of presentation, we assume that $\mathcal{C}_{11}=\{t_2,s_2\}$,
$\mathcal{C}_{12}=\{t_1, s_1\}$,
and $\mathcal{C}_{13}=\{1,\ldots,n\}/\{t_1,s_1,t_2,s_2\}$.
Noting that $\widetilde{S}_{\{i,j\}}^{(s_1,t_1)}$ can be rewritten as $U$-statistics, we have
\begin{align}\label{widetildeS11}
I_2-\widetilde{S}_{\{i,j\}}^{(s_1,t_1)}
& =\widetilde{G}(\mathcal{C}_{11}, \mathcal{C}_{11} )
+\widetilde{G}(\mathcal{C}_{11}, \mathcal{C}_{13}  )
+\widetilde{G}(\mathcal{C}_{13}, \mathcal{C}_{13}  ),
\end{align}
\begin{align}\label{widetildeS22}
I_2-\widetilde{S}_{\{i,j\}}^{(s_2,t_2)}
& =\widetilde{G}(\mathcal{C}_{12}, \mathcal{C}_{12} )
+\widetilde{G}(\mathcal{C}_{12}, \mathcal{C}_{13}  )
+\widetilde{G}(\mathcal{C}_{13}, \mathcal{C}_{13} ),
\end{align}
and hence,
{\begin{align*}
J_{231}(\nu_1,\nu_2|\{i,j\}, \mathcal{S}_1)
&= {E} \Big\{  \tilde{\bm{X}}_{i j;s_1}^T
\Big[\widetilde{G}(\mathcal{C}_{11}, \mathcal{C}_{11})
+\widetilde{G}(\mathcal{C}_{11}, \mathcal{C}_{13})
+\widetilde{G}(\mathcal{C}_{13}, \mathcal{C}_{13})\Big]^{\nu_1}
\tilde{\bm{X}}_{i j;t_1}  \,     \\
&~~~~~~~\times
\tilde{\bm{X}}_{i j;s_2}^T
\Big[
\widetilde{G}(\mathcal{C}_{12}, \mathcal{C}_{12})
+\widetilde{G}(\mathcal{C}_{12}, \mathcal{C}_{13})
+\widetilde{G}(\mathcal{C}_{13}, \mathcal{C}_{13})
\Big]^{\nu_2}
\tilde{\bm{X}}_{i j;t_2}^T
\Big\}   \\
&= \sum_{ \substack{ l_{11}+l_{12} \leq \nu_1 \\ l_{21}+l_{22} \leq \nu_2 }}
 \binom{\nu_1}{l_{11}}\binom{\nu_1-l_{11}}{l_{12}}
 \binom{\nu_2}{l_{21}}\binom{\nu_2-l_{21}}{l_{22}}
 J_{\{i,j\}}(\bm{\nu},\bm{l}_{1},\bm{l}_{2}|, \mathcal{S}_1),
\end{align*}}
where $\bm{\nu}=(\nu_1,\nu_2)$,
$\bm{l}_{1}=(l_{11},l_{21})$,
$\bm{l}_{2}=(l_{12},l_{22})$,
and
{\small
\begin{align*}
J_{\{i,j\}}(\bm{\nu},\bm{l}_{1},\bm{l}_{2}| \mathcal{S}_1)
&= {E} \Big\{
 \tilde{\bm{X}}_{i j;s_1}^T
\Big[\widetilde{G}(\mathcal{C}_{11}, \mathcal{C}_{11})\Big]^{l_{11}}
\Big[\widetilde{G}(\mathcal{C}_{11}, \mathcal{C}_{13})\Big]^{l_{12}}
\Big[\widetilde{G}(\mathcal{C}_{13},\mathcal{C}_{13})\Big]^{\nu_1-l_{11}-l_{12}}
\tilde{\bm{X}}_{i j;t_1}  \,  \nonumber \\
&~~~~~~~\times  \tilde{\bm{X}}_{i j;s_2}^T
\Big[\widetilde{G}(\mathcal{C}_{12}, \mathcal{C}_{12})\Big]^{l_{21}}
\Big[\widetilde{G}(\mathcal{C}_{12}, \mathcal{C}_{13})\Big]^{l_{22}}
\Big[\widetilde{G}(\mathcal{C}_{13},\mathcal{C}_{13})\Big]^{\nu_2-l_{21}-l_{22}}
\tilde{\bm{X}}_{i j;t_2}
\Big\}.
\end{align*}}

Note that for $l=1,2$,  $\widetilde{G}(\mathcal{C}_{1l}, \mathcal{C}_{1l})=O_p(n^{-2})$,
$\widetilde{G}(\mathcal{C}_{1l}, \mathcal{C}_{13})=O_p(n^{-1})$
and $\widetilde{G}(\mathcal{C}_{13}, \mathcal{C}_{13})=O_p(n^{-1/2})$.
We have
$$J_{\{i,j\}}(\bm{\nu},\bm{l}_{1},\bm{l}_{2}|\mathcal{S}_1)=
O\big(n^{-[(\nu_1+\nu_2)+3(l_{11}+l_{21})+(l_{12}+l_{22})]/2}\big).$$
On one hand, if $(\nu_1+\nu_2 )+3(l_{11}+l_{21})+(l_{12}+l_{22}) \geq  6$,
\begin{align}\label{lem61.0}
J_{\{i,j\}}(\bm{\nu},\bm{l}_{1},\bm{l}_{2}| \mathcal{S}_1)
=O(n^{-3}).
\end{align}
On the other hand, if $(\nu_1+\nu_2)+3(l_{11}+l_{21} )
+(l_{12}+l_{22} ) \leq 5$, we show that
\begin{align}\label{lem61}
J_{\{i,j\}}(\bm{\nu},\bm{l}_{1},\bm{l}_{2}| \mathcal{S}_1) =0.
\end{align}

Note also that $\nu_1, \nu_2 \geq 1$, $l_{11}+l_{12}\leq \nu_1$ and
$l_{21}+l_{22}\leq \nu_2$. Hence,
$(\nu_1+\nu_2  )+3(l_{11}+l_{21} )
+(l_{12}+l_{22}  )\leq 5$  can be decomposed into the following five scenarios:

\begin{itemize}
  \item[(a.1)] $\nu_1+\nu_2=2$, $l_{11}+l_{21}=0$,
  $l_{12}+l_{22}\leq 2$;
  \item[(a.2)] $\nu_1+\nu_2=2$, $l_{11}+l_{21}=1$,
  $l_{12}=l_{22}=0$;
  \item[(a.3)] $\nu_1+\nu_2=3$, $l_{11}+l_{21}=0$,
  $l_{12}+l_{22} \leq 2$;
  \item[(a.4)] $\nu_1+\nu_2=4$, $l_{11}+l_{21}=0$,
  $l_{12}+l_{22} \leq 1$;
  \item[(a.5)] $\nu_1+\nu_2=5$,
   $l_{12}+l_{22}=0$, $l_{12}+l_{22}= 0$.
\end{itemize}

%
We now show  that (\ref{lem61}) holds under (a.1).
Following the similar procedure, we can prove that (\ref{lem61}) holds under (a.2)--(a.5).
Firstly, if $l_{12}=l_{22}=0$, we have
\begin{align*}
 J_{\{i,j\}}(\bm{\nu},\bm{l}_{1},\bm{l}_{2}| \mathcal{S}_1)
&=   {E} \Big\{
\tilde{\bm{X}}_{i j;s_1}^T
\Big[\widetilde{G}(\mathcal{C}_{13},\mathcal{C}_{13})\Big]
\tilde{\bm{X}}_{i j;t_1}  \,   \tilde{\bm{X}}_{i j;s_2}^T
\Big[\widetilde{G}(\mathcal{C}_{13},\mathcal{C}_{13})\Big]
\tilde{\bm{X}}_{i j;t_2}
\Big\}   \nonumber \\ \,
&=  E \Big[
{E}\big(\tilde{\bm{X}}_{i j;s_1}^T\big)
 \widetilde{G}(\mathcal{C}_{13},\mathcal{C}_{13})
{E}\big(\tilde{\bm{X}}_{i j;t_1}\big)  \,
{E}\big(\tilde{\bm{X}}_{i j;s_2}^T\big)
 \widetilde{G}(\mathcal{C}_{13},\mathcal{C}_{13})
{E}\big(\tilde{\bm{X}}_{i j;t_2}\big)
\Big] \nonumber  \\
&=0,
\end{align*}where the last equality comes from the fact that $\widetilde{G}(\mathcal{C}_{13}, \mathcal{C}_{13})$
is independent with $\tilde{\bm{X}}_{i j;l}$, $l \in \{s_1,t_1,s_2,t_2\}$.

Secondly, if $l_{12}=1$ and $l_{22}=0$, then
\begin{align*}
J_{\{i,j\}}(\bm{\nu},\bm{l}_{1},\bm{l}_{2}| \mathcal{S}_1)
&=
{E} \Big\{
\tilde{\bm{X}}_{i j;s_1}^T
\Big[\widetilde{G}(\mathcal{C}_{11}, \mathcal{C}_{13})\Big]
\tilde{\bm{X}}_{i j;t_1}  \,   \tilde{\bm{X}}_{i j;s_2}^T
\Big[\widetilde{G}(\mathcal{C}_{13},\mathcal{C}_{13})\Big]
\tilde{\bm{X}}_{i j;t_2}
\Big\}    \nonumber \\ \,
&=
{E} \Big\{
E(\tilde{\bm{X}}_{i j;s_1}^T)
\Big[\widetilde{G}(\mathcal{C}_{11}, \mathcal{C}_{13})\Big]
E(\tilde{\bm{X}}_{i j;t_1})  \,
\tilde{\bm{X}}_{i j;s_2}^T
\Big[\widetilde{G}(\mathcal{C}_{13},\mathcal{C}_{13})\Big]
\tilde{\bm{X}}_{i j;t_2} \Big\} \\
&=0,  \nonumber
\end{align*}
where the second equality comes from that
$\tilde{\bm{X}}_{i j;s_1}$ and $\tilde{\bm{X}}_{i j;t_1}^T$
are  independent with $\widetilde{G}(\mathcal{C}_{11}, \mathcal{C}_{13})$,
$\widetilde{G}(\mathcal{C}_{13}, \mathcal{C}_{13})$,
$\tilde{\bm{X}}_{i j;s_2}$ and $\tilde{\bm{X}}_{i j;t_2}$.
Similarly, if $l_{12}=0$ and $l_{22}=1$, we have $J_{\{i,j\}}(\bm{\nu},\bm{l}_{1},\bm{l}_{2}| \mathcal{S}_1)=0$.

Thirdly, if $l_{12}= l_{22}=1$, we have
\begin{align*}
 J_{\{i,j\}}(\bm{\nu},\bm{l}_{1},\bm{l}_{2}| \mathcal{S}_1) &=
 {E} \big\{
 \tilde{\bm{X}}_{i j;s_1}^T
[\widetilde{G}(\mathcal{C}_{11}, \mathcal{C}_{13})]
\tilde{\bm{X}}_{i j;t_1}  \,   \tilde{\bm{X}}_{i j;s_2}^T
[\widetilde{G}(\mathcal{C}_{12},\mathcal{C}_{13})]
\tilde{\bm{X}}_{i j;t_2}
\big\}.
\end{align*}
Noting that
$\widetilde{G}(\mathcal{C}_{11}, \mathcal{C}_{13})=
\widetilde{G}(\{{s}_2 \}, \mathcal{C}_{13})+\widetilde{G}(\{{t}_2 \}, \mathcal{C}_{13})$,
and $\widetilde{G}(\mathcal{C}_{12}, \mathcal{C}_{13})=
\widetilde{G}(\{\tilde{s}_1 \}, \mathcal{C}_{13})+\widetilde{G}(\{\tilde{t}_1 \}, \mathcal{C}_{13}),
$
we have
\begin{align*}
 J_{\{i,j\}}(\bm{\nu},\bm{l}_{1},\bm{l}_{2}| \mathcal{S}_1)
 &= \sum_{\substack{ \tilde{t}_1 \in\{s_2,t_2\}  \\
                    \tilde{t}_2 \in \{s_1,t_1\} }
                     }
 {E} \Big\{
 \tilde{\bm{X}}_{i j;s_1}^T
\big[\widetilde{G}(\{\tilde{t}_1 \}, \mathcal{C}_{13})\big]
\tilde{\bm{X}}_{i j;t_1} \,
\tilde{\bm{X}}_{i j;s_2}^T
\big[\widetilde{G}(\{\tilde{t}_2 \}, \mathcal{C}_{13})\big]
\tilde{\bm{X}}_{i j;t_2} \,
\Big\}.
\end{align*}
Note  that for any $\tilde{t}_1 \in \mathcal{C}_{11}$
and $\tilde{t}_2 \in \mathcal{C}_{12}$,
$${E}\{\tilde{\bm{X}}_{i j;s_1}^T
[\widetilde{G}(\{\tilde{t}_1 \}, \mathcal{C}_{13})]
\tilde{\bm{X}}_{i j;t_1} \,
\tilde{\bm{X}}_{i j;s_2}^T
[\widetilde{G}(\{\tilde{t}_2 \}, \mathcal{C}_{13})]
\tilde{\bm{X}}_{i j;t_2} \,
\}=0.$$
For example, when $\tilde{t}_1=s_2$ and $\tilde{t}_2=t_1$, we have
\begin{align*}
& ~ {E} \big(
 \tilde{\bm{X}}_{i j;s_1}^T
[\widetilde{G}(\{\tilde{t}_1 \}, \mathcal{C}_{13})]
\tilde{\bm{X}}_{i j;t_1} \,
\tilde{\bm{X}}_{i j;s_2}^T
[\widetilde{G}(\{\tilde{t}_2 \}, \mathcal{C}_{13})]
\tilde{\bm{X}}_{i j;t_2}
\big)  \\
=&~{E}\big(
E(\tilde{\bm{X}}_{i j;s_1}^T)
\widetilde{G}(\{s_2 \}, \mathcal{C}_{13})
\tilde{\bm{X}}_{i j;t_1} \,
\tilde{\bm{X}}_{i j;s_2}^T
\widetilde{G}(\{t_1\}, \mathcal{C}_{13})
E(\tilde{\bm{X}}_{i j;t_2}) \,
\big) =0.
\end{align*}
This indicates that
$J_{\{i,j\}}(\bm{\nu},\bm{l}_{1},\bm{l}_{2}| \mathcal{S}_1)=0$.
Thus, we complete the proof for (\ref{lem61}).
Finally, by (\ref{lem61.0}) and (\ref{lem61}), we have
$J_{\{i,j\}}(\bm{\nu},\bm{l}_{1},\bm{l}_{2}| \mathcal{S}_1)
=O(n^{-3}).$
And consequently,
\begin{eqnarray}\label{lem6S1}
J_{231}(\nu_1,\nu_2|\{i,j\}, \mathcal{S}_1)
&=&\sum_{ \substack { l_{11}+l_{12} \leq \nu_1  \\ l_{21}+l_{22} \leq \nu_2 } }
 \binom{\nu_1}{l_{11}}\binom{\nu_1-l_{11}}{l_{12}}
 \binom{\nu_2}{l_{21}}\binom{\nu_2-l_{21}}{l_{22}}
J_{\{i,j\}}(\bm{\nu},\bm{l}_{1},\bm{l}_{2}| \mathcal{S}_1) \nonumber \\
&=&  O(n^{-3}).
\end{eqnarray}


\vskip12pt
\noindent{\bf(II)}
  Let $\mathcal{S}_2=\{(s_1, t_1, s_2, t_2) |
  s_1\neq t_1, s_2\neq t_2\}$ such that
there is only one common  element in $\{s_1, t_1\} \cap {\{s_2, t_2\}}$.
\vskip12pt

Since $\{s_1, t_1\}, \{s_2, t_2\}$ are symmetric,
 for simplicity, we only consider $t_1=s_2$.
Let $\mathcal{C}_{21}= \{t_2\}$,
 $\mathcal{C}_{22}=\{s_1\}$,
and $\mathcal{C}_{23}=\{1,2,\ldots,n\}/{ \{s_1,t_1,t_2\}}$.
Consequently,
\begin{eqnarray*}
&& J_{231}(\nu_1,\nu_2|\{i,j\},\mathcal{S}_2) \\
&=&  {E} \big\{  \tilde{\bm{X}}_{i j;s_1}^T
\big[
\widetilde{G}(\mathcal{C}_{21}, \mathcal{C}_{23})
+\widetilde{G}(\mathcal{C}_{23}, \mathcal{C}_{23})\big]^{\nu_1}
\tilde{\bm{X}}_{i j;t_1}  \tilde{\bm{X}}_{i j;s_2}^T
\big[\widetilde{G}(\mathcal{C}_{22}, \mathcal{C}_{23})
 +\widetilde{G}(\mathcal{C}_{23}, \mathcal{C}_{23})\big]^{\nu_2}
\tilde{\bm{X}}_{i j;t_2}^T
\big\}\\
&=& J_{\{i,j\}}(\bm{\nu} | \mathcal{S}_2  \cap{ \{ t_1=s_2\}}  ).
\end{eqnarray*}
Note that
$J_{\{i,j\}}(\bm{\nu} | \mathcal{S}_2  \cap{ \{ t_1=s_2\}}  )
= \sum_{ l_{11}=0}^{\nu_1}
\sum_{l_{21}=0}^{\nu_2}
J_{\{i,j\}}(\bm{\nu},l_{11}, l_{21}| \mathcal{S}_2 \cap {\{t_1=s_2\}} ),$
where {\small
\begin{align*}
J_{\{i,j\}}(\bm{\nu},l_{11}, l_{21}|\mathcal{S}_2\cap {\{t_1=s_2\}})
&=
{E} \Big\{
\tilde{\bm{X}}_{i j;s_1}^T
\big[\widetilde{G}(\mathcal{C}_{21}, \mathcal{C}_{23})\big]^{l_{11}}
\big[\widetilde{G}(\mathcal{C}_{23},\mathcal{C}_{23})\big]^{\nu_1-l_{11}}
\tilde{\bm{X}}_{i j;t_1}  \,     \\
&~~~~~~~\times \tilde{\bm{X}}_{i j;t_1}^T
\big[\widetilde{G}(\mathcal{C}_{22}, \mathcal{C}_{23})\big]^{l_{21}}
\big[\widetilde{G}(\mathcal{C}_{23},\mathcal{C}_{23})\big]^{\nu_2-l_{21}}
\tilde{\bm{X}}_{i j;t_2}
\Big\}.
\end{align*}}
Since
$\widetilde{G}(\mathcal{C}_{2l}, \mathcal{C}_{23})=O_p(n^{-1})$
and $\widetilde{G}(\mathcal{C}_{23}, \mathcal{C}_{23})=O_p(n^{-1/2})$ for $l=1,2$,
we have
$$  J_{\{i,j\}}(\bm{\nu},l_{11}, l_{21}|\mathcal{S}_2\cap {\{t_1=s_2\}}) =O\big(n^{-(\nu_1+\nu_2+l_{11}+l_{21})/2}\big).$$
On the one hand, if $\nu_1+\nu_2+l_{11}+l_{21}\geq  4$,
\begin{align}\label{lem63.0}
 J_{\{i,j\}}(\bm{\nu},l_{11}, l_{21}|\mathcal{S}_2\cap {\{t_1=s_2\}})  =O(n^{-2}).
\end{align}
On the other hand, if $\nu_1+\nu_2+l_{11}+l_{21}\leq 3$, we will show that
\begin{align}\label{lem63}
 J_{\{i,j\}}(\bm{\nu},l_{11}, l_{21}|\mathcal{S}_2\cap {\{t_1=s_2\}})  =0.
\end{align}
We then decompose $\nu_1+\nu_2+l_{11}+l_{21}\leq 3$ into  three scenarios
as follows:
\begin{itemize}
  \item[(b.1)] $\nu_1+\nu_2=3$, $l_{11}+l_{21}=0$;
  \item[(b.2)] $\nu_1+\nu_2=2 $, $l_{11}=0, l_{21}=1$;
  \item[(b.3)] $\nu_1+\nu_2=2 $, $l_{11}=1, l_{21}=0$.
\end{itemize}

For simplicity, we only demonstrate that
(\ref{lem63}) holds under the second scenario.
According to (b.2), we have $\nu_1=\nu_2=1$, $l_{11}=0$ and $l_{21}=1$.
Noting that  $\tilde{\bm{X}}_{i j;s_1}$, $\tilde{\bm{X}}_{i j;t_1}$ and
$\tilde{\bm{X}}_{i j;t_2}^T$ are independent with
$\widetilde{G}(\mathcal{C}_{23},\mathcal{C}_{23})$, we have
\begin{align*}
J_{\{i,j\}}(\bm{\nu},l_{11}, l_{21}|\mathcal{S}_2\cap {\{t_1=s_2\}})
&= {E} \big(
 \tilde{\bm{X}}_{i j;s_1}^T
\widetilde{G}(\mathcal{C}_{23},\mathcal{C}_{23})
\tilde{\bm{X}}_{i j;t_1}  \,
 \tilde{\bm{X}}_{i j;t_1}^T
\widetilde{G}(\mathcal{C}_{22},\mathcal{C}_{23})
\tilde{\bm{X}}_{i j;t_2}  \,
\big)   \\
&={E} \big(
\tilde{\bm{X}}_{i j;s_1}^T
\widetilde{G}(\mathcal{C}_{23},\mathcal{C}_{23})
E(\tilde{\bm{X}}_{i j;t_1}  \tilde{\bm{X}}_{i j;t_1}^T)
\widetilde{G}(\mathcal{C}_{22}, \mathcal{C}_{23})
E(\tilde{\bm{X}}_{i j;t_2})  \,
\big)  \\
&=0.
\end{align*}
Similarly,  for scenarios  (b.1) and (b.3),  we can obtain
$J_{\{i,j\}}(\bm{\nu},\bm{l}_{1},\bm{l}_{2}|\mathcal{S}_2\cap {\{t_1=s_2\}})=0$
 and hence prove (\ref{lem63}). Then, by (\ref{lem63.0}) and (\ref{lem63}), we have
\begin{align*}
J_{\{i,j\}}(\bm{\nu},l_{11}, l_{21}|\mathcal{S}_2  \cap {\{t_1=s_2\}})
=O(n^{-2}),
\end{align*}
and
{\small
\begin{align*}
  J_{\{i,j\}}(\bm{\nu},l_{11}, l_{21}| \mathcal{S}_2)
=&~  J_{\{i,j\}}(\bm{\nu},l_{11}, l_{21}| \mathcal{S}_2  \cap {\{s_1=s_2\}} )
+J_{\{i,j\}}(\bm{\nu},l_{11}, l_{21}| \mathcal{S}_2  \cap {\{s_1=t_2\}} ) \\
&+ J_{\{i,j\}}(\bm{\nu},l_{11}, l_{21}| \mathcal{S}_2  \cap {\{t_1=s_2\}} )
+J_{\{i,j\}}(\bm{\nu},l_{11}, l_{21}| \mathcal{S}_2  \cap {\{t_2=t_2\}} )\\
=&~O(n^{-2}).
\end{align*}}
Finally, we obtain that
\begin{align}\label{lem6S2}
J_{231}(\nu_1,\nu_2|\{i,j\}, \mathcal{S}_2)
=\sum_{ l_{11}=0}^{\nu_1}
\sum_{l_{21}=0}^{\nu_2}
J_{\{i,j\}}(\bm{\nu},l_{11}, l_{21}| \mathcal{S}_2)
= O(n^{-2}).
\end{align}


\noindent {\bf(III)} Let $\mathcal{S}_3=\{(s_1, t_1, s_2, t_2) |
  s_1\neq t_1, s_2\neq t_2\}$ such that
$\{s_1, t_1\} = {\{s_2, t_2\}}$.
\vskip12pt

Note that $\nu_1, \nu_2 \leq 1$, we have
\begin{align}\label{lem6S3}
J_{231}(\nu_1,\nu_2|\{i,j\}, \mathcal{S}_3)&= {E} \big(
(I-\widetilde{S}_{\{i,j\}}^{(s_1,t_1)})^{\nu_1} \,
(I-\widetilde{S}_{\{i,j\}}^{(s_2,t_2)})^{\nu_2}
\big)
=  O(n^{-1}).
\end{align}
Together with (\ref{lem6S1}), (\ref{lem6S2}) and (\ref{lem6S3}), we have
\begin{align}\label{lem6ord}
J_{231}(\nu_1,\nu_2|\{i,j\})
=&~  \frac{1}{n^2(n-1)^2}\sum_{s_1=1}^{n}\sum_{t_1\neq s_1}^{n}
\sum_{s_2=1}^{n}\sum_{t_2\neq s_2}^{n}
\big[
J_{\{i,j\}}(\bm{\nu}|(s_1,t_1,s_2,t_2) \in \mathcal{S}_1)   \nonumber \\
& +J_{\{i,j\}}(\bm{\nu}| (s_1,t_1,s_2,t_2) \in \mathcal{S}_2)
+ J_{\{i,j\}}(\bm{\nu}| (s_1,t_1,s_2,t_2) \in \mathcal{S}_3)
\big]     \nonumber \\
=&~ J_{\{i,j\}}(\bm{\nu}|(s_1,t_1,s_2,t_2) \in \mathcal{S}_1)
+ O(n^{-1})J_{\{i,j\}}(\bm{\nu}|(s_1,t_1,s_2,t_2) \in \mathcal{S}_2)  \nonumber \\
& +O(n^{-2})J_{ \{i,j\} }(\bm{\nu}| (s_1,t_1,s_2,t_2) \in \mathcal{S}_3)
= O(n^{-3}).
\end{align}
In addition, let
$M_{\{i,j\}}^{(1)}(k)=E[(1,0)\tilde{\bm{X}}_{i j}]^k$,
$M_{\{i,j\}}^{(2)}(k)=E[(0,1)\tilde{\bm{X}}_{i j}]^k$ and
$M_{\{i,j\}}^{(3)}(\nu_1,\nu_2)=E\{[(1,0)\tilde{\bm{X}}_{i j}]^{\nu_1}[(0,1)\tilde{\bm{X}}_{i j}]^{\nu_2}\}$,
where $\tilde{\bm{X}}_{i j}^T=[(X_i, X_j)-(\mu_i,\mu_j)]\Sigma_{ij}^{-\frac{1}{2}}$,
$(0,1)$ and $(1,0)$ are two dimensional row vectors.
Let
$$M_{ij}(\bm{h}^{(1)}, \bm{h}^{(2)}, \bm{h}^{(3)})=
\prod_{k_1=1}^{2m_0} \Big[M_{\{i,j\}}^{(1)}(k_1)\Big]^{h_{k_1}^{(1)}}
\prod_{k_2=1}^{2m_0} \Big[M_{\{i,j\}}^{(2)}(k_2)\Big]^{h_{k_2}^{(2)}}
\prod_{\nu_1+\nu_2 \leq 2 m_0}\Big[M_{\{i,j\}}^{(3)}(\nu_1,\nu_2)\Big]^{h_{\nu_1,\nu_2}^{(3)}},$$
where $h_{k_1}^{(1)}$, $h_{k_2}^{(2)}$, $h_{\nu_1,\nu_2}^{(3)}$ are nonnegative integers,
and
$\bm{h}^{(1)}=(h_{1}^{(1)}, \ldots, h_{2m_0}^{(1)})$,
$\bm{h}^{(2)}=(h_{1}^{(2)}, \ldots, h_{2m_0}^{(2)})$,
$\bm{h}^{(3)}=(h_{1,1}^{(3)},\ldots,h_{1,2m_0-1}^{(3)},
h_{2,1}^{(3)},\cdots, h_{2,2m_0-2}^{(3)},  \ldots , h_{2m_0-1,1}^{(3)})$
such that
$\sum_{k_1=1}^{2m_0} k_1 h_{k_1}^{(1)} +\sum_{k_2=1}^{2m_0} k_2 h_{k_2}^{(2)}
+\sum_{\nu_1+\nu_2 \leq 2m_0} (\nu_1+\nu_2) h_{\nu_1,\nu_2}^{(3)} \leq 4m_0+4.$

By (\ref{J231}),  (\ref{lem6E_1}) and (\ref{lem6ord}), we can show that
\begin{align*}
{\rm Var} \Big( \frac{1}{n(n-1)} \sum_{s=1}^{n}\sum_{t\neq s}^{n}
\tilde{\bm{X}}_{i j;s}^T \Xi_{\{i,j\}}^{(s,t)} \tilde{\bm{X}}_{i j;t}\Big)
& =\sum_{ (\bm{h}^{(1)}, \bm{h}^{(2)}, \bm{h}^{(3)}) \in \mathcal{D}}
\widetilde{\varphi}(\bm{h}^{(1)}, \bm{h}^{(2)}, \bm{h}^{(3)} )
 M_{ij}(\bm{h}^{(1)}, \bm{h}^{(2)}, \bm{h}^{(3)}).
\end{align*}
where  the summation is over the set
$\mathcal{D}=\{  \bm{h}^{(1)}, \bm{h}^{(2)}, \bm{h}^{(3)}  |
\sum_{k_1=1}^{2m_0} k_1 h_{k_1}^{(1)} +\sum_{k_2=1}^{2m_0} k_2 h_{k_2}^{(2)}
+\sum_{\nu_1+\nu_2 \leq 2m_0} (\nu_1+\nu_2) h_{\nu_1,\nu_2}^{(3)} \leq 4m_0+4 \}$, and
$\widetilde{\varphi}(\bm{h}^{(1)}, \bm{h}^{(2)}, \bm{h}^{(3)})$ is the coefficient that is not related to the index $\{i,j\}$, and only determined by $\bm{h}^{(1)}, \bm{h}^{(2)}, \bm{h}^{(3)}$.

Note that $M_{ij}(\bm{h}^{(1)}, \bm{h}^{(2)}, \bm{h}^{(3)} )$ are finite combination of higher order moments of $X_i$ and $X_j$.
By condition (C4),  $M_{ij}(\bm{h}^{(1)}, \bm{h}^{(2)}, \bm{h}^{(3)})$ are bounded uniformly.
And consequently, for any $(i,j)\in A_1$,
$${\rm Var} \Big( \frac{1}{n(n-1)} \sum_{s=1}^{n}\sum_{t\neq s}^{n}
\tilde{\bm{X}}_{i j;s}^T \Xi_{\{i,j\}}^{(s,t)} \tilde{\bm{X}}_{i j;t}\Big)^{2}=O(n^{-3})$$  hold uniformly.
We complete the proof that $E(U_{231}^2)=o(p n^{-2})$ as  $n\to\infty$.

\vskip12pt
\noindent { \bf Part II-1.2: Proof of ${ U_{232}=o_p(p^{1/2}n^{-1})}$ }
\vskip12pt
Note that
\begin{align*}
E\Big(\frac{2}{n(n-1)} \sum_{s=1}^{n}\sum_{t\neq s}^{n}
\tilde{\bm{X}}_{ij;s}^T
\Xi_{\{i,j\}}^{(s,t)}
\tilde{\bm{\mu}}_{i j}  \Big)=0.
\end{align*}
By condition (C2) and following the same procedure as in (\ref{alpha.ineq}),
we have
\begin{align*}
E(U_{232}^2)
&= \sum_{\substack{ (i_1,j_1) \in A_1 \\
                    (i_2,j_2) \in A_1 }}
\text{Cov}\Big(\frac{2}{n(n-1)} \sum_{s=1}^{n}\sum_{t\neq s}^{n}
\tilde{\bm{X}}_{i_1 j_1;s}^T \Xi_{\{i_1,j_1\}}^{(s,t)} \tilde{\bm{\mu}}_{i_1 j_1},\,
\\
&~~~~~~~~~~~~~~~~~~~~~~~\frac{2}{n(n-1)} \sum_{l=1}^{n}\sum_{m \neq l}^{n}
\tilde{\bm{X}}_{i_2 j_2;l}^T \Xi_{\{i_2,j_2\}}^{(l,m)} \tilde{\bm{\mu}}_{i_2 j_2}\Big)  \\
&\leq \Big(2+\frac{\varpi_0}{1-\exp(-1)} \Big)K_0^2p
\max_{(i,j)\in A_1} {\rm Var}\Big( \frac{2}{n(n-1)} \sum_{s=1}^{n}\sum_{t\neq s}^{n}
\tilde{\bm{X}}_{i j;s}^T \Xi_{\{i,j\}}^{(s,t)} \tilde{\bm{\mu}}_{i j}\Big).
\end{align*}
Consequently, by $(ii)$ in Lemma \ref{Lemmaeigen},
\begin{align*}
{E(U_{232}^2)}/(pn^{-2})
&=O(n^2)\max_{(i,j)\in A_1} {\rm Var}\Big( \frac{2}{n(n-1)} \sum_{s=1}^{n}\sum_{t\neq s}^{n}
\tilde{\bm{X}}_{i j;s}^T \Xi_{\{i,j\}}^{(s,t)} \tilde{\bm{\mu}}_{i j}\Big).
\end{align*}

If we can show that
\begin{eqnarray}\label{U232_prof}
\max_{(i,j)\in A_1} {\rm Var}\Big( \frac{1}{n(n-1)} \sum_{s=1}^{n}\sum_{t\neq s}^{n}
\tilde{\bm{X}}_{i j;s}^T \Xi_{\{i,j\}}^{(s,t)} \tilde{\bm{\mu}}_{i j}\Big)
= O(n^{-2}) \max_{(i,j)\in A_1}\tilde{\bm{\mu}}_{i j}^T \tilde{\bm{\mu}}_{i j},
\end{eqnarray}
then
$${E(U_{232}^2)}
=
O(pn^{-2})\max_{(i,j)\in A_1}\tilde{\bm{\mu}}_{i j}^T \tilde{\bm{\mu}}_{i j}
=
O(pn^{-2})\max_{(i,j)\in A_1}{\bm{\mu}}_{i j}^T \Sigma_{\{ij\}}^{-1}{\bm{\mu}}_{i j}.$$
As shown in the proof of Lemma \ref{Lemmaeigen}, the eigenvalues of
$\Sigma_{\{ij\}}^{-1} \in \mathbb{R}^{2 \times 2}$ are bounded uniformly over $(i,j)\in A_1$.
Then by condition (C5) we have
$$\max_{(i,j)\in A_1}{\bm{\mu}}_{i j}^T \Sigma_{\{ij\}}^{-1}{\bm{\mu}}_{i j}=O(n^{-1/2}),$$
and consequently, $U_{232}=o_p (pn^{-2})$ as $n \to \infty$.

In the following, we show the result in (\ref{U232_prof}).
Noting that
\begin{align*}
 & {\rm Var}\Big( \frac{1}{n(n-1)} \sum_{s=1}^{n}\sum_{t\neq s}^{n}
\tilde{\bm{X}}_{i j;s}^T \Xi_{\{i,j\}}^{(s,t)} \tilde{\bm{\mu}}_{i j}\Big) \\
= &~ \tilde{\bm{\mu}}_{i j}^T
E\Big(\frac{1}{n^2(n-1)^2}
\sum_{s_1=1}^{n}\sum_{t_1\neq s_1}^{n}
\sum_{s_2=1}^{n}\sum_{t_2\neq s_2}^{n}
\Xi_{\{i,j\}}^{(s_1,t_1)} \tilde{\bm{X}}_{i j;s_1}
\tilde{\bm{X}}_{i j;s_2}^T \Xi_{\{i,j\}}^{(s_2,t_2)}
\Big) \tilde{\bm{\mu}}_{i j},
\end{align*}
we then decompose $(s_1,t_1,s_2,t_2)$ into the following four cases:
\begin{itemize}
  \item[(c.1)] $s_1=s_2$;
  \item[(c.2)] $s_1 \neq s_2$ and $s_1=t_2$;
  \item[(c.3)] $s_1 \neq s_2$, $s_1 \neq t_2$, $t_1=s_2$ or  $t_1=t_2$;
  \item[(c.4)] $s_1 \neq s_2$, $s_1 \neq t_2$, $t_1 \neq s_2$ and $t_1 \neq t_2$.
\end{itemize}
Under case (c.1), by the fact that $\Xi_{{\{ij\}}}^{(s,t)}=O(n^{-1/2})$, we have
\begin{align}\label{c.1}
& \tilde{\bm{\mu}}_{i j}^T E\Big(\frac{1}{n^2(n-1)^2}
\sum_{s_1=1}^{n}\sum_{t_1\neq s_1}^{n}
\sum_{s_2=1}^{n}\sum_{t_2\neq s_2}^{n}
\Xi_{\{i,j\}}^{(s_1,t_1)}
\tilde{\bm{X}}_{i j;s_1}
\tilde{\bm{X}}_{i j;s_2}^T
\Xi_{\{i,j\}}^{(s_2,t_2)}
\Big) \tilde{\bm{\mu}}_{i j} \nonumber \\
=&~ \frac{1}{n^2(n-1)^2} \tilde{\bm{\mu}}_{i j}^T  E\Big(
\sum_{s_1=1}^{n}\sum_{t_1\neq s_1}^{n}\sum_{t_2\neq s_2}^{n}
\Xi_{\{i,j\}}^{(s_1,t_1)}
\tilde{\bm{X}}_{i j;s_1}
\tilde{\bm{X}}_{i j;s_1}^T
\Xi_{\{i,j\}}^{(s_1,t_2)}
\Big) \tilde{\bm{\mu}}_{i j} \nonumber \\
=&~ O(n^{-2}) \tilde{\bm{\mu}}_{i j}^T\tilde{\bm{\mu}}_{i j}.
\end{align}
Under cases (c.2) and (c.3),
\begin{eqnarray}\label{c.23}
 \tilde{\bm{\mu}}_{i j}^T E\Big(\frac{1}{n^2(n-1)^2}
\sum_{ \substack{ s_1=1 \\ t_1\neq s_1 }}^{n}
\sum_{\substack{ s_2=1 \\ t_2\neq s_2} }^{n}
\Xi_{\{i,j\}}^{(s_1,t_1)}
\tilde{\bm{X}}_{i j;s_1}
\tilde{\bm{X}}_{i j;s_2}^T
\Xi_{\{i,j\}}^{(s_2,t_2)}
\Big)\tilde{\bm{\mu}}_{i j}
= O(n^{-2}) \tilde{\bm{\mu}}_{i j}^T\tilde{\bm{\mu}}_{i j}.
\end{eqnarray}
Under case (c.4),
by  (\ref{G1fun}) and (\ref{G2fun})  we have
\begin{align*}
&\tilde{\bm{\mu}}_{i j}^T E\Big(
\Xi_{\{i,j\}}^{(s_1,t_1)}
\tilde{\bm{X}}_{i j;s_1}
\tilde{\bm{X}}_{i j;s_2}^T
\Xi_{\{i,j\}}^{(s_2,t_2)}
\Big) \tilde{\bm{\mu}}_{i j} \\
= &
\sum_{\nu_1=1}^{m_0}\sum_{\nu_2=1}^{m_0}
 \tilde{\bm{\mu}}_{i j}^T
 E\big\{
[
\widetilde{G}(\mathcal{C}_{11}, \mathcal{C}_{11})
+\widetilde{G}(\mathcal{C}_{11}, \mathcal{C}_{13})
+\widetilde{G}(\mathcal{C}_{13}, \mathcal{C}_{13})
]^{\nu_1}
\tilde{\bm{X}}_{i j;s_1} \\
& ~~~~~~~~~~~~~~~~~ \times \tilde{\bm{X}}_{i j;s_2}^T
[
\widetilde{G}(\mathcal{C}_{12}, \mathcal{C}_{12})
+\widetilde{G}(\mathcal{C}_{12}, \mathcal{C}_{13})
+\widetilde{G}(\mathcal{C}_{13}, \mathcal{C}_{13})
]^{\nu_2}
\big\} \tilde{\bm{\mu}}_{i j} \\
= &
\sum_{\nu_1+\nu_2 \leq 3}
\tilde{\bm{\mu}}_{i j}^T
E\big\{
[
\widetilde{G}(\mathcal{C}_{11}, \mathcal{C}_{11})
+\widetilde{G}(\mathcal{C}_{11}, \mathcal{C}_{13})
+\widetilde{G}(\mathcal{C}_{13}, \mathcal{C}_{13})
]^{\nu_1}
\tilde{\bm{X}}_{i j;s_1} \\
& ~~~~~~~~~~~~~~~
\times \tilde{\bm{X}}_{i j;s_2}^T
[
\widetilde{G}(\mathcal{C}_{12}, \mathcal{C}_{12})
+\widetilde{G}(\mathcal{C}_{12}, \mathcal{C}_{13})
+\widetilde{G}(\mathcal{C}_{13}, \mathcal{C}_{13})
]^{\nu_2}
\big\}\tilde{\bm{\mu}}_{i j}
+O(n^{-2})\tilde{\bm{\mu}}_{i j}^T \tilde{\bm{\mu}}_{i j}.
\end{align*}
The second term in the last equality is obtained by
using that
$\widetilde{G}(\mathcal{C}_{11}, \mathcal{C}_{11})
+\widetilde{G}(\mathcal{C}_{11}, \mathcal{C}_{13})
+\widetilde{G}(\mathcal{C}_{13}, \mathcal{C}_{13})
=O_p(n^{-1/2})$,
and
$\widetilde{G}(\mathcal{C}_{12}, \mathcal{C}_{12})
+\widetilde{G}(\mathcal{C}_{12}, \mathcal{C}_{13})
+\widetilde{G}(\mathcal{C}_{13}, \mathcal{C}_{13})
=O_p(n^{-1/2})$ (see, (\ref{widetildeS11}) and (\ref{widetildeS22})).

To verify
$\tilde{\bm{\mu}}_{i j}^T
E(\Xi_{\{i,j\}}^{(s_1,t_1)}
\tilde{\bm{X}}_{i j;s_1}
\tilde{\bm{X}}_{i j;s_2}^T
\Xi_{\{i,j\}}^{(s_2,t_2)})
\tilde{\bm{\mu}}_{i j}=O(n^{-2})\tilde{\bm{\mu}}_{i j}^T \tilde{\bm{\mu}}_{i j}$,
it suffices to show that
\begin{eqnarray}\label{6.37}
&& \tilde{\bm{\mu}}_{i j}^T
E\big\{[
\widetilde{G}(\mathcal{C}_{11}, \mathcal{C}_{11})
+\widetilde{G}(\mathcal{C}_{11}, \mathcal{C}_{13})
+\widetilde{G}(\mathcal{C}_{13}, \mathcal{C}_{13})
]  \tilde{\bm{X}}_{i j;s_1}   \nonumber \\
&& \times \tilde{\bm{X}}_{i j;s_2}^T
[
\widetilde{G}(\mathcal{C}_{12}, \mathcal{C}_{12})
+\widetilde{G}(\mathcal{C}_{12}, \mathcal{C}_{13})
+\widetilde{G}(\mathcal{C}_{13}, \mathcal{C}_{13})
]\big\}\tilde{\bm{\mu}}_{i j}  \nonumber \\
& = & O(n^{-2})\tilde{\bm{\mu}}_{i j}^T \tilde{\bm{\mu}}_{i j},
\end{eqnarray}
\begin{eqnarray}\label{6.38}
&& \tilde{\bm{\mu}}_{i j}^T
E\big\{ \big[
\widetilde{G}(\mathcal{C}_{11}, \mathcal{C}_{11})
+\widetilde{G}(\mathcal{C}_{11}, \mathcal{C}_{13})
+\widetilde{G}(\mathcal{C}_{13}, \mathcal{C}_{13})
\big]^2  \tilde{\bm{X}}_{i j;s_1}   \nonumber \\
&& \times \tilde{\bm{X}}_{i j;s_2}^T
\big[
\widetilde{G}(\mathcal{C}_{12}, \mathcal{C}_{12})
+\widetilde{G}(\mathcal{C}_{12}, \mathcal{C}_{13})
+\widetilde{G}(\mathcal{C}_{13}, \mathcal{C}_{13})
\big] \big\}\tilde{\bm{\mu}}_{i j}  \nonumber \\
&=& O(n^{-2})\tilde{\bm{\mu}}_{i j}^T \tilde{\bm{\mu}}_{i j},
\end{eqnarray}
and
\begin{eqnarray}\label{6.39}
&& \tilde{\bm{\mu}}_{i j}^T
E\big\{ \big[
\widetilde{G}(\mathcal{C}_{11}, \mathcal{C}_{11})
+\widetilde{G}(\mathcal{C}_{11}, \mathcal{C}_{13})
+\widetilde{G}(\mathcal{C}_{13}, \mathcal{C}_{13})
\big]
\tilde{\bm{X}}_{i j;s_1}   \nonumber \\
&& \times \tilde{\bm{X}}_{i j;s_2}^T
\big[
\widetilde{G}(\mathcal{C}_{12}, \mathcal{C}_{12})
+\widetilde{G}(\mathcal{C}_{12}, \mathcal{C}_{13})
+\widetilde{G}(\mathcal{C}_{13}, \mathcal{C}_{13})
\big]^2
\big\}  \tilde{\bm{\mu}}_{i j}    \nonumber \\
&=& O(n^{-2})\tilde{\bm{\mu}}_{i j}^T \tilde{\bm{\mu}}_{i j}.
\end{eqnarray}
Consequently, under the case (c.4), we have
\begin{eqnarray}\label{c.4}
&& \tilde{\bm{\mu}}_{i j}^T
E\Big(\frac{1}{n^2(n-1)^2}
\sum_{s_1=1}^{n}\sum_{t_1\neq s_1}^{n}
\sum_{s_2=1}^{n}\sum_{t_2\neq s_2}^{n}
\Xi_{\{i,j\}}^{(s_1,t_1)}
\tilde{\bm{X}}_{i j;s_1}
\tilde{\bm{X}}_{i j;s_2}^T
\Xi_{\{i,j\}}^{(s_2,t_2)}
\Big)\tilde{\bm{\mu}}_{i j}   \nonumber \\
&=& \frac{1}{n^2(n-1)^2} \sum_{s_1=1}^{n}\sum_{t_1 \neq s_1}^{n}
\sum_{s_2=1}^{n}\sum_{t_2 \neq s_1}^{n}
\tilde{\bm{\mu}}_{i j}^T
E\Big(
\Xi_{\{i,j\}}^{(s_1,t_1)}
\tilde{\bm{X}}_{i j;s_1}
\tilde{\bm{X}}_{i j;s_2}^T
\Xi_{\{i,j\}}^{(s_2,t_2)}
\Big) \tilde{\bm{\mu}}_{i j}   \nonumber \\
&=&  O(n^{-2})\tilde{\bm{\mu}}_{i j}^T \tilde{\bm{\mu}}_{i j}
\end{eqnarray}

Note that $\mathcal{C}_{11}=\{ s_2, t_2\}$, $\mathcal{C}_{12}=\{ s_1, t_1\}$ and $\mathcal{C}_{13}={\{1,\ldots,n \}}/\{ s_1, t_1,s_2,t_2\}$.
By (\ref{G1fun}) and (\ref{G2fun}), we have $\widetilde{G}(\mathcal{C}_{1l}, \mathcal{C}_{1l})=O_p(n^{-2})$ and
$\widetilde{G}(\mathcal{C}_{1l}, \mathcal{C}_{13})=O_p(n^{-1})$ for  $l=1,2$.
Then,
\begin{eqnarray*}
&& E\big\{
\big[
\widetilde{G}(\mathcal{C}_{11}, \mathcal{C}_{11})
+\widetilde{G}(\mathcal{C}_{11}, \mathcal{C}_{13})
+\widetilde{G}(\mathcal{C}_{13}, \mathcal{C}_{13})
\big]
\tilde{\bm{X}}_{i j;s_1}      \\
&&~~~\times
 \tilde{\bm{X}}_{i j;s_2}^T
\big[
\widetilde{G}(\mathcal{C}_{12}, \mathcal{C}_{12})
+\widetilde{G}(\mathcal{C}_{12}, \mathcal{C}_{13})
+\widetilde{G}(\mathcal{C}_{13}, \mathcal{C}_{13})
\big]\big\}                  \\
&=&  E\big\{
\big[
\widetilde{G}(\mathcal{C}_{11}, \mathcal{C}_{11})
+\widetilde{G}(\mathcal{C}_{11}, \mathcal{C}_{13})
\big]
\tilde{\bm{X}}_{i j;s_1}  \tilde{\bm{X}}_{i j;s_2}^T
\big[
\widetilde{G}(\mathcal{C}_{12}, \mathcal{C}_{12})
+\widetilde{G}(\mathcal{C}_{12}, \mathcal{C}_{13})
\big]\big\}  \\
&=&O(n^{-2}).
\end{eqnarray*}
This shows (\ref{6.37}). The proofs of (\ref{6.38}) and (\ref{6.39}) are similar and hence are omitted. Finally, by
(\ref{c.1}), (\ref{c.23}) and (\ref{c.4}), it yields  (\ref{U232_prof}).

\vskip12pt
\noindent { \bf Part II-1.3: Proof of $U_{233}=o_p(p^{1/2}n^{-1})$ }
\vskip12pt

Since
$
E\{ | \tilde{\bm{\mu}}_{ij}^T
((\widetilde{S}_{\{i,j\}}^{(s,t)})^{-1} -I_2 )
\tilde{\bm{\mu}}_{ij} | \}
\leq
E ( \| (\widetilde{S}_{\{i,j\}}^{(s,t)})^{-1} -I_2   \| )
\times \tilde{\bm{\mu}}_{ij}^T\tilde{\bm{\mu}}_{ij}
=
O(n^{-1/2}) \tilde{\bm{\mu}}_{ij}^T\tilde{\bm{\mu}}_{ij},
$
as  $n \to \infty$ we have
\begin{align*}
E( |U_{233}|)
& \leq
\frac{1}{n(n-1)} \sum_{ (i,j) \in A_1 }
\sum_{s=1}^{n}\sum_{t\neq s}^{n}
E\big\{  \big| \tilde{\bm{\mu}}_{ij}^T
\big((\widetilde{S}_{\{i,j\}}^{(s,t)})^{-1} -I_2 \big)
\tilde{\bm{\mu}}_{ij} \big|\big\}  \\
&\leq
O(n^{-1/2}) \sum_{ (i,j) \in A_1 } \tilde{\bm{\mu}}_{ij}^T\tilde{\bm{\mu}}_{ij}       \\
&=  O(n^{-1/2})\sum_{ (i,j) \in A_1 } {\bm{\mu}_{ij}}^T \Sigma_{ij}^{-1}{\bm{\mu}_{ij}} \\
&=
O(n^{-1/2}) {\bm{\mu}}^T P_{\mathcal{O}} {\bm{\mu}} \\
&=o(p^{1/2}n^{-1}).
\end{align*}
The last equality is from condition (C5).

To summarize, from the conclusions of  Parts II-1.1, II-1.2 and II-1.3,
we complete the proof of Part II-1.

\vskip12pt
\noindent { \bf Part II-2: Proof of $U_{n24}=o_p(p^{1/2}n^{-1})$ }
\vskip12pt

Let $\widetilde{X}_{sj}=({X}_{sj}-\mu_j)/\sigma_{jj}^{1/2}$,
$\widetilde{\mu}_j=\mu_j/\sigma_{jj}$, $\widetilde{s}_{jj}^{(s,t)}={s}_{jj}^{(s,t)}/\sigma_{jj}$,
and
$\Xi_{jj}^{(s,t)}=(1-\widetilde{s}_{jj}^{(s,t)})+(1-\widetilde{s}_{jj}^{(s,t)})^2
+\cdots+ (1-\widetilde{s}_{jj}^{(s,t)})^{m_0}$.
Since
$|((\widetilde{s}_{jj}^{(s,t)})^{-1}-1) - \Xi_{jj}^{(s,t)} |
=O_p(n^{-(m_0+1)/2}),$
we have
\begin{align*}
U_{n24}
&=\frac{1}{n(n-1)}\sum_{s=1}^{n}\sum_{t\neq s}^{n}{\bm{X}}_{s}^T
\Big(\sum_{i \in {A}_2 }P_{i}^T ( P_{i}S^{(s,t)} P_{i}^T)^{-1} P_{i}
-\sum_{i \in A_2} P_{i}^T ( P_{i} \Sigma P_{i}^T)^{-1} P_{i}\Big){\bm{X}}_{t}  \\
&=\frac{1}{n(n-1)}\sum_{s=1}^{n}\sum_{t\neq s}^{n}
\sum_{j \in {A}_2 }
\widetilde{X}_{sj} \widetilde{X}_{tj}\Xi_{jj}^{(s,t)}
+\frac{2}{n(n-1)}\sum_{s=1}^{n}\sum_{t\neq s}^{n}
\sum_{j \in {A}_2 }
\widetilde{\mu}_{j} \widetilde{X}_{tj}\Xi_{jj}^{(s,t)} \\
&~~~+\frac{2}{n(n-1)}\sum_{s=1}^{n}\sum_{t\neq s}^{n}
\sum_{j \in {A}_2 }
\widetilde{\mu}_{j}^2\big((\widetilde{s}_{jj}^{(s,t)})^{-1}-1\big)
+\text{card}(A_2)O_p\big(n^{-{m_0}/{2}}\big)  \\
&=U_{241}+U_{242}+U_{243}+\text{card}(A_2)O_p\big(n^{-(m_0+1)/2}\big).
\end{align*}

Note that for $m_0\geq 4$,
$\text{card}(A_2)O_p\big(n^{-(m_0+1)/2}\big)=O_p(pn^{-(m_0+1)/2})=o_p(p^{1/2}n^{-1})$ as $n \to \infty$.
We now show that $U_{241}=o_p(p^{1/2}n^{-1})$,
$U_{242}=o_p (p^{1/2}n^{-1})$ and $U_{241}=o_p(p^{1/2}n^{-1})$ as $n \to \infty$
in Part II-2.1, Part II-2.2 and Part II-2.3, respectively.

\vskip12pt
\noindent { \bf Part II-2.1: Proof of $U_{241}=o_p(p^{1/2}n^{-1})$ }
\vskip12pt

Noting that $E(U_{241})=0$, and following the $\rho$-mixing inequality and condition (C2), we have
\begin{align*}
E(U_{241}^2)
&=\sum_{j_1 \in {A}_2 } \sum_{j_2 \in {A}_2 }
\text{Cov}\Big(
\frac{1}{n(n-1)}\sum_{s=1}^{n}\sum_{t\neq s}^{n}
\widetilde{X}_{s j_1} \widetilde{X}_{t j_1 }\Xi_{j_1 j_1}^{(s,t)},
\frac{1}{n(n-1)}\sum_{s=1}^{n}\sum_{t\neq s}^{n}
\widetilde{X}_{s j_2} \widetilde{X}_{t j_2 }\Xi_{j_2 j_2}^{(s,t)}
\Big)   \\
&\leq \varpi_0 \sum_{j_1 \in {A}_2 } \sum_{j_2 \in {A}_2 }
\exp(-|j_1-j_2|)
\max_{j\in A_2} \text{Var}\Big(
\frac{1}{n(n-1)}\sum_{s=1}^{n}\sum_{t\neq s}^{n}
\widetilde{X}_{s j} \widetilde{X}_{t j }\Xi_{j j}^{(s,t)} \Big)   \\
&\leq  \frac{\varpi_0 p}{1-\exp(-1)}
\max_{j\in A_2} \text{Var}\Big(
\frac{1}{n(n-1)}\sum_{s=1}^{n}\sum_{t\neq s}^{n}
\widetilde{X}_{s j} \widetilde{X}_{t j }\Xi_{j j}^{(s,t)} \Big).
\end{align*}

Following the similar proof for (\ref{asyU231}),
it can be shown that
$$\text{Var}\Big(
\frac{1}{n(n-1)}\sum_{s=1}^{n}\sum_{t\neq s}^{n}
\widetilde{X}_{s j} \widetilde{X}_{t j }\Xi_{j j}^{(s,t)} \Big)=O(n^{-3})$$
hold uniformly for $j \in A_2$.
Thus, $E(U_{241}^2)=O(pn^{-3})$ as $n \to \infty$.  This implies that
$U_{241}=o_p(p^{1/2}n^{-1})$.

\vskip12pt
\noindent { \bf Part II-2.2: Proof of $ U_{242}=o_p(p^{1/2}n^{-1})$ }
\vskip12pt
Note that
\begin{align*}
E\Big(\frac{2}{n(n-1)} \sum_{s=1}^{n}\sum_{t\neq s}^{n}
\tilde{X}_{j;s}^T \Xi_{jj}^{(s,t)} \tilde{\mu}_{j} \Big)=0.
\end{align*}
By the $\rho$-mixing inequality and
 condition (C2), we have
\begin{align*}
E(U_{242}^2)
&= \sum_{\substack{ j_1 \in A_2 \\
                    j_2 \in A_2 }}
\text{Cov}\Big(\frac{2}{n(n-1)} \sum_{s=1}^{n}\sum_{t\neq s}^{n}
\tilde{X}_{j_1;s}^T \Xi_{j_1 j_1}^{(s,t)} \tilde{\mu}_{j_1},\,
\frac{2}{n(n-1)}
\sum_{s=1}^{n}\sum_{t\neq s}^{n}
\tilde{X}_{j_2;s}^T \Xi_{j_2 j_2}^{(s,t)} \tilde{\mu}_{j_2} \Big)  \\
&\leq
\frac{\varpi_0 p}{1-\exp(-1)}
\max_{j \in A_2} {\rm Var}\Big( \frac{2}{n(n-1)} \sum_{s=1}^{n}\sum_{t\neq s}^{n}
\tilde{X}_{j;s}^T \Xi_{jj}^{(s,t)} \tilde{\mu}_j \Big).
\end{align*}
Then,
\begin{align*}
{E(U_{242}^2)}/(pn^{-2})
&=
O(n^2)\max_{j \in A_2} {\rm Var}\Big( \frac{2}{n(n-1)} \sum_{s=1}^{n}\sum_{t\neq s}^{n}
\tilde{X}_{j;s}^T \Xi_{jj}^{(s,t)} \tilde{\mu}_{j}\Big).
\end{align*}

If we can show that
\begin{eqnarray}\label{U242_prof}
\max_{j \in A_2} {\rm Var}\Big(
\frac{1}{n(n-1)} \sum_{s=1}^{n}\sum_{t\neq s}^{n}
\tilde{X}_{j;s} \Xi_{jj}^{(s,t)} \tilde{\mu}_{j}
\Big)
= O(n^{-2}) \max_{j \in A_2 }\tilde{\mu}_{j}^2,
\end{eqnarray}
then
$${E(U_{242}^2)}=
O(pn^{-2})\max_{j\in A_2}\tilde{\mu}_{j}^2=O(pn^{-2})\max_{j\in A_2}{\mu}_{j}^2/\sigma_{jj}.$$

As in the proof of Lemma \ref{Lemmaeigen}, we have shown that
$\sigma_{jj}$ for $j=1,\ldots,p$ are bounded uniformly, then
by condition (C5) we have
$\max_{j\in A_2}{\mu}_{j}^2/\sigma_{jj}=O(n^{-1/2})$.
And consequently, $U_{233}=o_p (p^{1/2}n^{-1})$ as $n \to \infty$.

Following the similar proof for (\ref{U232_prof}), we can show
 (\ref{U242_prof}) also hold.

\vskip12pt
\noindent { \bf Part II-2.3: Proof of $U_{243}=o_p(p^{1/2}n^{-1})$ }
\vskip12pt
By condition (C5), $E|(\widetilde{s}_{jj}^{(s,t)})^{-1}-1|=O(n^{-1/2})$
for $j=1,\ldots,p$. Then, as $n\to\infty$
\begin{align*}
E\big(|U_{243}|\big)
&\leq   \sum_{j \in {A}_2 }\widetilde{\mu}_{j}^2
E\Big(\Big| \frac{2}{n(n-1)}\sum_{s=1}^{n}\sum_{t\neq s}^{n}
   \big((\widetilde{s}_{jj}^{(s,t)})^{-1}-1\big)\Big| \Big)   \\
&\leq   2 \sum_{j \in {A}_2 }\widetilde{\mu}_{j}^2
E\Big(\Big| (\widetilde{s}_{jj}^{(s,t)})^{-1} -1  \Big|\Big)   \\
&=  O(n^{-1/2}) \sum_{j \in {A}_2 } {\mu}_{j}^2/\sigma_{jj} \\
&=  O(n^{-1/2}) \Big(\sum_{j \in {A}_2 } {\mu}_{j}^2/\sigma_{jj}
    +\sum_{ (i,j) \in A_1 } {\bm{\mu}_{ij}}^T \Sigma_{ij}^{-1}{\bm{\mu}_{ij}} \Big)
    \\
&=   O(n^{-1/2})  {\bm{\mu}}^T P_{\mathcal{O}}{\bm{\mu}} \\
&=o(p^{1/2}n^{-1}).
\end{align*}
The last equality is based on condition (C5).

As a summary, from the conclusions of Parts II-2.1, II-2.2 and II-2.3, we show
that $U_{n24}=o_p (p^{1/2}n^{-1})$ as $n \to \infty$.
This completes the proof of (\ref{asy_Un2}).
\hfill$\Box$

\subsection{Proof of Lemma \ref{lemma1.1}}\label{Prof.lemma1}

Let
\begin{align*}
&{P}_{\mathcal{O}}^{(s,t)}
=
\sum_{(i,j) \in {A}_1 } P_{ij}^T (P_{ij} S^{(s,t)} P_{ij}^T)^{-1}P_{ij}
+\sum_{i \in {A}_2 } P_{i}^T ( P_{i}S^{(s,t)} P_{i}^T)^{-1} P_{i},    \\
&L_{11}
=\frac{1}{n(n-1)} \sum_{s=1}^{n}\sum_{t \neq s}^{n}
({\bm X_{s}} -{\bar{\bm X}^{(s,t)}})^T
{P}_{\mathcal{O}}^{(s,t)} {\bm X}_{t}  \,
({\bm X_{t}} - {\bar{\bm X}^{(s,t)}})^T
{P}_{\mathcal{O}}^{(s,t)}  {\bm X}_{s}.
\end{align*}
Since $(\{\hat{A}_1=A_1\}\cap \{\hat{A}_2=A_2\} )
\subseteq \{ {P}_{\mathcal{O}}^{(s,t)}=\widehat{P}_{\mathcal{O}}^{(s,t)} \}
\subseteq \{ L_{11}=\widehat{\text{tr}(\Lambda_1^2)}  \}$,
then for any $\epsilon_1>0$, as $n \to \infty$ we have
$$P\big(\big|L_{11}-\widehat{\text{tr}(\Lambda_1^2)}\big|
>\epsilon_1 {\text{tr}(\Lambda_1^2)} \big)
\leq P\big( \{\hat{A}_1\neq  A_1\} \big)+P\big( \{\hat{A}_2\neq A_2\} \big) \to 0.  $$
This indicates that
${\widehat{\text{tr}(\Lambda_1^2)}}/{\text{tr}(\Lambda_1^2)}
-{L_{11}}/{\text{tr}(\Lambda_1^2)} \stackrel{P}{\longrightarrow}0$  as $n \to \infty.$
Hence to prove Lemma \ref{lemma1.1}, it  is equivalent  to verifying that
\begin{align}\label{Prof.L11}
\frac{L_{11}}{\text{tr}(\Lambda_1^2)} \stackrel{P}{\longrightarrow} 1
{\rm ~~as~~} n \to \infty.
\end{align}

For simplicity, let  $L_{11}= B_{1}+B_{2}+B_{3}$, where
\begin{align*}
B_{1}=&~ \frac{1}{n(n-1)} \sum_{s=1}^{n}\sum_{t \neq s}^{n}
({\bm X_{s}} -{\bar{\bm X}^{(s,t)}})^T  P_{\mathcal{O}}  {\bm X}_{t}  \,
({\bm X_{t}} - {\bar{\bm X}^{(s,t)}})^T  P_{\mathcal{O}}  {\bm X}_{s}   \\
=&~ \frac{1}{n(n-1)} \sum_{s=1}^{n}\sum_{t \neq s}^{n}
(\breve{\bm X}_{s} -\breve{\bar{{\bm X}}}^{(s,t)})^T  \breve{\bm X}_{t}
(\breve{\bm X}_{t} -\breve{\bar{{\bm X}}}^{(s,t)})^T  \breve{\bm X}_{s}, \\
B_{2}=&~ \frac{2}{n(n-1)} \sum_{s=1}^{n}\sum_{t \neq s}^{n}
({\bm X_{s}} -{\bar{\bm X}^{(s,t)}})^T P_{\mathcal{O}}{\bm X}_{t}  \,
({\bm X_{t}}-{\bar{\bm X}^{(s,t)}})^T({P}_{\mathcal{O}}^{(s,t)}-P_{\mathcal{O}}) {\bm X}_{s},  \\
B_{3}=&~ \frac{1}{n(n-1)} \sum_{s=1}^{n}\sum_{t \neq s}^{n}
({\bm X_{s}} -{\bar{\bm X}^{(s,t)}})^T({P}_{\mathcal{O}}^{(s,t)} - P_{\mathcal{O}}){\bm X}_{t} \,
({\bm X_{t}} - {\bar{\bm X}^{(s,t)}})^T({P}_{\mathcal{O}}^{(s,t)} - P_{\mathcal{O}})  {\bm X}_{s},
\end{align*}
where  $\breve{\bm X}_{s}= P_{\mathcal{O}}^{1/2}{\bm X}_{s}$ and
$\breve{\bar{{\bm X}}}^{(s,t)}=P_{\mathcal{O}}^{1/2}{\bar{{\bm X}}}^{(s,t)}$.
By Lemma \ref{Lemmaeigen}, $P_{\mathcal{O}}$ is  a positive definite matrix with eigenvalues bounded uniformly
away from $0$ and $\infty$.  Then, under the linear model (\ref{model1}), we have
$\breve{\bm X}_{s}=P_{\mathcal{O}}^{1/2} C {\bm Z}_s + \breve{{\bm \mu}}$,
where $\breve{{\bm \mu}}=P_{\mathcal{O}}^{1/2}{\bm \mu}$ and
$\text{Var}(\breve{\bm X}_{s})=\Lambda_1$.
Together with (\ref{con1}) and  $(ii)$ in Lemma \ref{Lemmaeigen},
we can see that $(\breve{\bm X}_{1},\ldots,\breve{\bm X}_{n})$
satisfies the conditions of Theorem 2 in \citet{chen2010}.
Hence,
\begin{align}\label{asy.B1}
\frac{B_1}{{\rm tr}(\Lambda_1^2)} \stackrel{P}{\longrightarrow} 1
{\rm~~as~~} n \to \infty.
\end{align}
It remains to show that ${B_2}/{{\rm tr}(\Lambda_1^2)}=o_p(1)$, and
${B_3}/{{\rm tr}(\Lambda_1^2)}=o_p(1)$ as $n \to \infty$. By (\ref{ord_trace}),
it is equivalent  to verifying that   $B_{2}=o_p(p)$
and $B_{3}=o_p(p)$.

\vskip12pt
\noindent{\bf Part-I:  Proof of} ${B_2=o_p(p)}$
\vskip12pt

For  $s\neq t$, let
\begin{align*}
B_{21}^{(s,t)} &=
\sum_{ (i,j) \in A_1 }
(\tilde{\bm{X}}_{ij;s}-   \tilde{\bar{\bm{X}}}_{ij}^{(s,t)}  )^T
(  ( \widetilde{S}_{\{i,j\}}^{(s,t)} )^{-1}-I_2 )
(\tilde{\bm{X}}_{ij;t}+\tilde{\bm{\mu}}_{ij}),  \\
B_{22}^{(s,t)} &=
\sum_{ j \in A_2 }
(\tilde{X}_{tj}-  \tilde{\bar{X}}_{j}^{(s,t)})
(1/\tilde{s}_{jj}^{(s,t)} -1 )
(\tilde{X}_{sj}+\tilde{\mu}_{j}),
\end{align*}
where
$\tilde{\bar{\bm{X}}}_{ij}^{(s,t)}=\sum_{k \neq s,t}^{n} \tilde{\bm{X}}_{ij;k}/(n-2)$,
$\tilde{\bar{{X}}}_{j}^{(s,t)}=\sum_{k \neq s,t}^{n} \tilde{{X}}_{kj}/(n-2)$
and $\tilde{s}_{jj}^{(s,t)}$ is the sample variance for $\{ \tilde{{X}}_{kj}\}_{k\neq s,t}$.
Then,
\begin{align*}
E(|B_2|)
&\leq
\frac{2}{n(n-1)} \sum_{s=1}^{n}\sum_{t \neq s}^{n}
E\Big(\big|
({\bm X_{s}} -{\bar{\bm X}^{(s,t)}})^T P_{\mathcal{O}}{\bm X}_{t}
\big| \, \big|  ({\bm X_{t}}-{\bar{\bm X}^{(s,t)}})^T
({P}_{\mathcal{O}}^{(s,t)}-P_{\mathcal{O}})
{\bm X}_{s}  \big| \Big) \\
&=
2 E\Big(\big|
({\bm X_{1}} -{\bar{\bm X}^{(1,2)}})^T P_{\mathcal{O}}{\bm X}_{2}
\big| \, \big|  ({\bm X_{2}}-{\bar{\bm X}^{(1,2)}})^T
({P}_{\mathcal{O}}^{(1,2)}-P_{\mathcal{O}})
{\bm X}_{1}  \big| \Big).
\end{align*}
Note that ${\bar{\bm X}^{(s,t)}}={\bar{\bm X}^{(t,s)}}$
and ${P}_{\mathcal{O}}^{(s,t)}={P}_{\mathcal{O}}^{(t,s)}$.
This leads to
\begin{align}\label{B2.expand}
& ~~({\bm X_{2}}-{\bar{\bm X}^{(1,2)}})^T
({P}_{\mathcal{O}}^{(1,2)}-P_{\mathcal{O}})
{\bm X}_{1} \nonumber\\
=&
\sum_{(i,j)\in A_1}
({\bm X_{ij;2}}-{\bar{\bm X}_{ij}^{(2,1)}})^T
\left( ( S_{\{ij\}}^{(2,1)} )^{-1} - \Sigma_{\{ij\}}^{-1} \right)
{\bm X}_{ij;1}
+\sum_{j\in A_2}
( X_{2j}-{\bar{X}}_{j}^{(2,1)} )^T
\left( \frac{1}{s_{jj}^{(2,1)} } - \frac{1}{\sigma_{jj}} \right)
{X}_{1j}    \nonumber\\
=&
\sum_{(i,j)\in A_1}
({\bm X_{ij;2}}-{\bar{\bm X}_{ij}^{(2,1)}})^T
\Sigma_{\{ij\}}^{-1/2}
\left(\Sigma_{\{ij\}}^{1/2} ( S_{\{ij\}}^{(2,1)} )^{-1} \Sigma_{\{ij\}}^{1/2} - I_2 \right)
\Sigma_{\{ij\}}^{-1/2}
{\bm X}_{ij;1}   \nonumber\\
&+\sum_{j\in A_2}
\frac{ ( X_{1j}-{\bar{X}}_{j}^{(2,1)} )^T } {\sigma_{jj}^{1/2}}
\left( \frac{ \sigma_{jj} }{ s_{jj}^{(2,1)} } - 1  \right)
\frac{ {X}_{1j} }{ \sigma_{jj}^{1/2} }  \nonumber\\
=& ~ B_{21}^{(2,1)}+B_{22}^{(2,1)}.
\end{align}
The last equality is based on the facts that
$\widetilde{S}_{\{i,j\}}^{(s,t)}= \Sigma_{\{ij\}}^{-1/2}  S_{\{ij\}}^{(s,t)}  \Sigma_{\{ij\}}^{-1/2}$,
$\tilde{\bm{X}}_{ij;t}=\Sigma_{\{ij\}}^{-1/2} ({\bm X}_{ij;t}-{\bm \mu}_{ij})$,
$\widetilde{X}_{tj}=({X}_{tj}-\mu_j)/\sigma_{jj}^{1/2}$ and
 $\widetilde{s}_{jj}^{(s,t)}={s}_{jj}^{(s,t)}/\sigma_{jj}$.

Note also that
$E( ({\bm X_{s}} -{\bar{\bm X}^{(s,t)}})^T P_{\mathcal{O}}{\bm X}_{t}) =0$
and
$\text{Var}( ({\bm X_{s}} -{\bar{\bm X}^{(s,t)}})^T P_{\mathcal{O}}{\bm X}_{t})
=E( ({\bm X_{s}} -{\bar{\bm X}^{(s,t)}})^T P_{\mathcal{O}}{\bm X}_{t})^2
= {(n-2)} \text{tr}(\Lambda_1^2)/(n-1)=O(p)$.  Then,
\begin{align*}
E(|B_2|)
&\leq
2 E\Big(\big|
({\bm X_{1}} -{\bar{\bm X}^{(1,2)}})^T P_{\mathcal{O}}{\bm X}_{2}
\big| \,
\big( \big| B_{21}^{(2,1)}\big| + \big| B_{22}^{(2,1)} \big| \big) \Big) \\
&\leq
2 \Big[ E\big| ({\bm X_{1}} -{\bar{\bm X}^{(1,2)}})^T P_{\mathcal{O}}{\bm X}_{2}
\big|^2 \Big]^{\frac{1}{2}}\,
\left\{ \big[ E \big( B_{21}^{(2,1)} \big)^2 \big]^{\frac{1}{2}}
+ \big[ E \big( B_{22}^{(2,1)} \big)^2 \big]^{\frac{1}{2}} \right\}  \\
&=
O(p^{1/2}) \left\{ \big[ E \big( B_{21}^{(2,1)} \big)^2 \big]^{\frac{1}{2}}
+ \big[ E \big( B_{22}^{(2,1)} \big)^2 \big]^{\frac{1}{2}} \right\}.
\end{align*}
Next, we show that $E(B_{21}^{(s,t)})^2=o(p)$ for any $s\neq t$.
By letting
\begin{align*}
B_{211}^{(s,t)}&=
\sum_{ (i,j) \in A_1 }
(\tilde{\bm{X}}_{ij;s}-   \tilde{\bar{\bm{X}}}_{ij}^{(s,t)}  )^T
\big(  ( \widetilde{S}_{\{i,j\}}^{(s,t)} )^{-1}-I_2 \big)
\tilde{\bm{X}}_{ij;t},  \\
B_{212}^{(s,t)}&=
\sum_{ (i,j) \in A_1 }
(\tilde{\bm{X}}_{ij;s}-   \tilde{\bar{\bm{X}}}_{ij}^{(s,t)}  )^T
\big(  ( \widetilde{S}_{\{i,j\}}^{(s,t)} )^{-1}-I_2 \big)
\tilde{\bm{\mu}}_{ij},
\end{align*}
we have $E \big( B_{21}^{(s,t)} \big)^2=E(B_{211}^{(s,t)})^2+ E(B_{212}^{(s,t)})^2.$
In the following, we  show
 $E(B_{211}^{(s,t)})^2=o(p)$ and  $E(B_{212}^{(s,t)})^2=o(p)$ as $n \to \infty$, respectively.

Noting that
$E [
(\tilde{\bm{X}}_{ij;s}-   \tilde{\bar{\bm{X}}}_{ij}^{(s,t)}  )^T
(  ( \widetilde{S}_{\{i,j\}}^{(s,t)} )^{-1}-I_2 )
\tilde{\bm{X}}_{i j;t} ]=0$, we have
\begin{align*}
E(B_{211}^{(s,t)})^2
&=
\sum_{\substack{ (i_1,j_1) \in A_1 \\
                    (i_2,j_2) \in A_1 }}
\text{Cov}\Big(
(\tilde{\bm{X}}_{i_1 j_1;s}-  \tilde{\bar{\bm{X}}}_{i_1 j_1}^{(s,t)}  )^T
\big(  ( \widetilde{S}_{\{i_1,j_1\}}^{(s,t)} )^{-1}-I_2 \big)
\tilde{\bm{X}}_{i_1 j_1;t},   \nonumber \\
&~~~~~~~~~~ ~~~~~~~~~~~~~~~~~~~~
(\tilde{\bm{X}}_{i_2 j_2;s}-  \tilde{\bar{\bm{X}}}_{i_2 j_2}^{(s,t)}  )^T
\big(  ( \widetilde{S}_{\{i_2,j_2\}}^{(s,t)} )^{-1}-I_2 \big)
\tilde{\bm{X}}_{i_2 j_2;t} \Big)   \\
&\leq
\Big(2+\frac{\varpi_0}{1-\exp(-1)} \Big)K_0^2p
\max_{(i,j)\in A_1} {\rm Var}\Big(
\big(\tilde{\bm{X}}_{i j;s}-  \tilde{\bar{\bm{X}}}_{i j}^{(s,t)}  \big)^T
\big(  ( \widetilde{S}_{\{i,j\}}^{(s,t)})^{-1}-I_2  \big)
\tilde{\bm{X}}_{i j;t} \Big),
\end{align*}
where the last inequality is
based on the $\rho$-mixing inequality, and the upper bound can be obtained by following
the same procedure as  (\ref{alpha.ineq}).
Note also that
\begin{align*}
&~~~~{\rm Var}\Big(
\big(\tilde{\bm{X}}_{i j;s}-  \tilde{\bar{\bm{X}}}_{i j}^{(s,t)}  \big)^T
\big(  ( \widetilde{S}_{\{i,j\}}^{(s,t)} )^{-1}-I_2  \big)
\tilde{\bm{X}}_{i j;t} \Big)  \\
&=
E\Big(\big(\tilde{\bm{X}}_{i j;s}-  \tilde{\bar{\bm{X}}}_{i j}^{(s,t)}  \big)^T
\big(  ( \widetilde{S}_{\{i,j\}}^{(s,t)} )^{-1}-I_2  \big)^2
\big(\tilde{\bm{X}}_{i j;s}-  \tilde{\bar{\bm{X}}}_{i j}^{(s,t)}  \big) \Big)   \\
&\leq
E\Big(
\big\| \big(  ( \widetilde{S}_{\{i,j\}}^{(s,t)} )^{-1}-I_2  \big)
\big(  \tilde{\bm{X}}_{i j;s}- \tilde{\bar{\bm{X}}}_{i j}^{(s,t)}  \big)
\big\|^2 \Big)    \\
&\leq
\Big( E \big\|  ( \widetilde{S}_{\{i,j\}}^{(s,t)} )^{-1}-I_2 \big\|^4 \Big)^{\frac{1}{2}}
\Big( E \big\| \tilde{\bm{X}}_{i j;s}- \tilde{\bar{\bm{X}}}_{i j}^{(s,t)}\big\|^4 \Big)^{\frac{1}{2}}.
\end{align*}
By $(ii)$ in Lemma \ref{boundlemma5},
$E \big\| ( \widetilde{S}_{\{i,j\}}^{(s,t)} )^{-1}-I_2 \big\|^4=O(n^{-2})$
hold uniformly  over $(i,j)\in A_1$.
In addition,  $E\big\| \tilde{\bm{X}}_{i j;s}- \tilde{\bar{\bm{X}}}_{i j}^{(s,t)}
\big\|^4$ are finite combinations of higher order moments, where the highest terms
are $E(\tilde{X}_{ki}^{4})$ and $E(\tilde{X}_{kj}^{4})$ for $k\neq s,t$,
and hence are bounded uniformly over $(i,j)\in A_1$.
Consequently, we have
$${\rm Var}\Big(
\big(\tilde{\bm{X}}_{i j;s}-  \tilde{\bar{\bm{X}}}_{i j}^{(s,t)}  \big)^T
\big(  (\widetilde{S}_{\{i,j\}}^{(s,t)})^{-1}-I_2  \big)
\tilde{\bm{X}}_{i j;t} \Big)=O(n^{-1}),$$
which also hold uniformly over $(i,j)\in A_1$.
This shows that $E(B_{211}^{(s,t)})^2=o(p)$.

In addition,
\begin{align*}
E(B_{211}^{(s,t)})^2
&\leq
\sum_{ (i_1,j_1) \in A_1 } \sum_{ (i_2,j_2) \in A_1 }
\big\| \tilde{\bm{\mu}}_{i_1 j_1}^T \big\|
E   \Big(
\big\| (\widetilde{S}_{\{i_1 j_1 \}}^{(s,t)} )^{-1}-I_2  \big\|
\times  \big\| \tilde{\bm{X}}_{i_1 j_1;s}- \tilde{\bar{\bm{X}}}_{i_1 j_1}^{(s,t)} \big\|   \\
&~~~~\times \big\| \tilde{\bm{X}}_{i_2 j_2;s}- \tilde{\bar{\bm{X}}}_{i_2 j_2}^{(s,t)}  \big\|
\times \big\|  ( \widetilde{S}_{\{i_2 j_2 \}}^{(s,t)} )^{-1}-I_2 \big\|
\Big) \big\| \tilde{\bm{\mu}}_{i_2 j_2}\big\|.
\end{align*}
By ${(ii)}$ in Lemma \ref{boundlemma5},
{\small
\begin{align*}
&E  \left(
\big\| ( \widetilde{S}_{\{i_1 j_1 \}}^{(s,t)} )^{-1}-I_2  \big\|
\times  \big\| \tilde{\bm{X}}_{i_1 j_1;s}- \tilde{\bar{\bm{X}}}_{i_1 j_1}^{(s,t)} \big\|
\times \big\| \tilde{\bm{X}}_{i_2 j_2;s}- \tilde{\bar{\bm{X}}}_{i_2 j_2}^{(s,t)}  \big\|
\times \big\|  (\widetilde{S}_{\{i_2 j_2 \}}^{(s,t)} )^{-1}-I_2 \big\|
\right) \\
\leq
 &\left[ E \big(
\big\| ( \widetilde{S}_{\{i_1 j_1 \}}^{(s,t)} )^{-1}-I_2  \big\|^2
  \big\| \tilde{\bm{X}}_{i_1 j_1;s}- \tilde{\bar{\bm{X}}}_{i_1 j_1}^{(s,t)} \big\|^2
\big) \right]^{\frac{1}{2}}
\left[ E \big( \big\| \tilde{\bm{X}}_{i_2 j_2;s}- \tilde{\bar{\bm{X}}}_{i_2 j_2}^{(s,t)}  \big\|^2
 \big\|  (\widetilde{S}_{\{i_2 j_2 \}}^{(s,t)} )^{-1}-I_2 \big\|^2 \big)
\right]^{\frac{1}{2}}   \\
\leq &
\Big[ E\Big(
\big\| ( \widetilde{S}_{\{i_1 j_1 \}}^{(s,t)} )^{-1}-I_2  \big\|^4 \Big)  \Big]^{\frac{1}{4}}
\Big[ E\Big(  \big\| \tilde{\bm{X}}_{i_1 j_1;s}- \tilde{\bar{\bm{X}}}_{i_1 j_1}^{(s,t)} \big\|^4
\Big) \Big]^{\frac{1}{4}}
\Big[ E\Big(
\big\| ( \widetilde{S}_{\{i_2 j_2 \}}^{(s,t)})^{-1}-I_2  \big\|^4 \Big)  \Big]^{\frac{1}{4}}  \\
&\times
\Big[ E \Big(  \big\| \tilde{\bm{X}}_{i_2 j_2;s}- \tilde{\bar{\bm{X}}}_{i_2 j_2}^{(s,t)} \big\|^4 \Big) \Big]^{\frac{1}{4}} \\
= &~O(n^{-1})
\end{align*}}{hold} uniformly for any  $(i_1,j_1)$ and $(i_2,j_2)\in A_1$.
Thus, there exists  a constant  $K_{01}>0$ such that
\begin{align*}
E(B_{211}^{(s,t)})^2
&\leq
\Big( \sum_{ (i_1,j_1) \in A_1 }\big\| \tilde{\bm{\mu}}_{i_1 j_1} \big\|  \Big)
\Big( \sum_{ (i_2,j_2) \in A_1 } \big\| \tilde{\bm{\mu}}_{i_2 j_2} \big\|   \Big) K_{01}/n \\
&\leq
\sqrt{\text{card}(A_1)}
\Big( \sum_{ (i_1,j_1) \in A_1 } \big\| \tilde{\bm{\mu}}_{i_1 j_1}\big\|^2  \Big)^{1/2}
\sqrt{\text{card}(A_1)}
\Big( \sum_{ (i_2,j_2) \in A_1 } \big\| \tilde{\bm{\mu}}_{i_2 j_2} \big\|^2   \Big)^{1/2} K_{01}/n  \\
&=
O({pn^{-1}}) \Big(
\sum_{ (i_1,j_1) \in A_1 } \big\| \tilde{\bm{\mu}}_{i_1 j_1}\big\|^2  \Big) \\
&=
O({pn^{-1}})  {\bm \mu}^T P_{\mathcal{O}}{\bm \mu},
\end{align*}
where the second inequality is based on the Cauchy-Schwarz inequality,
and the last equality is based on the fact that
$\sum_{ (i,j) \in A_1 } \big\| \tilde{\bm{\mu}}_{i j}\big\|^2
=\sum_{ (i,j) \in A_1} {\bm{\mu}}_{i j}^T \Sigma_{\{ij\}}^{-1} {\bm{\mu}}_{i j}
\leq  {\bm{\mu}}^T P_{\mathcal{O}}{\bm{\mu}}$.
By condition (C5) and the assumption that  $p/n^3=o(1)$, we have
$E(B_{212}^{(s,t)})^2/{p}=O(n^{-1}){\bm{\mu}}^T P_{\mathcal{O}}{\bm{\mu}} = o\big(p^{1/2}n^{-3/2}\big)
=o(1)$ as $n \to \infty$. This indicates that $E(B_{212}^{(s,t)})^2=o({p})$.
Consequently, we have  $E(B_{21}^{(s,t)})^2=o(p)$ as $n \to \infty$.
Following the similar procedure, we can prove that $E(B_{22}^{(s,t)})^2=o(p)$ as $n \to \infty$
for $s\neq t$.

\vskip12pt
\noindent{\bf Part-II:  Proof of ${B_3=o_p(p)}$ }
\vskip12pt

By (\ref{B2.expand}), we have
\begin{align}
E(|B_3|)
&=
E\Big(
\big| ( {\bm X_{1}} -\bar{\bm X}^{(1,2)} )^T
({P}_{\mathcal{O}}^{(1,2)} - P_{\mathcal{O}})
{\bm X}_{2} \big|\,
\big|({\bm X_{2}} - {\bar{\bm X}^{(1,2)}})^T
({P}_{\mathcal{O}}^{(1,2)} - P_{\mathcal{O}})
 {\bm X}_{1}\big|
\Big)  \nonumber  \\
&=
E\big(
\big| B_{21}^{(1,2)}+B_{22}^{(1,2)} \big|\,
\big|B_{21}^{(2,1)} +B_{22}^{(2,1)}\big|
\big)   \nonumber  \\
&\leq
\Big[ E \big( B_{21}^{(1,2)}\big)^2 \Big]^{\frac{1}{2}}
\Big[ E \big( B_{21}^{(2,1)}\big)^2 \Big]^{\frac{1}{2}}
+\Big[ E \big( B_{21}^{(1,2)}\big)^2 \Big]^{\frac{1}{2}}
\Big[ E \big( B_{22}^{(2,1)} \big)^2 \Big]^{\frac{1}{2}}    \nonumber \\
&~~~~+\Big[ E \big( B_{22}^{(1,2)}\big)^2 \Big]^{\frac{1}{2}}
\Big[ E \big( B_{21}^{(2,1)} \big)^2 \Big]^{\frac{1}{2}}
+\Big[ E \big( B_{22}^{(1,2)}\big)^2 \Big]^{\frac{1}{2}}
\Big[ E \big( B_{22}^{(2,1)} \big)^2 \Big]^{\frac{1}{2}}.
\end{align}
Note that
$E\big( B_{21}^{(1,2)}\big)^2=E \big( B_{21}^{(2,1)}\big)^2$
and
$E\big( B_{22}^{(1,2)}\big)^2=E \big( B_{22}^{(2,1)}\big)^2$.
Also in the proof of Part-I, we have shown that
$E\big( B_{21}^{(t,s)} \big)^2=o(p)$
and
$E\big( B_{22}^{(t,s)} \big)^2=o(p)$  as $n \to \infty$ for any $t \neq s$.
Thus, $E(B_3)=o(p)$ as $n \to \infty$ and so we complete the proof of Lemma~\ref{lemma1.1}.

 \hfill$\Box$

\end{appendices}

\newpage

\begin{appendices}
\renewcommand{\thesection}{\Alph{section}:}
\section{ Proofs of Theorem 3.1, Theorem 3.2 and Lemma 2}
\renewcommand{\thesection}{\Alph{section}}
\subsection{Proof of Theorem \ref{pro2sam}} \label{Prof.pro2sam}
Noting that
\begin{align*}
&~~~~\big\{ \hat{\tau}_{ij} <\tau_0 \big|  \tau_{ij} >\tau_0 \big\} \\
&\subseteq
\big\{|\hat{\tau}_{ij} -\tau_{ij}| >\epsilon_0 \big|  \tau_{ij} >\tau_0 \big\}\\
&\subseteq
\big\{|\hat{\tau}_{ij,1} -\tau_{ij}| >{n_1 \epsilon_0}/{(n_1+n_2)} \big|  \tau_{ij} >\tau_0 \big\}
\cup \big\{|\hat{\tau}_{ij,2} -\tau_{ij}| >{n_2 \epsilon_0}/{(n_1+n_2)}  \big|  \tau_{ij} >\tau_0 \big\}\\
&\subseteq
\big\{|\hat{r}_{ij,1} -r_{ij}| >{n_1 \epsilon_0}/{(n_1+n_2)}  \big|  \tau_{ij} >\tau_0 \big\}
\cup \big\{|\hat{r}_{ij,2} -r_{ij}| >{n_2 \epsilon_0}/{(n_1+n_2)}  \big|  \tau_{ij} >\tau_0 \big\},
\end{align*}
we have
\begin{align*}
&P\big(\{\hat{\tau}_{ij}<\tau_0 | \tau_{ij}>\tau_0\}\big)\\
\leq&~
P\big(\{|\hat{r}_{ij,1} -r_{ij}| >n_1\epsilon_0/(n_1+n_2) | \tau_{ij}>\tau_0\}\big)
 + P\big(\{|\hat{r}_{ij,2} -r_{ij}| >n_2 \epsilon_0/(n_1+n_2) | \tau_{ij}>\tau_0\}\big)\\
\leq&~
2 \Big( \exp\Big(-\frac{n_1^3  \epsilon_0^2}{4(n_1+n_2)^2}\Big)
 +\exp\Big(-\frac{n_2^3 \epsilon_0^2}{4 (n_1+n_2)^2 }\Big)
 \Big).
\end{align*}
Similarly, we can show that
$$P\big(\{\hat{\tau}_{ij}>\tau_0 | \tau_{ij}<\tau_0\}\big)
\leq
2 \Big( \exp\Big(-\frac{n_1^3  \epsilon_0^2}{4(n_1+n_2)^2}\big)
 +\exp\Big(-\frac{n_2^3 \epsilon_0^2}{4 (n_1+n_2)^2 }\Big)
 \Big).$$

Then according to the same decomposition of
$\{\hat{A}_1 \neq A_1\}$ and $\{\hat{A}_2 \neq A_2\}$ in
 (\ref{set1}),  (\ref{ineqA_2.1}) and  (\ref{ineqA_2}), we have
\begin{align*}\label{2samA_1}
& P\Big(\{\hat{A}_1 \neq A_1\}\Big)
\leq  2p(p-1)  \Big( \exp\Big(-\frac{n_1^3  \epsilon_0^2}{4(n_1+n_2)^2}\Big)
 +\exp\Big(-\frac{n_2^3 \epsilon_0^2}{4 (n_1+n_2)^2 }\Big)
 \Big)
\end{align*}
 and
\begin{align*}
& P\Big(\{\hat{A}_2\neq A_2\}\Big)
\leq  2p(p-1)  \Big( \exp\Big(-\frac{n_1^3  \epsilon_0^2}{4(n_1+n_2)^2}\Big)
 +\exp\Big(-\frac{n_2^3 \epsilon_0^2}{4 (n_1+n_2)^2 }\Big)
 \Big).
\end{align*}
Consequently, as $N \to \infty$ we have
$P\big(\{\hat{A}_1 \neq A_1\}\big)\to 0$ and
$P\big(\{\hat{A}_2 \neq A_2\}\big)\to 0$.

\hfill$\Box$

\subsection{Proof of Theorem \ref{th2_1}} \label{Prof.th2_1}

Let
$T_{21}=T_{211}+ 2 T_{212}+ ({\bm \mu}_1-{\bm \mu}_2)^T P_{\mathcal{O}}({\bm \mu}_1-{\bm \mu}_2)$ where
{
\begin{align*}
T_{211}
&=
\frac{1}{n_1(n_1-1)}\sum_{s=1}^{n_1}\sum_{t\neq s}^{n_1}
({\bm{X}}_{s}-{\bm \mu}_1)^T P_{\mathcal{O}} ({\bm{X}}_{t}-{\bm \mu}_1) \\
&~~~~+\frac{1}{n_2(n_2-1)}\sum_{s=1}^{n_2}\sum_{t\neq s}^{n_2}
({\bm{Y}}_{s}-{\bm \mu}_2)^T
P_{\mathcal{O}}
({\bm{Y}}_{t}-{\bm \mu}_2)       \\
&~~~~ -\frac{2}{n_1 n_2}\sum_{s=1}^{n_1}\sum_{t=1}^{n_2}
({\bm{X}}_{s}-{\bm \mu}_1)^T
P_{\mathcal{O}}
({\bm{Y}}_{t}-{\bm \mu}_2), \\
T_{212} &=
\frac{1}{n_1} \sum_{t=1}^{n_1}
({\bm \mu}_1-{\bm \mu}_2)^T  P_{\mathcal{O}} ({\bm{X}}_{t}-{\bm \mu}_1)
+\frac{1}{n_2}\sum_{t= 1}^{n_2}({\bm \mu}_1-{\bm \mu}_2)^T P_{\mathcal{O}}
({\bm{Y}}_{t}-{\bm \mu}_2).
\end{align*}}
Let also $T_{22}= T_{221}+T_{222}-2 T_{223}$ where
\begin{align*}
T_{221}&=\frac{1}{n_1(n_1-1)}\sum_{s=1}^{n_1}\sum_{t\neq s}^{n_1}
{\bm{X}}_{s}^T \big( \widehat{P}_{1,\mathcal{O}}^{(s,t)}-P_{\mathcal{O}} \big)
{\bm{X}}_{t}, \\
T_{222}&=\frac{1}{n_2(n_2-1)}\sum_{s=1}^{n_2}\sum_{t\neq s}^{n_2}
{\bm{Y}}_{s}^T \big(\widehat{P}_{2,\mathcal{O}}^{(s,t)}-P_{\mathcal{O}}\big)
{\bm{Y}}_{t}, \\
T_{223}&=\frac{1}{n_1 n_2}\sum_{s=1}^{n_1}\sum_{t=1}^{n_2}{\bm{X}}_{s}^T
\big( \widehat{P}_{12,\mathcal{O}}^{(s,t)}-P_{\mathcal{O}} \big){\bm{Y}}_{t}.
\end{align*} Then  $T_2(\tau_0)=T_{21} +T_{22}$.
Hence to show Theorem \ref{th2_1}, it suffices to show that
\begin{eqnarray}\label{asy_T21}
\frac{T_{21}-({\bm \mu}_1-{\bm \mu}_2)^T P_{\mathcal{O}} ({\bm \mu}_1-{\bm \mu}_2) }
{\sqrt{\phi(n_1,n_2){\rm tr}(\Lambda_1^2)}} \stackrel{D}{\longrightarrow}N(0,1) {\rm~ ~as~~ } N\to\infty,
\end{eqnarray}
and
\begin{eqnarray}\label{asy_T22}
\frac{T_{22}}{\sqrt{\phi(n_1,n_2){\rm tr}(\Lambda_1^2)}}\stackrel{P}{\longrightarrow}0
{\rm~ ~as~~ } N\to\infty.
\end{eqnarray}

\vskip 12pt
\noindent {\it  {\bf Part I: Proof of (\ref{asy_T21})}}
\vskip 12pt

First of all, we show that
\begin{equation}\label{asy.T212}
\frac{T_{212}}{\sqrt{\phi(n_1,n_2){\rm tr}(\Lambda_1^2)}}\stackrel{P}{\longrightarrow}0
{\rm~ ~as~~ } N\to\infty.
\end{equation}
Since $n_1/N \to \varphi_0 \in (0,1)$ and $\phi(n_1,n_2)=O(N^{-2})$, then by (\ref{ord_trace}),
we only need to show $T_{212}=o_p(p^{1/2}N^{-1}).$
Note that $E(T_{212})=0$ and
$E(T_{212})^2=(n_1^{-1}+n_2^{-1})({\bm \mu}_1-{\bm \mu}_2)^T
P_{\mathcal{O}}\Sigma P_{\mathcal{O}} ({\bm \mu}_1-{\bm \mu}_2).$
By $(iii)$ in Lemma \ref{Lemmaeigen}, we have
$E(T_{212})^2=o(pN^{-2}).$
This indicates that $T_{212}=o_p(p^{1/2}N^{-1})$.

Secondly, we show that
\begin{equation}\label{asy.T211}
\frac{T_{211}}{\sqrt{\phi(n_1,n_2){\rm tr}(\Lambda_1^2)}}\stackrel{D}{\longrightarrow}N(0,1)
{\rm~ ~as~~ } N\to\infty.
\end{equation}
Let $\breve{{\bm{X}}}_{s}=P_{\mathcal{O}}^{1/2}({\bm{X}}_{s}-{\bm \mu}_1)$ for $s=1,\ldots,n_1$,
and $\breve{{\bm{Y}}}_{t}=P_{\mathcal{O}}^{1/2}({\bm{Y}}_{t}-{\bm \mu}_2)$ for $t=1,\ldots,n_2$.
Then,
$$T_{211}=\frac{1}{n_1(n_1-1)}\sum_{s=1}^{n_1}\sum_{t\neq s}^{n_1} \breve{{\bm{X}}}_{s}^T \breve{{\bm{X}}}_{t}
+\frac{1}{n_2(n_2-1)}\sum_{s=1}^{n_2}\sum_{t\neq s}^{n_2}\breve{{\bm{Y}}}_{s}^T \breve{{\bm{Y}}}_{t}
-\frac{2}{n_1n_2}\sum_{s=1}^{n_1} \sum_{t=1}^{n_2} \breve{{\bm{X}}}_{s}^T \breve{{\bm{Y}}}_{t}.$$
Note that
$\text{Var}(\breve{\bm{X}}_{s})=\text{Var}(\breve{\bm{Y}}_{t})=P_{\mathcal{O}}^{1/2} \Sigma P_{\mathcal{O}}^{1/2}$
and $\text{Var}(T_{211})=\phi(n_1,n_2)\text{tr}(\Lambda_1^2)$.
After direct verification, we can see that the random samples $\breve{{\bm{X}}}_{1},\ldots,\breve{{\bm{X}}}_{n_1}$,  $\breve{{\bm{Y}}}_{1},\ldots,\breve{{\bm{Y}}}_{n_2}$
and the common covariance matrix satisfies the conditions in Theorem 1 in \citet{chen2010}.
Consequently,
${T_{211}}/ {\sqrt{ \text{Var}(T_{211}) }} \stackrel{D}{\longrightarrow}N(0,1)$ as $N \to \infty$
and this completes the proof of (\ref{asy.T211}).

\vskip 12pt
\noindent {\it  {\bf Part II: Proof of (\ref{asy_T22})}}
\vskip 12pt
Under conditions (C1) and (C2$'$)--(C5$'$),
to show (\ref{asy_T22}), it suffices to verify that
${T_{22l}}/{\sqrt{\phi(n_1,n_2){\rm tr}(\Lambda_1^2)}}\stackrel{P}{\longrightarrow}0$
as $N \to \infty$
for $l=1,2,3$.
To save space, we only  prove the result for the case with $l=3$.
The proofs for $l=1,2$ are nearly the same as
(\ref{asy_Un2}) in Section \ref{Prof.th1}, and hence are omitted.
Let
\begin{align*}
&T_{XY,1}=
\frac{1}{n_1 n_2} \sum_{s=1}^{n_1} \sum_{t=1}^{n_2}
{\bm{X}}_{s}^T \Big( \sum_{(i,j) \in \hat{A}_1 }
P_{ij}^T (P_{ij} S_{12,*}^{(s,t)} P_{ij}^T)^{-1}P_{ij}
-\sum_{(i,j) \in {A}_1} P_{ij}^T (P_{ij} S_{12,*}^{(s,t)} P_{ij}^T)^{-1}P_{ij}
\Big) {\bm{Y}}_{t},  \\
&T_{XY,2}=
\frac{1}{n_1 n_2} \sum_{s=1}^{n_1} \sum_{t=1}^{n_2}
{\bm{X}}_{s}^T
\Big( \sum_{i \in \hat{A}_2}P_{i}^T ( P_{i}S_{12,*}^{(s,t)} P_{i}^T)^{-1}P_{i}
-\sum_{i \in {A}_2 }P_{i}^T ( P_{i}S_{12,*}^{(s,t)} P_{i}^T)^{-1} P_{i}
\Big)
{\bm{Y}}_{t},    \\
&T_{XY,3}=\frac{1}{n_1 n_2}\sum_{s=1}^{n_1} \sum_{t=1}^{n_2} {\bm{X}}_{s}^T
\Big(
\sum_{ (i,j) \in A_1 } P_{ij}^T (P_{ij} S_{12,*}^{(s,t)} P_{ij}^T)^{-1}P_{ij}
-\sum_{ (i,j) \in A_1 } P_{ij}^T (P_{ij} \Sigma P_{ij}^T)^{-1}P_{ij} \Big){\bm{Y}}_{t},    \\
&T_{XY,4}=\frac{1}{n_1 n_2}\sum_{s=1}^{n_1} \sum_{t=1}^{n_2}
{\bm{X}}_{s}^T \Big(
\sum_{i \in {A}_2 } P_{i}^T ( P_{i}S_{12,*}^{(s,t)} P_{i}^T)^{-1} P_{i}
-\sum_{i \in A_2}   P_{i}^T ( P_{i} \Sigma P_{i}^T)^{-1} P_{i}
\Big){\bm{Y}}_{t}.
\end{align*}
Then, we have
$T_{223}=T_{XY,1}+T_{XY,2}+T_{XY,3}+T_{XY,4}.$
Note also that for any $\epsilon_1>0$,
$\{ |T_{XY,1}| > \epsilon_1 \sqrt{\phi(n_1,n_2) \text{tr}(\Lambda_1^2)} \}
\subseteq \{ \hat{A}_1 \neq A_1 \}$.
By Theorem \ref{pro2sam}, we have
$P( |T_{XY,1}| > \epsilon_1 \sqrt{\phi(n_1,n_2) \text{tr}(\Lambda_1^2)} )
\leq P ( \hat{A}_1 \neq A_1 )\to 0$ as $N \to \infty$.
Similarly,
$P( |T_{XY,2}| > \epsilon_1 \sqrt{\phi(n_1,n_2) \text{tr}(\Lambda_1^2)} )
\leq P ( \hat{A}_2 \neq A_2 )\to 0$ as $N \to \infty$.
This indicates that as $N \to \infty$, $T_{XY,1}/\sqrt{\phi(n_1,n_2) \text{tr}(\Lambda_1^2)}=o_p(1)$
and $T_{XY,2}/\sqrt{\phi(n_1,n_2) \text{tr}(\Lambda_1^2)}=o_p(1)$.
It remains to show that $T_{XY,3}/\sqrt{\phi(n_1,n_2) \text{tr}(\Lambda_1^2)}=o_p(1)$
and $T_{XY,4}/\sqrt{\phi(n_1,n_2) \text{tr}(\Lambda_1^2)}=o_p(1)$ as $N \to \infty$.
By (\ref{ord_trace}) and the fact that $\phi(n_1,n_2)=O(N^{-2})$,
we only need to verify Parts II-1 and II-2, respectively.

\vskip 12pt
\noindent {\it  {\bf Part II-1: Proof of }
 ${ T_{XY,3}=o_p(p^{1/2}N^{-1})}$}
\vskip 12pt

First of all, we have
\begin{align*}
T_{XY,3}
&=
\frac{1}{n_1 n_2}\sum_{ (i,j) \in A_1 }  \sum_{s=1}^{n_1} \sum_{t=1}^{n_2} {\bm{X}}_{ij;s}^T
\big(  ( S_{12,\{i,j\} }^{(s,t)} )^{-1} - \Sigma_{ \{i,j\} }^{-1} \big)
{\bm{Y}}_{ij;t}   \\
&=
\frac{1}{n_1 n_2}\sum_{ (i,j) \in A_1 }  \sum_{s=1}^{n_1} \sum_{t=1}^{n_2}
{\bm{X}}_{ij;s}^T \Sigma_{ \{i,j\} }^{-1/2}
\big( \Sigma_{ \{i,j\} }^{1/2}  ( S_{12,\{i,j\} }^{(s,t)} )^{-1}
\Sigma_{ \{i,j\} }^{1/2}
-  I_2 \big) \Sigma_{ \{i,j\} }^{-1/2}{\bm{Y}}_{ij;t}   \\
&=\frac{1}{n_1 n_2}\sum_{ (i,j) \in A_1 }  \sum_{s=1}^{n_1} \sum_{t=1}^{n_2}
  (\tilde{\bm{X}}_{ij;s}+\tilde{\bm{\mu}}_{1,ij} )^T
  \big( (\widetilde{S}_{12,\{i,j\} }^{(s,t)} )^{-1}-I_2 \big)
   (\tilde{\bm{Y}}_{ij;t} +\tilde{\bm{\mu}}_{2,ij} ).
\end{align*}
Let $\Xi_{12,\{i,j\}}^{(s,t)}=(I_2-\widetilde{S}_{12,\{i,j\}}^{(s,t)})
+(I_2-\widetilde{S}_{12,\{i,j\}}^{(s,t)})^2
+\cdots+\big(I_2-\widetilde{S}_{12,\{i,j\}}^{(s,t)}\big)^{m_0}$.
Noting that $\widetilde{S}_{12, \{i,j\}}^{(s,t)}$  is a
$\sqrt{n}$-consistent estimator of $I_{2\times2}$,
hence by Taylor expansion for matrix functions,
$ \| ((\widetilde{S}_{12,\{i,j\}}^{(s,t)})^{-1} - I_2 )
-\Xi_{\{i,j\}}^{(s,t)} \|=O_p(N^{-(m_0+1)/2}).$
Let also
\begin{align*}
T_{XY,31}
&=\frac{1}{n_1 n_2}\sum_{ (i,j) \in A_1 }  \sum_{s=1}^{n_1} \sum_{t=1}^{n_2}
\tilde{\bm{X}}_{ij;s}^T \Xi_{12, \{i,j\}}^{(s,t)} \tilde{\bm{Y}}_{ij;t},  \\
T_{XY,32}
&=\frac{1}{n_1 n_2} \sum_{ (i,j) \in A_1 }  \sum_{s=1}^{n_1} \sum_{t=1}^{n_2}
\tilde{\bm{X}}_{ij;s}^T \Xi_{12, \{i,j\}}^{(s,t)} \tilde{\bm{\mu}}_{2,ij},   \\
T_{XY,33}
&=\frac{1}{n_1 n_2} \sum_{ (i,j) \in A_1 }  \sum_{s=1}^{n_1} \sum_{t=1}^{n_2}
\tilde{\bm{Y}}_{ij;t}^T \Xi_{12, \{i,j\}}^{(s,t)} \tilde{\bm{\mu}}_{1,ij},     \\
T_{XY,34}
&=\frac{1}{n_1 n_2} \sum_{ (i,j) \in A_1 }  \sum_{s=1}^{n_1} \sum_{t=1}^{n_2}
  \tilde{\bm{\mu}}_{1,ij}^T   \big( ( \widetilde{S}_{12,\{i,j\} }^{(s,t)} )^{-1}-I_2 \big) \tilde{\bm{\mu}}_{2,ij}.
\end{align*}
Then, under conditions (C4$'$) and (C5$'$), we have
\begin{align*}
T_{XY,3}
= T_{XY,31}+T_{XY,32}+T_{XY,33}+ T_{XY,34}+\text{card}(A_1) O_p(N^{-(m_0+1)/{2}}).
\end{align*}
By condition (C$3'$), we have $\text{card}(A_1) \leq K_0 p$. Thus for
 $m_0\geq 4$, $\text{card}(A_1) O_p(N^{-(m_0+1)/2})=O_p(pN^{-(m_0+1)/2})
=o_p (p^{1/2}N^{-1})$ as $N \to \infty$.

\vskip 12pt
\noindent {\it  {\bf Part II-1.1: Proof of }
 ${ T_{XY,31}=o_p(p^{1/2}N^{-1})}$}
\vskip 12pt
Since
$E(T_{XY,31})=0$, we only need to show  $E(T_{XY,31}^2)=o(pN^{-2})$ as $N \to \infty$.
Note that
\begin{align*}
& E(T_{XY,31}^2) \\
=&
\sum_{\substack{ (i_1,j_1) \in A_1 \\
                    (i_2,j_2) \in A_1 }}
\text{Cov}\Big(
 \frac{1}{n_1 n_2} \sum_{s=1}^{n_1}\sum_{t=1}^{n_2}
\tilde{\bm{X}}_{i_1 j_1;s}^T \Xi_{12,\{i_1,j_1\}}^{(s,t)}  \tilde{\bm{Y}}_{i_1 j_1;t},\,
 \frac{1}{n_1 n_2} \sum_{l=1}^{n_1}\sum_{m =1}^{n_2}
\tilde{\bm{X}}_{i_2 j_2;l}^T \Xi_{12, \{i_2,j_2\}}^{(l,m)} \tilde{\bm{Y}}_{i_2 j_2;m}\Big).
\end{align*}
By the $\rho$-mixing inequality and following the same procedure as in (\ref{alpha.ineq}),
we have
\begin{align*}
&\Big|\text{Cov}\Big(\frac{1}{n_1(n_2)} \sum_{s=1}^{n_1}\sum_{t=1}^{n_2}
\tilde{\bm{X}}_{i_1 j_1;s}^T \Xi_{12, \{i_1,j_1\}}^{(s,t)} \tilde{\bm{Y}}_{i_1 j_1;t},\,
\frac{1}{n_1 n_2} \sum_{l=1}^{n_1}\sum_{m=1}^{n_2}
\tilde{\bm{X}}_{i_2 j_2;l}^T
\Xi_{\{i_2,j_2\}}^{(l,m)}
\tilde{\bm{X}}_{i_2 j_2;m}
\Big)\Big|   \nonumber  \\
& \leq \varpi_0 \exp\Big(-{\text{dist}(\{i_1,j_1\},\{i_2,j_2\})}\Big)
 \max_{(i,j)\in A_1}
 {\rm Var}\Big( \frac{1}{n_1 n_2} \sum_{s=1}^{n_1}\sum_{t=1}^{n_2}
\tilde{\bm{X}}_{i j; s}^T \Xi_{12,\{i,j\}}^{(s,t)} \tilde{\bm{Y}}_{i j;t}\Big),
\end{align*}
where ${\text{dist}(\{i_1,j_1\},\{i_2,j_2\})}=\min\{|i_1-i_2|,|i_1-j_2|,
|j_1-i_2|, |j_1-j_2|\}$.
Then by condition (C3$'$), we have
\begin{align*}
E(T_{XY,31}^2)
&\leq \Big(2+\frac{\varpi_0}{1-\exp(-1)} \Big)K_0^2p
\max_{(i,j)\in A_1} {\rm Var}\Big( \frac{1}{n_1 n_2} \sum_{s=1}^{n_1}\sum_{t=1}^{n_2}
\tilde{\bm{X}}_{i j;s}^T \Xi_{12, \{i,j\}}^{(s,t)} \tilde{\bm{Y}}_{i j;t}\Big).
\end{align*}

To show $E(T_{XY,31}^2)=o(pN^{-2})$, it suffices to verify that
\begin{equation}\label{asyT_31}
{\rm Var}\Big( \frac{1}{n_1 n_2 } \sum_{s=1}^{n_1}\sum_{t=1}^{n_2}
\tilde{\bm{X}}_{i j;s}^T \Xi_{12, \{i,j\}}^{(s,t)} \tilde{\bm{Y}}_{i j;t}\Big)=O(N^{-3})
\end{equation}
hold uniformly over $(i,j)\in A_1$.
For ease of notation, we let
\begin{align}\label{JXY}
J_{\{i,j\}}^{X,Y}(\nu_1,\nu_2)
= \frac{1}{n_1^2 n_2^2} \sum_{\substack{ s_1 =1  \\ s_2= 1 }}^{n_1}
\sum_{\substack{ t_1=1  \\ t_2=1 } }^{n_2} M_{\{i,j\}}^{X,Y}(s_1,t_1,\nu_1,s_2,t_2,\nu_2),
\end{align}
where
$$M_{\{i,j\}}^{X,Y}(s_1,t_1,\nu_1,s_2,t_2,\nu_2)=
{E}\Big(
\tilde{\bm{X}}_{i j;s_1}^T
(I_2-\widetilde{S}_{12, \{i,j\}}^{(s_1,t_1)})^{\nu_1}
\tilde{\bm{Y}}_{i j;t_1}
\times
\tilde{\bm{X}}_{i j;s_2}^T
(I_2-\widetilde{S}_{12, \{i,j\}}^{(s_2,t_2)})^{\nu_2}
\tilde{\bm{Y}}_{i j;t_2} \Big).$$
Then,
\begin{align}\label{lem6E}
{\rm Var} \Big( \frac{1}{n_1 n_2}
\sum_{s_1 =1}^{n_1}
\sum_{t_1=1 }^{n_2}
\tilde{\bm{X}}_{i j; s}^T \Xi_{12, \{i,j\}}^{(s,t)} \tilde{\bm{Y}}_{i j; t}\Big)
&=
{E} \Big( \frac{1}{n_1 n_2} \sum_{s=1}^{n_1}\sum_{t =1 }^{n_2}
\tilde{\bm{X}}_{ij;s}^T \Xi_{12, \{i,j\}}^{(s,t)} \tilde{\bm{Y}}_{i j;t}\Big)^{2}   \nonumber \\
&=
\sum_{\nu_1=1}^{m_0}
\sum_{\nu_2=1}^{m_0}
J_{\{i,j\}}^{X,Y}(\nu_1,\nu_2).
\end{align}
We further  decompose $J_{\{i,j\}}^{X,Y}(\nu_1,\nu_2)$ into three exclusive sets:
\begin{itemize}
  \item[(I)]
  $\mathcal{S}_1=\{(s_1, t_1, s_2, t_2) |
  s_1 = s_2, t_1= t_2 \}$;

  \item[{(II)}]
 $\mathcal{S}_2=\{(s_1, t_1, s_2, t_2) |
  s_1= s_1, s_2\neq t_2 \}
  \cup \{(s_1, t_1, s_2, t_2) |
  s_1\neq t_1, s_2= t_2 \}$;

  \item[{(III)}]
 $\mathcal{S}_3=\{(s_1, t_1, s_2, t_2) \Big|
  s_1\neq s_2, t_1\neq t_2 \}$.
\end{itemize}

Let also
\begin{align*}
J_{\{i,j\}}^{X,Y}(\nu_1,\nu_2 | \mathcal{S}_1)
=
\frac{1}{n_1^2 n_2^2}
\sum_{ (s_1, t_1, s_2, t_2) \in \mathcal{S}_1 }
M_{\{i,j\}}^{X,Y}(s_1,t_1,\nu_1,s_2,t_2,\nu_2),  \\
J_{\{i,j\}}^{X,Y}(\nu_1,\nu_2 | \mathcal{S}_2)
=
\frac{1}{n_1^2 n_2^2}
\sum_{ (s_1, t_1, s_2, t_2) \in \mathcal{S}_2 }
M_{\{i,j\}}^{X,Y}(s_1,t_1,\nu_1,s_2,t_2,\nu_2),     \\
J_{\{i,j\}}^{X,Y}(\nu_1,\nu_2 | \mathcal{S}_3)
=
\frac{1}{n_1^2 n_2^2}
\sum_{ (s_1, t_1, s_2, t_2) \in \mathcal{S}_3 }
M_{\{i,j\}}^{X,Y}(s_1,t_1,\nu_1,s_2,t_2,\nu_2).
\end{align*}
Then,
$J_{\{i,j\}}^{X,Y}(\nu_1,\nu_2)=
J_{\{i,j\}}^{X,Y}(\nu_1,\nu_2 | \mathcal{S}_1)
+J_{\{i,j\}}^{X,Y}(\nu_1,\nu_2 | \mathcal{S}_2)
+J_{\{i,j\}}^{X,Y}(\nu_1,\nu_2 | \mathcal{S}_3).$
Following the similar proof for (\ref{J_ij}),
we can show that  if $m_0\geq 4$,
  $J_{\{i,j\}}^{X,Y}(\nu_1,\nu_2 | \mathcal{S}_k )=O(N^{-3})$
hold uniformly over $(i,j)\in A_1$ for $k=1,2,3$, respectively.
To save space, we only prove the case for $k=1$.
Note that
\begin{align*}
&~J_{\{i,j\}}^{X,Y}(\nu_1, \nu_2| \mathcal{S}_1) \\
=&~\frac{1}{n_1^2 n_2^2}
\sum_{s_1=s_2=1}^{n_1}\sum_{t_1=t_2=1}^{n_2}
{\rm tr} \big\{ {E} \big[
E( \tilde{\bm{X}}_{i j;s_2} \tilde{\bm{X}}_{i j;s_1}^T )
(I_2-\widetilde{S}_{12, \{i,j\}}^{(s_1,t_1)})^{\nu_1}
E( \tilde{\bm{Y}}_{i j;t_1} \tilde{\bm{Y}}_{i j;t_2}^T )
(I_2-\widetilde{S}_{12, \{i,j\}}^{(s_2,t_2)})^{\nu_2}
\big]\big\}  \\
=&~\frac{1}{n_1^2 n_2^2}
\sum_{s_1=1}^{n_1}\sum_{t_1=1}^{n_2}
{\rm tr} \big\{ {E} \big[
(I_2-\widetilde{S}_{12, \{i,j\}}^{(s_1,t_1)})^{\nu_1+\nu_2}
\big]\big\}.
\end{align*}
By $(iii)$ of Lemma \ref{boundlemma5.2sam}, we have
$J_{\{i,j\}}^{X,Y}(\nu_1, \nu_2| \mathcal{S}_1)=  {O(N^{-(\nu_1+\nu_2)/2})}/{(n_1 n_2)} =
O(N^{-3})$  hold uniformly over $(i,j)\in A_1$.

\vskip 12pt
\noindent {\it  {\bf Part II-1.2: Proof of }
 $T_{XY,32}=o_p(p^{1/2}N^{-1}) $}
\vskip 12pt

Noting that $E(T_{XY,32})=0$, we only need to show $E(T_{XY,32}^2)=o( pN^{-2})$ as $N \to \infty$.
By condition (C2$'$) and following the similar procedure as in (\ref{alpha.ineq}), we have
\begin{align*}
E(T_{XY,32}^2)
& =\sum_{\substack{ (i_1,j_1) \in A_1 \\
                    (i_2,j_2) \in A_1 }}
\text{Cov}\Big(
\frac{1}{n_1 n_2}\sum_{ (i_1,j_1) \in A_1 }  \sum_{s=1}^{n_1} \sum_{t=1}^{n_2}
   \tilde{\bm{X}}_{i_1 j_1;s}^T \Xi_{12, \{i_1,j_1\}}^{(s,t)} \tilde{\bm{\mu}}_{2,i_1 j_1}, \\
&~~~~~~~~~~~~~~~~~~~~~~~~~~~~~~~~~~~~
   \frac{1}{n_1 n_2}\sum_{ (i_2,j_2) \in A_1 }  \sum_{s=1}^{n_1} \sum_{t=1}^{n_2}
   \tilde{\bm{X}}_{i_1 j_2;s}^T \Xi_{12, \{i_2,j_2\}}^{(s,t)} \tilde{\bm{\mu}}_{2,i_2 j_2}
   \Big)                                                                                     \\
&\leq \Big(2+\frac{\varpi_0}{1-\exp(-1)} \Big)K_0^2 p
\max_{(i,j)\in A_1} {\rm Var}\Big( \frac{1}{n_1 n_2} \sum_{s=1}^{n_1}\sum_{t=1}^{n_2}
\tilde{\bm{X}}_{i j;s}^T \Xi_{12, \{i,j\}}^{(s,t)} \tilde{\bm{\mu}}_{2,i j}\Big).
\end{align*}
Consequently, by $(ii)$ in Lemma \ref{Lemmaeigen},
\begin{align*}
{E(T_{XY,32}^2)}/(p N^{-2})
=
O(N^{2})
\max_{(i,j)\in A_1} {\rm Var}\Big( \frac{1}{n_1 n_2} \sum_{s=1}^{n_1}\sum_{t=1}^{n_2}
\tilde{\bm{X}}_{i j;s}^T \Xi_{12, \{i,j\}}^{(s,t)} \tilde{\bm{\mu}}_{2,i j}\Big).
\end{align*}
Following  the similar procedure as the proof for (\ref{U232_prof}), we can show that
\begin{eqnarray}\label{TXY32_prof}
\max_{(i,j)\in A_1} {\rm Var}\Big( \frac{1}{n_1 n_2} \sum_{s=1}^{n_1}\sum_{t=1}^{n_2}
\tilde{\bm{X}}_{i j;s}^T \Xi_{12, \{i,j\}}^{(s,t)} \tilde{\bm{\mu}}_{2,i j}\Big)
= O(N^{-2}) \max_{(i,j)\in A_1}\tilde{\bm{\mu}}_{2,i j}^T \tilde{\bm{\mu}}_{2, i j}.
\end{eqnarray}
Consequently,
${E(T_{XY,32}^2)}
= O(p N^{-2})\max_{(i,j)\in A_1}\tilde{\bm{\mu}}_{i j}^T \tilde{\bm{\mu}}_{i j}
=O(p N^{-2})\max_{(i,j)\in A_1}{\bm{\mu}}_{i j}^T \Sigma_{\{ij\}}^{-1}{\bm{\mu}}_{i j}$
as $N \to \infty$.
As shown in the proof of Lemma \ref{Lemmaeigen}, the eigenvalues of
$\Sigma_{\{ij\}}^{-1} \in R^{2 \times 2}$ are  bounded uniformly over $(i,j)\in A_1$.
Then by condition (C5$'$), we have
$\max_{(i,j)\in A_1}{\bm{\mu}}_{i j}^T \Sigma_{\{ij\}}^{-1}{\bm{\mu}}_{i j}=O(N^{-1/2}).$
This shows that  $E(T_{XY,32}^2)=o( pN^{-2})$ as $N \to \infty$.

\vskip 12pt
\noindent {\it  {\bf Part II-1.3: Proof of }
 $T_{XY,33}=o_p(p^{1/2}N^{-1}) $}
\vskip 12pt

The proof is nearly the same as that for $T_{XY,32}$ in Part II-1.2 and is hence omitted.

\vskip 12pt
\noindent {\it  {\bf Part II-1.4: Proof of }
 ${ T_{XY,34}=o_p(p^{1/2}N^{-1})}$}
\vskip 12pt

Note that
\begin{align*}
E\big( \big| \tilde{\bm{\mu}}_{1,ij}^T
\big((\widetilde{S}_{12, \{i,j\}}^{(s,t)})^{-1} -I_2 \big)
\tilde{\bm{\mu}}_{2,ij} \big|\big)
&\leq
E\big(\big\| (\widetilde{S}_{12, \{i,j\}}^{(s,t)})^{-1} -I_2   \big\| \big)  \,
\times  \big\| \tilde{\bm{\mu}}_{1,ij} \big\| \,
\times  \big\| \tilde{\bm{\mu}}_{2, ij}\big\|   \\
&=
O(N^{-1/2}) \big\| \tilde{\bm{\mu}}_{1,ij} \big\|
\times  \big\| \tilde{\bm{\mu}}_{2, ij}\big\|,
\end{align*}
and
$$ \sum_{ (i,j) \in A_1 }
\| \tilde{\bm{\mu}}_{1,ij} \|   \| \tilde{\bm{\mu}}_{2, ij} \|
\leq
( \sum_{ (i,j) \in A_1 }  {\bm{\mu}_{1, ij}}^T \Sigma_{ij}^{-1}{\bm{\mu}_{1, ij}} )^{1/2}
( \sum_{ (i,j) \in A_1 }  {\bm{\mu}_{2, ij}}^T \Sigma_{ij}^{-1}{\bm{\mu}_{2, ij}} )^{1/2}.$$
We have
\begin{align*}
E( |T_{XY,34}| )
&=
O(N^{-1/2})\Big( \sum_{ (i,j) \in A_1 }  {\bm{\mu}_{1, ij}}^T \Sigma_{ij}^{-1}{\bm{\mu}_{1, ij}} \Big)^{1/2}
\Big( \sum_{ (i,j) \in A_1 }  {\bm{\mu}_{2, ij}}^T \Sigma_{ij}^{-1}{\bm{\mu}_{2, ij}} \Big)^{1/2} \\
&= O(N^{-1/2}) \big( {\bm{\mu}_1}^T P_{\mathcal{O}} {\bm{\mu}_1} \big)^{1/2}
 \big( {\bm{\mu}_2}^T P_{\mathcal{O}} {\bm{\mu}_2} \big)^{1/2}.
\end{align*}
By  condition (C5$'$), and noting that
${\bm{\mu}_2}^T P_{\mathcal{O}} {\bm{\mu}_2}=o(p^{1/2}N^{-1/2})$, we have
$E(|T_{XY,34}|)=o(p^{1/2}N^{-1})$ as $N \to \infty$, and hence $T_{XY,34}=o_p(p^{1/2}N^{-1})$ as $N \to \infty$.

\vskip 12pt
\noindent {\it  {\bf Part II-2: Proof of }
 ${ T_{XY,4}=o_p(p^{1/2}N^{-1})}$}
\vskip 12pt

Let $\widetilde{X}_{sj}=({X}_{sj}-\mu_{1j})/\sigma_{jj}$,
$\widetilde{Y}_{sj}=({Y}_{sj}-\mu_{2j} )/\sigma_{jj}$,
$\widetilde{\mu}_{1j}=\mu_{1j}/\sigma_{jj}$,
$\widetilde{\mu}_{2j}=\mu_{2j}/\sigma_{jj}$,
$\widetilde{s}_{12; jj}^{(s,t)}={s}_{12;jj}^{(s,t)}/\sigma_{jj}$
and
$\Xi_{12;jj}^{(s,t)}=(1-\widetilde{s}_{12;jj}^{(s,t)})+(1-\widetilde{s}_{12;jj}^{(s,t)})^2
+\cdots+ (1-\widetilde{s}_{12;jj}^{(s,t)})^{m_0}$,
where $s_{12; jj}^{(s,t)}$ is the $j$th diagonal component of $S_{12,*}^{(s,t)}$.
Since
$|((\widetilde{s}_{12;jj}^{(s,t)})^{-1}-1) - \Xi_{12;jj}^{(s,t)} |
=O_p(N^{-(m_0+1)/2}),$
 by Taylor expansion we have
\begin{align*}
T_{XY,4}
&=
T_{XY,41} + T_{XY,42} + T_{XY,43} +  T_{XY,44}
+ \text{card}(A_2)O_p(N^{-{(m_0+1)}/{2}}),
\end{align*}
where
\begin{align*}
T_{XY,41}
&=\frac{1}{n_1 n_2}\sum_{s=1}^{n_1} \sum_{t=1}^{n_2}
\sum_{j \in {A}_2 }
\widetilde{X}_{sj} \widetilde{X}_{tj}\Xi_{12;jj}^{(s,t)},   \\
T_{XY,42}
&=\frac{1}{n_1 n_2} \sum_{s=1}^{n_1} \sum_{t=1}^{n_2}
\sum_{j \in {A}_2 }
\widetilde{\mu}_{2j} \widetilde{X}_{sj}\Xi_{12;jj}^{(s,t)},\\
T_{XY,43}
&=\frac{1}{n_1 n_2}\sum_{s=1}^{n_1} \sum_{t=1}^{n_2}
\sum_{j \in {A}_2 }
\widetilde{\mu}_{1j} \widetilde{Y}_{tj}\Xi_{12; jj}^{(s,t)},   \\
T_{XY,44}
&=\frac{1}{n_1 n_2}\sum_{s=1}^{n_1} \sum_{t=1}^{n_2}
\sum_{j \in {A}_2 }
\widetilde{\mu}_{1j}\widetilde{\mu}_{2j}((\widetilde{s}_{12;jj}^{(s,t)})^{-1}-1).
\end{align*}
Note that
$\text{card}(A_2)O_p\big(N^{-(m_0+1)/2}\big)=O_p(pN^{-(m_0+1)/2})$ as $N \to \infty$.
Thus for $m_0 \geq 4$, $\text{card}(A_2)O_p\big(N^{-{(m_0+1)}/{2}}\big)=o_p(pN^{-2})$.
In what follows,
we show that
$T_{XY,41}=o_p( p^{1/2}N^{-1} )$,
$T_{XY,42}=o_p(p^{1/2}N^{-1})$,
$T_{XY,43}=o_p(p^{1/2}N^{-1})$,
and $T_{XY,44}=o_p(p^{1/2}N^{-1})$ as $N \to \infty$, respectively.

\vskip 12pt
\noindent {\it  {\bf Part II-2.1: Proof of }
 $T_{XY,41}=o_p(p^{1/2}N^{-1})$}
\vskip 12pt

Note that $E(T_{XY,41})=0$.  By the $\rho$-mixing inequality and
 condition (C2$'$),
\begin{align*}
E(T_{XY,41}^2)
&=
\sum_{j_1 \in {A}_2 } \sum_{j_2 \in {A}_2 }
\text{Cov}\Big(
\frac{1}{n_1 n_2}\sum_{s=1}^{n_1} \sum_{t=1}^{n_2}
\widetilde{X}_{s j_1} \widetilde{Y}_{t j_1 }\Xi_{12; j_1 j_1}^{(s,t)},
\frac{1}{n_1 n_2}\sum_{s=1}^{n_1} \sum_{t=1}^{n_2}
\widetilde{X}_{s j_2} \widetilde{Y}_{t j_2 }\Xi_{12; j_2 j_2}^{(s,t)}
\Big)   \\
&\leq
\varpi_0 \sum_{j_1 \in {A}_2 } \sum_{j_2 \in {A}_2 }
\exp(-|j_1-j_2|)
\max_{j\in A_2} \text{Var}\Big(
\frac{1}{n_1 n_2}\sum_{s=1}^{n_1} \sum_{t=1}^{n_2}
\widetilde{X}_{s j} \widetilde{Y}_{t j }\Xi_{12; j j}^{(s,t)} \Big)   \\
&\leq
\frac{\varpi_0 p}{1-\exp(-1)}
\max_{j\in A_2} \text{Var}\Big(
\frac{1}{n_1 n_2}\sum_{s=1}^{n_1} \sum_{t=1}^{n_2}
\widetilde{X}_{s j} \widetilde{Y}_{t j }\Xi_{12; j j}^{(s,t)} \Big).
\end{align*}
Following the similar proof for (\ref{asyT_31}),
we can show that
$$\text{Var}\Big(
\frac{1}{n_1 n_2}\sum_{s=1}^{n_1} \sum_{t=1}^{n_2}
\widetilde{X}_{s j} \widetilde{Y}_{t j }\Xi_{12; j j}^{(s,t)} \Big)=O(N^{-3})$$
hold uniformly over $j \in A_2$.
This indicates that $E(T_{XY,41}^2)=O(pN^{-3})$ as $N \to \infty$,
and hence $T_{XY,41}=o_p(p^{1/2}N^{-1})$ as $N \to \infty$.

\vskip 12pt
\noindent {\it  {\bf Part II-2.2: Proof of }
 $T_{XY,42}=o_p(p^{1/2}N^{-1})$ }
\vskip 12pt


Note that
$ E(\sum_{s=1}^{n_1} \sum_{t=1}^{n_2}
\tilde{X}_{sj} \Xi_{12; jj}^{(s,t)} \tilde{\mu}_{2j} /(n_1 n_2) )=0.$ Then, $E(T_{XY,42})=0$.
By the $\rho$-mixing inequality and condition (C2$'$),
\begin{align*}
E(T_{XY,42}^2)
&=
\sum_{\substack{ j_1 \in A_2 \\ j_2 \in A_2 }}
\text{Cov}\Big(
\frac{1}{n_1 n_2}\sum_{s=1}^{n_1} \sum_{t=1}^{n_2}
\tilde{X}_{s j_1} \Xi_{12; j_1 j_1}^{(s,t)} \tilde{\mu}_{2 j_1},\,
\frac{1}{n_1 n_2}\sum_{s=1}^{n_1} \sum_{t=1}^{n_2}
\tilde{X}_{s j_2} \Xi_{12; j_2 j_2}^{(s,t)} \tilde{\mu}_{2 j_2} \Big)  \\
&\leq
\frac{\varpi_0 p}{1-\exp(-1)}
\max_{j \in A_2} {\rm Var}\Big(
\frac{1}{n_1 n_2}\sum_{s=1}^{n_1} \sum_{t=1}^{n_2}
\tilde{X}_{s j} \Xi_{12; jj}^{(s,t)} \tilde{\mu}_{2j}
\Big).
\end{align*}
Further by $(ii)$ in Lemma \ref{Lemmaeigen}, we have
\begin{align*}
\frac{E(T_{XY,42}^2)}{p N^{-2}}
&=
O(N^2)\max_{j \in A_2}
{\rm Var}\Big( \frac{1}{n_1 n_2}\sum_{s=1}^{n_1} \sum_{t=1}^{n_2}
\tilde{X}_{s j} \Xi_{12; jj}^{(s,t)} \tilde{\mu}_{2j}\Big).
\end{align*}
Following the similar proof for (\ref{TXY32_prof}), we can show that
\begin{eqnarray}\label{TXY42_prof}
\max_{j \in A_2} {\rm Var}\Big(
\frac{1}{n_1 n_2}\sum_{s=1}^{n_1} \sum_{t=1}^{n_2}
\tilde{X}_{sj} \Xi_{12; jj}^{(s,t)} \tilde{\mu}_{j}
\Big)
= O(N^{-2}) \max_{j \in A_2 }\tilde{\mu}_{2j}^2.
\end{eqnarray}
Consequently,
${E(T_{XY,42}^2)}
=
O(p N^{-2}) \max_{j\in A_2}\tilde{\mu}_{2j}^2
= O(p N^{-2}) \max_{j\in A_2} {  {\mu}_{2j}^2 }/{ \sigma_{jj} }$
as $N \to \infty$.
Also in the proof of Lemma \ref{Lemmaeigen}, we have shown that
$\sigma_{jj}$ are  bounded uniformly for $j=1,\ldots,p$. Then
by condition (C5$'$), we have
$\max_{j\in A_2}{\mu}_{2j}^2/\sigma_{jj}=O(N^{-1/2})$.
This indicates that $E(T_{XY,42}^2)=o(pN^{-2})$  as $N \to \infty$, and hence
$T_{XY,42}=o_p(p^{1/2}N^{-1})$ as $N \to \infty$.

\vskip 12pt
\noindent {\it  {\bf Part II-2.3: Proof of }
 ${ T_{XY,43}=o_p(p^{1/2}N^{-1})}$}
\vskip 12pt

The proof is similar as that for $T_{XY,42}$ in Part II-2.2 and is hence omitted.

\vskip12pt
\noindent {\bf Part II-2.4: Proof of
$T_{XY,44}=o_p(p^{1/2}N^{-1})$}
\vskip12pt

Note that $E\big|(\widetilde{s}_{12; jj}^{(s,t)})^{-1}-1\big|=O(N^{-1/2})$
for $j=1,\ldots,p$. We have
\begin{align*}
E\big(|T_{XY,44}|\big)
&\leq
\sum_{j \in {A}_2 } | \widetilde{\mu}_{1j} | | \widetilde{\mu}_{2j} |
E\Big(\Big| \frac{1}{n_1 n_2}\sum_{s=1}^{n_1} \sum_{t=1}^{n_2}
\big((\widetilde{s}_{12; jj}^{(s,t)})^{-1}-1\big)\Big| \Big)   \\
&=
O(N^{-1/2})\Big(\sum_{j \in {A}_2 } \frac{ {\mu}_{1j}^2 }{ \sigma_{jj} }
+\sum_{ (i,j) \in A_1 } {\bm{\mu}_{1;ij}}^T \Sigma_{ij}^{-1}{\bm{\mu}_{1; ij}} \Big)
   \\
&~~~+O(N^{-1/2})\Big(\sum_{j \in {A}_2 }
 \frac{ {\mu}_{2j}^2 }{\sigma_{jj} }
+\sum_{ (i,j) \in A_1 } {\bm{\mu}_{2;ij}}^T \Sigma_{ij}^{-1}{\bm{\mu}_{2; ij}} \Big)
\\
&=
O(N^{-1/2}) ( {\bm{\mu}_1}^T P_{\mathcal{O}}{\bm{\mu}_1}+{\bm{\mu}_2}^T P_{\mathcal{O}}{\bm{\mu}_2} ).
\end{align*}
Note also that $\bm{\mu}_1^T  P_{\mathcal{O}} \bm{\mu}_1 =o(p^{1/2}N^{-1/2})$
and $(\bm{\mu}_1- \bm{\mu}_2)^T  P_{\mathcal{O}} (\bm{\mu}_1- \bm{\mu}_2) =o(p^{1/2}N^{-1/2})$
by condition (C5$'$). We have  $\bm{\mu}_2^T  P_{\mathcal{O}} \bm{\mu}_2 =o(p^{1/2}N^{-1/2})$,
and consequently,
$$\frac{E(|T_{XY,44}|)}{p^{1/2}N^{-1}}
 =O(N^{1/2}p^{-1/2})
({{\bm{\mu}_1}^T P_{\mathcal{O}} {\bm{\mu}_1}
+{\bm{\mu}_2}^T P_{\mathcal{O}} {\bm{\mu}_2}})=o(1)
{\rm ~~as~~} N\to\infty.$$
This indicates that
$T_{XY,44}=o_p(p^{1/2}N^{-1})$ as $N \to \infty$.

\hfill$\Box$

%

\subsection{Proof of Lemma \ref{lemma2_1}}\label{Prof.lemma2_1}

Let
\begin{align*}
L_{x1}
&=
\frac{1}{n_1 (n_1-1)} \sum_{s=1}^{n_1} \sum_{t\neq s}^{n_1}
( {\bm X_{s}} -{\bar{\bm X}^{(s,t)}})^T \widehat{P}_{1,\mathcal{O}}^{(s,t)} {\bm X}_{t}
( {\bm X_{t}} -{\bar{\bm X}^{(s,t)}})^T \widehat{P}_{1,\mathcal{O}}^{(s,t)} {\bm X}_{s},  \\
L_{x2}
&=
\frac{1}{n_1 (n_1-1)} \sum_{s=1}^{n_1} \sum_{t\neq s}^{n_1}
( {\bm X_{s}} -{\bar{\bm X}^{(s,t)}})^T {P}_{1,\mathcal{O}}^{(s,t)} {\bm X}_{t}
( {\bm X_{t}} -{\bar{\bm X}^{(s,t)}})^T {P}_{1,\mathcal{O}}^{(s,t)} {\bm X}_{s},
\end{align*}
where
$${P}_{1,\mathcal{O}}^{(s,t)}=\sum_{(i,j) \in {A}_1}
P_{ij}^T (P_{ij} S_{1*}^{(s,t)} P_{ij}^T)^{-1}P_{ij}
+\sum_{i \in {A}_2}P_{i}^T ( P_{i} S_{1*}^{(s,t)} P_{i}^T)^{-1} P_{i}.$$
Note that $(\{\hat{A}_1=A_1\}\cap \{\hat{A}_2=A_2\} )
\subseteq \{ {P}_{1,\mathcal{O}}^{(s,t)}=\widehat{P}_{1,\mathcal{O}}^{(s,t)} \}
\subseteq \{ L_{x1}=L_{x2}  \}$. Then
for any $\epsilon_1>0$,  we have
$$P(|L_{x1}-L_{x2}| >\epsilon_1 {{\rm tr}(\Lambda_1^2)} )
\leq P( \{\hat{A}_1\neq  A_1\} )+P( \{\hat{A}_2\neq A_2\} ) \to 0 {\rm ~~as~~} N \to \infty.$$
This shows that
${L_{x1}}/{{\rm tr}(\Lambda_1^2)}
- {L_{x2}}/{{\rm tr}(\Lambda_1^2)}\stackrel{P}{\longrightarrow}0$
as $N \to \infty$.
Hence to prove Lemma~\ref{lemma2_1}, it is equivalent  to verifying that
${L_{x2}}/{{\rm tr}(\Lambda_1^2)}\stackrel{P}{\longrightarrow} 1$
as $N \to \infty$.


Note that $L_{x2}= B_{x1}+B_{x2}+B_{x3}$, where
\begin{align*}
B_{x1}
&= \frac{1}{n_1(n_1-1)} \sum_{s=1}^{n_1}\sum_{t \neq s}^{n_1}
({\bm X_{s}} -{\bar{\bm X}^{(s,t)}})^T  P_{\mathcal{O}}  {\bm X}_{t}  \,
({\bm X_{t}} - {\bar{\bm X}^{(s,t)}})^T  P_{\mathcal{O}}  {\bm X}_{s}  \\
&= \frac{1}{n_1(n_1-1)} \sum_{s=1}^{n_1}\sum_{t \neq s}^{n_1}
(\breve{\bm X}_{s} -\breve{\bar{{\bm X}}}^{(s,t)})^T  \breve{\bm X}_{t}
(\breve{\bm X}_{t} -\breve{\bar{{\bm X}}}^{(s,t)})^T  \breve{\bm X}_{s}, \\
B_{x2}
&= \frac{2}{n_1(n_1-1)} \sum_{s=1}^{n_1}\sum_{t \neq s}^{n_1}
({\bm X_{s}} -{\bar{\bm X}^{(s,t)}})^T P_{\mathcal{O}}{\bm X}_{t}  \,
({\bm X_{t}}-{\bar{\bm X}^{(s,t)}})^T({P}_{1,\mathcal{O}}^{(s,t)}-P_{\mathcal{O}}) {\bm X}_{s},  \\
B_{x3}
&= \frac{1}{n_1(n_1-1)} \sum_{s=1}^{n_1}\sum_{t \neq s}^{n_1}
({\bm X_{s}} -{\bar{\bm X}^{(s,t)}})^T({P}_{1,\mathcal{O}}^{(s,t)} - P_{\mathcal{O}}){\bm X}_{t} \,
({\bm X_{t}} - {\bar{\bm X}^{(s,t)}})^T({P}_{1,\mathcal{O}}^{(s,t)} - P_{\mathcal{O}})  {\bm X}_{s},
\end{align*}
where $\breve{\bm X}_{s}= P_{\mathcal{O}}^{1/2}{\bm X}_{s}$ and
$\breve{\bar{{\bm X}}}^{(s,t)}=P_{\mathcal{O}}^{1/2}{\bar{{\bm X}}}^{(s,t)}$.
Following the similar proof as that for (\ref{asy.B1}), we can show that
${B_{x1}} / {{\rm tr}(\Lambda_1^2)} \stackrel{P}{\longrightarrow} 1$ as $N \to \infty$.
In what follows, we show that ${B_{x2}}/{{\rm tr}(\Lambda_1^2)}=o_p(1)$ and
${B_{x3}}/{{\rm tr}(\Lambda_1^2)}=o_p(1)$ as $N \to \infty$.  By (\ref{ord_trace}), it is equivalent  to showing that
 $B_{x2}=o_p(p)$ and  $B_{x3}=o_p(p)$.

\vskip12pt
\noindent{\bf Part-I:  Proof of $B_{x2}=o_p(p)$ }
\vskip12pt

For $s \neq t$, let
\begin{align*}
B_{x21}^{(s,t)} &=
B_{x211}^{(s,t)} +B_{x212}^{(s,t)}, \\
B_{x22}^{(s,t)} &=
\sum_{ j \in A_2 }
(\tilde{X}_{sj}-  \tilde{\bar{X}}_{j}^{(s,t)})
(1/\tilde{s}_{1,jj}^{(s,t)}-1)
(\tilde{X}_{tj}+\tilde{\mu}_{1,j}),
\end{align*}
where
$B_{x211}^{(s,t)}
=\sum_{ (i,j) \in A_1 }
(\tilde{\bm{X}}_{ij;s}-   \tilde{\bar{\bm{X}}}_{ij}^{(s,t)}  )^T
(  ( \widetilde{S}_{1,\{ij\}}^{(s,t)} )^{-1}-I_2 )
\tilde{\bm{X}}_{ij;t}, $
$B_{x212}^{(s,t)}
=\sum_{ (i,j) \in A_1 }
(\tilde{\bm{X}}_{ij;s}-   \tilde{\bar{\bm{X}}}_{ij}^{(s,t)}  )^T
(  ( \widetilde{S}_{1,\{ij\}}^{(s,t)} )^{-1}-I_2 )
\tilde{\bm{\mu}}_{1, ij},$
$\tilde{\bar{\bm{X}}}_{ij}^{(s,t)}=\sum_{k \neq s,t}^{n_1} \tilde{\bm{X}}_{ij;k}/(n_1-2)$,
$\tilde{\bar{{X}}}_{j}^{(s,t)}=\sum_{k \neq s,t}^{n_1} \tilde{{X}}_{kj}/(n_1-2)$,
and $\tilde{s}_{jj}^{(s,t)}$ is the sample variance of $\{ \tilde{{X}}_{kj}\}_{k\neq s,t}$.
Then,
\begin{align*}
E(|B_{x2}|)
&\leq
\frac{2}{n_1(n_1-1)} \sum_{s=1}^{n_1}\sum_{t \neq s}^{n_1}
E\Big(
\big|({\bm X_{s}} -{\bar{\bm X}^{(s,t)}})^T P_{\mathcal{O}}{\bm X}_{t} \big| \,
\big| ({\bm X_{t}}-{\bar{\bm X}^{(s,t)}})^T
\big({P}_{1,\mathcal{O}}^{(s,t)}-P_{\mathcal{O}})
{\bm X}_{s} \big| \Big) \\
&=
2 E\Big(\big|
({\bm X_{1}} -{\bar{\bm X}^{(1,2)}})^T P_{\mathcal{O}}{\bm X}_{2}
\big| \, \big|  ({\bm X_{2}}-{\bar{\bm X}^{(1,2)}})^T
({P}_{1,\mathcal{O}}^{(1,2)}-P_{\mathcal{O}})
{\bm X}_{1}  \big| \Big).
\end{align*}
Note  that ${\bar{\bm X}^{(s,t)}}={\bar{\bm X}^{(t,s)}}$
and ${P}_{1,\mathcal{O}}^{(s,t)}={P}_{1,\mathcal{O}}^{(t,s)}$.
This leads to
\begin{align}\label{B.X2.expand}
({\bm X_{2}}-{\bar{\bm X}^{(1,2)}})^T
({P}_{1,\mathcal{O}}^{(1,2)}-P_{\mathcal{O}})
{\bm X}_{1}
=&~
\sum_{(i,j)\in A_1}
({\bm X_{ij;2}}-{\bar{\bm X}_{ij}^{(1,2)}})^T
( ( S_{1,\{ij\}}^{(2,1)} )^{-1} - \Sigma_{\{ij\}}^{-1}  )
{\bm X}_{ij;1}   \nonumber\\
&~+\sum_{j\in A_2}
( X_{2j}-{\bar{X}}_{j}^{(2,1)} )^T
( 1/ s_{1,jj}^{(2,1)} - 1/\sigma_{jj} )
{X}_{1j}    \nonumber\\
=& ~ B_{x21}^{(2,1)}+B_{x22}^{(2,1)},
\end{align}
where the last equality is obtained by a direct calculation as (\ref{B2.expand}).

Note also that $E( ({\bm X_{s}} -{\bar{\bm X}^{(s,t)}})^T P_{\mathcal{O}}{\bm X}_{t}) =0$
and
$E( ({\bm X_{s}} -{\bar{\bm X}^{(s,t)}})^T P_{\mathcal{O}}{\bm X}_{t})^2
= {\rm tr}(\Lambda_1^2){(n_1-2)}/(n_1-1)=O(p)$ as $N \to \infty$.
Then,
\begin{align*}
E(|B_{x2}|)
&\leq
2 E\Big(\big|
({\bm X_{1}} -{\bar{\bm X}^{(1,2)}})^T P_{\mathcal{O}}{\bm X}_{2}
\big| \,
\big( \big| B_{x21}^{(2,1)}\big| + \big| B_{x22}^{(2,1)} \big| \big) \Big) \\
&\leq
2 \big[ E\big( ({\bm X_{1}} -{\bar{\bm X}^{(1,2)}})^T P_{\mathcal{O}}{\bm X}_{2}
\big)^2 \big]^{\frac{1}{2}}\,
\left\{ \big[ E \big( B_{x21}^{(2,1)} \big)^2 \big]^{\frac{1}{2}}
+ \big[ E \big( B_{x22}^{(2,1)} \big)^2 \big]^{\frac{1}{2}} \right\}  \\
&=
O(p^{1/2}) \left\{ \big[ E \big( B_{x21}^{(1,2)} \big)^2 \big]^{\frac{1}{2}}
+ \big[ E \big( B_{x22}^{(1,2)} \big)^2 \big]^{\frac{1}{2}} \right\},
\end{align*}
where the last equality is based on the fact that
$E \big( B_{x21}^{(1,2)} \big)^2= E \big( B_{x21}^{(2,1)} \big)^2$ and $E \big( B_{x22}^{(1,2)} \big)^2= E \big( B_{x22}^{(2,1)} \big)^2$.

Next, we  show that $ E \big( B_{x21}^{(s,t)} \big)^2=o(p)$ as $N \to \infty$ for $s\neq t$.
Note that $E(B_{x21}^{(s,t)})^2= E(B_{x211}^{(s,t)})^2+ E(B_{x212}^{(s,t)})^2$.
In what follows, we show that
$E(B_{x211}^{(s,t)})^2=o(p)$ and  $E(B_{x212}^{(s,t)})^2=o(p)$ as $N \to \infty$, respectively.

Noting that $E[
(\tilde{\bm{X}}_{ij;s}-   \tilde{\bar{\bm{X}}}_{ij}^{(s,t)}  )^T
(  ( \widetilde{S}_{1, \{ij\}}^{(s,t)} )^{-1}-I_2 )
\tilde{\bm{X}}_{ij;t}]=0$, we have
\begin{align*}
E(B_{x211}^{(s,t)})^2
&=
\sum_{\substack{ (i_1,j_1) \in A_1 \\
                    (i_2,j_2) \in A_1 }}
\text{Cov}\Big(
\big(\tilde{\bm{X}}_{i_1 j_1;s}-  \tilde{\bar{\bm{X}}}_{i_1 j_1}^{(s,t)} \big)^T
\big(  ( \widetilde{S}_{1, \{i_1 j_1\}}^{(s,t)} )^{-1}-I_2 \big)
\tilde{\bm{X}}_{i_1 j_1;t},   \nonumber \\
&~~~~~~~~~~ ~~~~~~~~~~~~~~~~~~~~
(\tilde{\bm{X}}_{i_2 j_2;s}-  \tilde{\bar{\bm{X}}}_{i_2 j_2}^{(s,t)}  )^T
\big(  ( \widetilde{S}_{1, \{i_2 j_2\}}^{(s,t)} )^{-1}-I_2 \big)
\tilde{\bm{X}}_{i_2 j_2;t} \Big)   \\
&\leq
\Big(2+\frac{\varpi_0}{1-\exp(-1)} \Big)K_0^2p
\max_{(i,j)\in A_1} {\rm Var}\Big(
\big(\tilde{\bm{X}}_{i j;s}-  \tilde{\bar{\bm{X}}}_{i j}^{(s,t)}  \big)^T
\big(  ( \widetilde{S}_{1, \{ij\}}^{(s,t)} )^{-1}-I_2  \big)
\tilde{\bm{X}}_{i j;t} \Big),
\end{align*}
where the last inequality is
based on the $\rho$-mixing inequality, and the upper bound can be obtained by following
the same procedure as  (\ref{alpha.ineq}).
Note also that
\begin{align*}
&~~~~{\rm Var}\Big(
\big(\tilde{\bm{X}}_{i j;s}-  \tilde{\bar{\bm{X}}}_{i j}^{(s,t)}  \big)^T
\big(  ( \widetilde{S}_{1, \{ij\}}^{(s,t)} )^{-1}-I_2  \big)
\tilde{\bm{X}}_{i j;t} \Big)  \\
&=
E\Big(\big(\tilde{\bm{X}}_{i j;s}-  \tilde{\bar{\bm{X}}}_{i j}^{(s,t)}  \big)^T
\big(  (\widetilde{S}_{1, \{ij\}}^{(s,t)} )^{-1}-I_2  \big)
E(\tilde{\bm{X}}_{i j;t} \tilde{\bm{X}}_{i j;t}^T)
\big(  (\widetilde{S}_{1, \{ij\}}^{(s,t)} )^{-1}-I_2  \big)
\big(\tilde{\bm{X}}_{i j;s}-  \tilde{\bar{\bm{X}}}_{i j}^{(s,t)}  \big) \Big)   \\
&=
E\Big(\big(\tilde{\bm{X}}_{i j;s}-  \tilde{\bar{\bm{X}}}_{i j}^{(s,t)}  \big)^T
\big(  (\widetilde{S}_{1, \{ij\}}^{(s,t)} )^{-1}-I_2  \big)^2
\big(\tilde{\bm{X}}_{i j;s}-  \tilde{\bar{\bm{X}}}_{i j}^{(s,t)}  \big) \Big)   \\
&\leq
E\Big(
\Big\| \big(  ( \widetilde{S}_{1, \{ij\}}^{(s,t)} )^{-1}-I_2  \big)
\big(  \tilde{\bm{X}}_{i j;s}- \tilde{\bar{\bm{X}}}_{i j}^{(s,t)}  \big)
\Big\|^2 \Big)    \\
&\leq
\Big( E \Big\|  ( \widetilde{S}_{1, \{ij\}}^{(s,t)} )^{-1}-I_2 \Big\|^4 \Big)^{\frac{1}{2}}
\Big( E \Big\| \tilde{\bm{X}}_{i j;s}- \tilde{\bar{\bm{X}}}_{i j}^{(s,t)}\Big\|^4 \Big)^{\frac{1}{2}},
\end{align*}
where the second equality is based on the fact that $E(\tilde{\bm{X}}_{i j;t} \tilde{\bm{X}}_{i j;t}^T)=I_2$.
By ${(iii)}$ in Lemma \ref{boundlemma5.2sam},
$E \big\|  ( \widetilde{S}_{1, \{ij\}}^{(s,t)} )^{-1}-I_2 \big\|^4=O(n_1^{-2})$
hold uniformly  over $(i,j)\in A_1$.
In addition,    $E\big\| \tilde{\bm{X}}_{i j;s}- \tilde{\bar{\bm{X}}}_{i j}^{(s,t)}
\big\|^4$  for $(i,j)\in A_1$ are finite combinations of higher order moments with the highest terms
 $E(\tilde{X}_{ki}^{4})$ and $E(\tilde{X}_{kj}^{4})$ for $k=1,2,\ldots,n_1$,
and hence are bounded uniformly.
Consequently, we have
$${\rm Var}\Big(
\big(\tilde{\bm{X}}_{i j;s}-  \tilde{\bar{\bm{X}}}_{i j}^{(s,t)}  \big)^T
\big(  ( \widetilde{S}_{1, \{ij\}}^{(s,t)} )^{-1}-I_2  \big)
\tilde{\bm{X}}_{i j;t} \Big)=O(n_1^{-1})$$
 hold uniformly over $(i,j)\in A_1$.
This shows that $E(B_{x211}^{(s,t)})^2=o(p)$.

In addition,
\begin{align*}
E(B_{x212}^{(s,t)})^2
&=
E \Big(\sum_{ (i,j) \in A_1 }
(\tilde{\bm{X}}_{ij;s}-   \tilde{\bar{\bm{X}}}_{ij}^{(s,t)}  )^T
\big(  (\widetilde{S}_{1, \{ij\}}^{(s,t)} )^{-1}-I_2 \big)
\tilde{\bm{\mu}}_{1,ij} \Big)^2   \\
&\leq
\sum_{ (i_1,j_1) \in A_1 } \sum_{ (i_2,j_2) \in A_1 }
\big\| \tilde{\bm{\mu}}_{1, i_1 j_1}^T \big\|
E   \Big(
\| ( \widetilde{S}_{1, \{i_1 j_1 \}}^{(s,t)} )^{-1}-I_2  \|
\times  \| \tilde{\bm{X}}_{i_1 j_1;s}- \tilde{\bar{\bm{X}}}_{i_1 j_1}^{(s,t)} \|   \\
&~~~~\times \| \tilde{\bm{X}}_{i_2 j_2;s}- \tilde{\bar{\bm{X}}}_{i_2 j_2}^{(s,t)} \|
\times \|  ( \widetilde{S}_{1, \{i_2 j_2 \}}^{(s,t)} )^{-1}-I_2 \|
\Big)  \big\| \tilde{\bm{\mu}}_{1, i_2 j_2}\big\|.
\end{align*}
Also by ${(iii)}$ in Lemma \ref{boundlemma5.2sam},
{\small
\begin{align*}
&E  \Big(
\big\| ( \widetilde{S}_{1, \{i_1 j_1 \}}^{(s,t)} )^{-1}-I_2 \big\|
\times  \big\| \tilde{\bm{X}}_{i_1 j_1;s}- \tilde{\bar{\bm{X}}}_{i_1 j_1}^{(s,t)}\big\|
\times \big\| \tilde{\bm{X}}_{i_2 j_2;s}- \tilde{\bar{\bm{X}}}_{i_2 j_2}^{(s,t)}  \big\|
\times \big\|  ( \widetilde{S}_{1, \{i_2 j_2 \}}^{(s,t)} )^{-1}-I_2 \big\|   \Big)  \\
\leq &
\Big[ E\Big(
\big\| ( \widetilde{S}_{1, \{i_1 j_1 \}}^{(s,t)} )^{-1}-I_2  \big\|^4 \Big)  \Big]^{\frac{1}{4}}
\Big[ E\Big(  \big\| \tilde{\bm{X}}_{i_1 j_1;s}- \tilde{\bar{\bm{X}}}_{i_1 j_1}^{(s,t)} \big\|^4
\Big) \Big]^{\frac{1}{4}}
\Big[ E\Big(
\big\| ( \widetilde{S}_{1, \{i_2 j_2 \}}^{(s,t)})^{-1}-I_2  \big\|^4 \Big)  \Big]^{\frac{1}{4}}  \\
&\times
\Big[ E\Big(  \big\| \tilde{\bm{X}}_{i_2 j_2;s}- \tilde{\bar{\bm{X}}}_{i_2 j_2}^{(s,t)} \big\|^4
\Big) \Big]^{\frac{1}{4}} \\
=&~O(n_1^{-1})
\end{align*}}{hold} uniformly  for any  $(i_1,j_1)$ and $(i_2,j_2)\in A_1$.
Thus, there exists  a constant $K_{02}>0$ such that
\begin{align*}
E(B_{x211}^{(s,t)})^2
&\leq
\Big( \sum_{ (i_1,j_1) \in A_1 }\big\| \tilde{\bm{\mu}}_{1, i_1 j_1} \big\|  \Big)
\Big( \sum_{ (i_2,j_2) \in A_1 } \big\| \tilde{\bm{\mu}}_{1, i_2 j_2} \big\|   \Big) \frac{K_{02}} {n_1}  \\
&\leq
\sqrt{\text{card}(A_1)}
\Big( \sum_{ (i_1,j_1) \in A_1 } \big\| \tilde{\bm{\mu}}_{1, i_1 j_1}\big\|^2  \Big)^{1/2}
\sqrt{\text{card}(A_1)}
\Big( \sum_{ (i_2,j_2) \in A_1 } \big\| \tilde{\bm{\mu}}_{1, i_2 j_2} \big\|^2   \Big)^{1/2} \frac{K_{02}} {n_1}  \\
&=
O({p n_1^{-1}}) \Big(
\sum_{ (i_1,j_1) \in A_1 } \big\| \tilde{\bm{\mu}}_{1, i_1 j_1}\big\|^2  \Big) \\
&=
O({p n_1^{-1}})  {\bm \mu}_1^T P_{\mathcal{O}}{\bm \mu}_1,
\end{align*}
where the second inequality is based on the Cauchy-Schwarz inequality,
and the last equality is based on the fact that
$\sum_{ (i,j) \in A_1 } \big\| \tilde{\bm{\mu}}_{1, ij}\big\|^2
=\sum_{ (i,j) \in A_1} {\bm{\mu}}_{1, i j}^T \Sigma_{1, \{ij\}}^{-1} {\bm{\mu}}_{1, i j}
\leq  {\bm{\mu}_1}^T P_{\mathcal{O}}{\bm{\mu}_1}$.
By   condition (C5$'$) and noting that $n_1/N \to \varphi_0 \in (0,1)$ as $N \to \infty$, we have
$E(B_{x212}^{(s,t)})^2/{p}=O(n_1^{-1}){\bm{\mu}}_1^T P_{\mathcal{O}}{\bm{\mu}}_1 = o\big(p^{1/2}n_1^{-3/2}\big)
=o(1)$ as $N \to \infty$.
Consequently, we have  $ E( B_{x21}^{(s,t)} )^2=o(p)$ as $N \to \infty$.
Following the similar procedure, we can show that
$ E( B_{x22}^{(s,t)} )^2=o(p)$ as $N \to \infty$ for $s \neq t$.

\vskip12pt
\noindent{\bf Part-II:  Proof of ${B_{x3}=o_p(p)}$ }
\vskip12pt

By (\ref{B.X2.expand}), we have
\begin{align}
E(|B_{x3}|)
&=
E\big(
\big| ({\bm X_{1}} -{\bar{\bm X}^{(1,2)}})^T
( {P}_{1,\mathcal{O}}^{(1,2)} - P_{\mathcal{O}})
{\bm X}_{2} \big|\,
\big|({\bm X_{2}} - {\bar{\bm X}^{(1,2)}})^T
( {P}_{1,\mathcal{O}}^{(1,2)} - P_{\mathcal{O}})
 {\bm X}_{1}\big|
\big)   \nonumber  \\
&\leq
E\big(
\big| B_{x21}^{(1,2)}+B_{x22}^{(1,2)} \big|\,
\big|B_{x21}^{(2,1)} +B_{x22}^{(2,1)}\big|
\big)   \nonumber  \\
&\leq
\Big[ E \big( B_{x21}^{(1,2)}\big)^2 \Big]^{\frac{1}{2}}
\Big[ E \big( B_{x21}^{(2,1)}\big)^2 \Big]^{\frac{1}{2}}
+\Big[ E \big( B_{x21}^{(1,2)}\big)^2 \Big]^{\frac{1}{2}}
\Big[ E \big( B_{x22}^{(2,1)} \big)^2 \Big]^{\frac{1}{2}}    \nonumber \\
&~~~~+\Big[ E \big( B_{x22}^{(1,2)}\big)^2 \Big]^{\frac{1}{2}}
\Big[ E \big( B_{x21}^{(2,1)} \big)^2 \Big]^{\frac{1}{2}}
+\Big[ E \big( B_{x22}^{(1,2)}\big)^2 \Big]^{\frac{1}{2}}
\Big[ E \big( B_{x22}^{(2,1)} \big)^2 \Big]^{\frac{1}{2}}.
\end{align}
Note that
$E\big( B_{x21}^{(1,2)}\big)^2=E \big( B_{x21}^{(2,1)}\big)^2$
and
$E\big( B_{x22}^{(1,2)}\big)^2=E \big( B_{x22}^{(2,1)}\big)^2$.
Also in the proof of Part-I,  we have
$E\big( B_{x21}^{(s,t)} \big)^2=o(p)$
and
$E\big( B_{x22}^{(s,t)} \big)^2=o(p)$ as $N \to \infty$.
This shows that $B_{x3}=o_p(p)$ as $N \to \infty$.


Following the similar procedure as that for $L_{x1}$, we can show that
\begin{align*}
L_{y1}
=
\frac{1}{n_2 (n_2-1)} \sum_{s=1}^{n_2} \sum_{t\neq s}^{n_2}
( {\bm Y_{s}} -{\bar{\bm Y}^{(s,t)}})^T \widehat{P}_{2,\mathcal{O}}^{(s,t)} {\bm Y}_{t}
( {\bm Y_{t}} -{\bar{\bm Y}^{(s,t)}})^T \widehat{P}_{2,\mathcal{O}}^{(s,t)} {\bm Y}_{s}
\end{align*}
is a ratio consistent estimator of ${\rm tr}(\Lambda_1^2)$ as $N \to \infty$.
This completes the proof of Lemma \ref{lemma2_1}.
\hfill$\Box$
\end{appendices}

\newpage

%
%
%


\end{document}